\documentclass{iopart}

\usepackage[english]{babel}
\usepackage{graphicx}
\usepackage{amsfonts}
\usepackage{amsbsy}
\usepackage{amssymb}
\usepackage[latin1]{inputenc}

\usepackage{xcolor}
\definecolor{red}{rgb}{0.7,0,0}
\definecolor{green}{rgb}{0.,0.35,0.}
\definecolor{blue}{rgb}{0.2,0.2,0.7}

\newcommand{\ket}[1]{|{#1}\rangle}
\newcommand{\bra}[1]{\langle{#1}|}
\newcommand{\braket}[2]{\langle{#1}|{#2}\rangle}
\newcommand{\mymod}[1]{\left|{#1}\right|}
\newcommand{\myvec}[1]{{\bf {#1}}}

\newcommand{\ud}{\mathrm{d}}
\providecommand{\openone}{\leavevmode\hbox{\small1\kern-3.8pt\normalsize1}}

\eqnobysec

\begin{document}

\title[Ultracold Dipolar Gases in Optical Lattices]{Ultracold Dipolar Gases in Optical Lattices}

\author{C. Trefzger$^{1}$, C. Menotti$^{1,2}$, B. Capogrosso-Sansone$^3$, and M. Lewenstein$^{1,4}$}

\address{$^1$ ICFO-Institut de Ciencies Fotoniques, Mediterranean Technology Park, 08860 Castelldefels (Barcelona), Spain}
\address{$^2$ INO-CNR BEC Center and Dipartimento di Fisica, Universit\`a di Trento, 38123 Povo, Italy}
\address{$^3$ Institute for Theoretical Atomic, Molecular and Optical Physics, Harvard-Smithsonian Center of Astrophysics, Cambridge, MA, 02138}
\address{$^4$ ICREA-Instituci\'o Catalana de Recerca i Estudis Avan\c cats, Lluis Companys 23, 08010 Barcelona, Spain}

\ead{christian.trefzger@lkb.ens.fr}

\begin{abstract}
This tutorial is a theoretical work, in which we study the physics
of ultra-cold dipolar bosonic gases in optical lattices. Such
gases consist of bosonic atoms or molecules that interact via
dipolar forces, and  that are cooled below the quantum degeneracy
temperature, typically in the nK range. When such a degenerate
quantum gas is loaded into an optical lattice produced by standing
waves of laser light, new kinds of physical phenomena occur. These
systems realize then extended Hubbard-type models, and can be
brought to a strongly correlated regime. The physical properties
of such gases, dominated by the long-range, anisotropic
dipole-dipole interactions, are discussed using the mean-field
approximations, and exact Quantum Monte Carlo techniques (the Worm
algorithm).
\end{abstract}

\submitto{Journal of Physics B}
\maketitle
\tableofcontents

\section{Introduction}
In 1989, M. Fisher {\it et. al.} predicted that the homogeneous
Bose-Hubbard model (BH)
exhibits the Superfluid-Mott insulator
(SF-MI) quantum phase transition \cite{BB:Fisher}. In 2002 the
transition between these two phases were observed experimentally
for the first time with cold atomic gases in the group of I.
Bloch, T. Esslinger and T. H\" ansch \cite{BB:Bloch00}. The
experimental realization of a dipolar Bose-Einstein condensate
(BEC) of Chromium by the group of T. Pfau \cite{BB:Pfau00,
BB:Pfau01, BB:Kaz01}, and the recent progresses in trapping and
cooling of dipolar molecules by the groups of D. Jin and J. Ye
\cite{BB:Molecule01, BB:Molecule02, BB:Molecule03}, have opened
the path towards ultra-cold quantum gases with dominant dipole
interactions. A natural evolution, and present challenge, on the
experimental side is then to load dipolar BECs into optical
lattices and study strongly correlated ultracold dipolar lattice
gases.

Studies of BH models with interactions extended to nearest
neighbors had pointed out that novel quantum phases, like
supersolid (SS) and checker board phases (CB) are expected
\cite{BB:Goral02, BB:Kovrizhin, BB:Batrouni, BB:Sengupta}. Due to
the long-range character of the dipole-dipole interaction, which
decays as the inverse cubic power of the distance, it is necessary
to include more than one nearest neighbor to have a faithful
quantitative description of dipolar systems. In fact, longer-range
interactions tend to allow for and stabilize more novel phases.

In this work we first study BH models with dipolar interactions,
going beyond the ground state search. We consider a
two-dimensional (2D) lattice where the dipoles are polarized
perpendicularly to the 2D plane, resulting in an isotropic
repulsive interaction. We use the mean-field approximations and a
Gutzwiller Ansatz which are quite accurate and suitable to
describe this system. We find that dipolar bosonic gas in 2D
lattices exhibits a multitude of insulating metastable states,
often competing with the ground state, similarly to a disordered
system. We study in detail the fate of these metastable states, in
particular what is their lifetime due to tunneling.
Moreover, we find that the ground state is characterized by
insulating checkerboard-like states with fractional filling
factors $\nu$ (average number of particles per site) that depend
on the cut-off used for the interaction range. We confirm this
prediction by studying the same system with Quantum Monte Carlo
methods (the Worm algorithm). In this case no cut-off for the
dipolar interaction is used, and we find evidence for a Devil' s
staircase in the ground state, i.e. insulating phases which appear
at all rational $\nu$ of the underlying lattice. We also find
regions of parameters where the ground state is a supersolid,
obtained by doping the solids either with particles or vacancies.
Recently \cite{BB:Cooper}, a complete devil' s staircase has been
predicted in the phase diagram of a one-dimensional dipolar Bose
gas.

In this work, we also investigate how the previous scenario
changes by considering a multi-layer structure. We focus on the
simplest situation composed of two 2D layers in which the dipoles
are polarized perpendicularly to the planes; the dipolar
interaction is then repulsive for particles laying on the same
plane, while it is attractive for particles at the same lattice
site on different layers. Instead we consider inter-layer
tunneling to be suppressed, which makes the system analogous to a
bosonic mixture in a 2D lattice. Our calculations show that
particles pair into composites, and demonstrate the existence of
the novel Pair Super Solid (PSS) quantum phase.

Moreover, we study a 2D lattice where the dipoles are free to
point in both directions perpendicularly to the plane, which
results in a nearest neighbor repulsive (attractive) interaction
for aligned (anti-aligned) dipoles. We find regions of parameters
where the ground state of the system exhibits insulating phases
with ferromagnetic or anti-ferromagnetic ordering, as well as with
rational values of the average magnetization. Evidence for the
existence of a Counterflow Super Solid (CSS) quantum phase is also
presented.

Our predictions have direct experimental consequences, and we hope
that they will be soon checked in experiments with ultracold
dipolar atomic and molecular gases.

Although this paper reports on many novel results and predictions,
it is written in a tutorial form. The reader will have a
possibility to learn first the basic introduction to dilute Bose
gases, and in  particular dipolar Bose gases. We will explain very
carefully various mean field methods that can be applied to
describe approximately diluted dipolar Bose gases in 2D and 3D.
These methods are elementary, but become more and more complex
for the extended Hubbard models, and in particular for the models
with infinite range interactions. In the chapters about multiple
layers and up-down mixtures we will present and explain in detail
methods  of deriving  effective low-energy Hamiltonians using
second order degenerate perturbation theory. Last, but not least,
we will present a very pedagogical introduction to QMC methods:
path integral Monte Carlo method and the Worm algorithm.

\section{Dipolar Bose gas in optical lattices}
\label{CH:Dipolar1}

\subsection{Optical lattices}
\label{SEC:OL} An optical lattice is an artificial crystal of
light, resulting from the interference pattern of two or more
counter propagating laser beams \cite{BB:Bloch}. The wavelengths
$\lambda_i$ of the laser beams determine the spatial periodicity
of the crystal; for example, two lasers of equal wavelengths
$\lambda_x$ propagating along $x$ but in opposite directions,
produce a standing wave with an intensity pattern $I(x)$ which is
spatially periodic with periodicity $\lambda_x/2$. An optical
lattice can trap neutral atoms by exploiting the energy shift
induced by the radiation on the atomic internal energy levels.

The electric field $\myvec{E}(\myvec{r},t) =
2E_0\cos\left(\myvec{k}\cdot\myvec{r} - \omega_L t\right)$ of a
monochromatic laser oscillating with frequency $\omega_L$,
interacts with a neutral atom, of spatial dimensions much smaller
compared to the wavelengths of the light $\lambda_i = 2\pi/k_i,\;
(i=x,y,z)$, through the Hamiltonian
\begin{equation}
\label{EQ:H_int}
\hat{H}_{\rm int}(t) = -\myvec{d} \cdot \myvec{E}(\myvec{r},t),
\end{equation}
where $\myvec{d} = -e\sum_i\myvec{r}_i$ is the electric dipole
moment of the atom, $\myvec{r}_i$ the positions of the atomic
electrons of charge $e$. With Hamiltonian (\ref{EQ:H_int}), one
can easily calculate the energy correction to the ground state of
the atom, by means of perturbation theory. The first order
correction vanishes because the dipole operator is odd with
respect to space inversion ($\myvec{r}_i \rightarrow -
\myvec{r}_i$), therefore the first non zero contribution is given
by the second order correction
\begin{equation}
\Delta E(\myvec{r}) = -\frac{1}{2}\alpha(\omega_L) \langle \myvec{E}(\myvec{r},t)^2\rangle_t
\end{equation}
where
\begin{equation}
\label{EQ:Polarizability}
\alpha(\omega_L) = \sum_\gamma \mymod{\bra{\gamma}\myvec{d}\cdot \myvec{\hat{\epsilon}} \ket{g}}^2\left(\frac{1}{E_\gamma - E_g + \hbar\omega_L} + \frac{1}{E_\gamma - E_g - \hbar\omega_L}\right),
\end{equation}
is the atomic polarizability \cite{BB:Pethick,BB:Maciek}, and
$\langle \cdots \rangle_t$ denotes a time average over one
oscillation period of the electric field \footnote{more
specifically $\langle \cdots \rangle_t = \frac{1}{t}\int_0^t
\cdots \ud t$ where $t = n\pi/\omega_L$, $n=1,2,\cdots$}. In the
last expression the energies in the denominators are the
unperturbed energies of the atom, where $g$ is the ground state,
the sum runs over all excited states $\gamma$, and
$\myvec{\hat{\epsilon}}$ is the unit vector in the direction of
the electric field. In a typical experiment the laser light is far
off resonance, which means that the laser frequency is close to
one of the unperturbed excited states (e.g $E_e = \hbar
\omega_e$), but does not induce any real transition. In such a
situation, one can take only the smallest of the denominators
(\ref{EQ:Polarizability}), and the polarizability becomes
inversely proportional to the laser detuning from resonance
$\hbar\Delta = \hbar\omega_L - (E_e - E_g)$
\begin{equation}
\label{EQ:OmegaL}
\alpha(\omega_L) \simeq - \frac{\mymod{\bra{e}\myvec{d}\cdot \myvec{\hat{\epsilon}} \ket{g}}^2}{\hbar\Delta}.
\end{equation}
In this situation, the energy shift is given by
\begin{equation}
\label{EQ:Shift}
\Delta E(\myvec{r}) = -\frac{1}{2} \alpha(\omega_L) \langle \myvec{E}(\myvec{r},t)^2 \rangle_t \propto \frac{I(\myvec{r})}{\hbar\Delta},
\end{equation}
where $I(\myvec{r})$ is the intensity of the laser. In the dressed
atom picture, the energy shift (\ref{EQ:Shift}) is interpreted as
an effective potential $V_{\rm opt}(\myvec{r}) = \Delta
E(\myvec{r})$, that follows the spatial pattern of the laser field
intensity, in which the atom moves. In this picture, the atom then
feels a force
\begin{equation}
\myvec{F}_{\rm dipole} = -\myvec{\nabla} V_{\rm opt}(\myvec{r}),
\end{equation}
that attracts it towards the regions of high intensity for the so
called red-detuned lasers (i.e. $\Delta < 0$), while a
blue-detuned light (i.e. $\Delta > 0$) pushes the atom out of the
regions of high intensity. In the literature this force is called
the dipole force, as it is the resulting interaction of the
induced atomic dipole moment with the spatially varying electric
field of the light. Note that in order to reduce heating caused by
inelastic scattering, i.e. photon absorption and spontaneous
emission processes, a large detuning is required because the
photon scattering rate scales as $I(\myvec{r})/\Delta^2$. In the
limit of large detuning an optical lattice is therefore
non-dissipative, which makes it a basic tool to manipulate cold
neutral atoms.

\begin{center}
\begin{figure}[h]
\begin{center}
\includegraphics[width=0.6\linewidth]{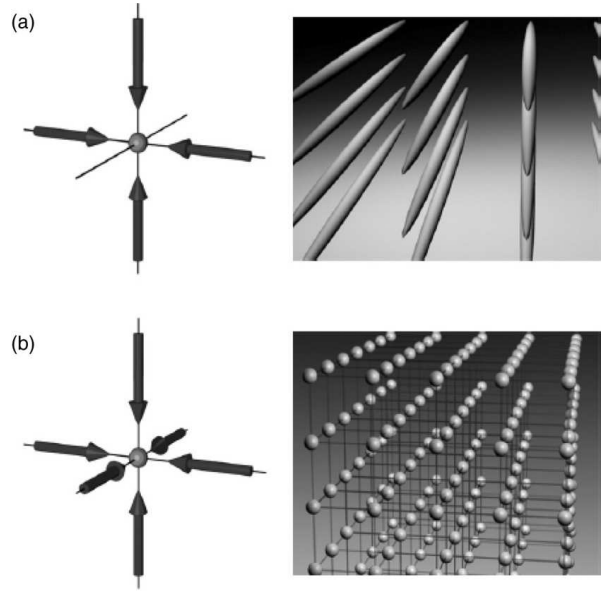}
\end{center}
\caption{Picture of an optical potential. (a) 2D square lattice of quasi 1D traps; (b) a 3D cubic lattice, picture courtesy of I. Bloch \cite{BB:Bloch}.}
\label{FIG:Lattice}
\end{figure}
\end{center}

For example, the simplest case of a one-dimensional lattice is
obtained by the superposition of two lasers propagating in
opposite directions, with electric fields linearly polarized, say
in the $z$ direction,and given by
\begin{eqnarray}
E_z(x,t) &=& 2E_0\cos\left(k_x x - \omega_L t\right) + 2E_0\cos \left(-k_x x - \omega_L t\right) \nonumber \\
         &=& 4E_0\cos\left(k_x x\right)\cos(\omega_L t).
\end{eqnarray}
The time average over one period of oscillation of the electric field gives then $\langle E_z(x,t)^2 \rangle_t = 2E_0\cos^2(k_x x)$, which yields
to the spatially varying optical potential
\begin{equation}
V_{\rm opt}(x) = V_{0,x}\cos^2(k_x x),
\end{equation}
with periodicity $\lambda_x/2 = \pi/k_x$, and $V_{0,x}=2E_0\alpha(\omega_L)$ from Eq. (\ref{EQ:OmegaL}).
The generalization to the two dimensional (2D) or three dimensional case (3D) is straightforward (see e.g. \cite{BB:Pethick}). For
example in Fig. \ref{FIG:Lattice} two different geometries are shown.

\subsection{Theory of dilute Bose gases}
\label{SEC:TDBG}
In this section we recall some basic theory of a dilute gas of neutral bosonic particles at temperature $T$ well below the degeneracy temperature. The type of particles we consider here can be atoms or molecules.

For a dilute gas, the interparticle separation (typically of the
order of $10^2$ nm for alkali atoms \cite{BB:Pethick}) is an order
of magnitude larger than the length scales associated with the
atom-atom interaction. In other words, a dilute gas of density $n$
is a very rarefied gas in which the "spatial extension" of an atom
is much smaller than the average volume per particle $n^{-1}$.
Because of this condition, the two-body interaction dominates the
physics while three-body or more are very unlikely and essentially
not important. The two-body interatomic potential $V(\myvec{r})$
depends on the type of particles one considers, the relative
distance between the atoms $\myvec{r} = \myvec{r}_1 - \myvec{r}_2$
and their internal states. For alkali atoms, the potential is
strongly repulsive for small atomic separations while for large
atomic distances it is dominated by the van der Waals attractions
that decay as $-C_6/\myvec{r}^6$, where the coefficient $C_6$
depends on the atomic species.

Here we will consider only elastic scattering, where the internal
states of the two atoms do not change in the collision process. If
the temperature of the gas is very low, i.e. $T \rightarrow 0$,
the kinetic energy of the particles is very small compared to the
centrifugal barrier and only $s$-wave scattering takes place.
Therefore, the only important parameter is the scattering length
given by
\begin{equation}
\label{EQ:ScatteringL}
a_s = \frac{m}{4\pi\hbar^2}\int \ud^3r V(\myvec{r}),
\end{equation}
with $m$ being the mass of the atoms. This quantity has the
dimensions of a length, and for $a_s > 0$ has the physical
interpretation of the radius the atoms would have if they were
considered to be perfect billiard balls. The condition for the
diluteness of the gas then reads
\begin{equation}
\label{EQ:GasParameter}
n|a_s|^3 \ll 1
\end{equation}
where $n$ is the density of the gas and $n|a_s|^3$ is called the
gas parameter. One can invert expression (\ref{EQ:ScatteringL})
and think of an effective contact interaction between the two
particles proportional to the scattering length, and given by
\begin{equation}
\label{EQ:ContactI}
V_{\rm eff}(\myvec{r}) =  g\, \delta^{(3)}(\myvec{r}),
\end{equation}
where $g$ is defined as
\begin{equation}
\label{EQ:g}
g = \frac{4\pi\hbar^2 a_s}{m},
\end{equation}
and $\delta$ is the Dirac delta function. Note also that since the
effective interaction depends only on the scattering length, it is
repulsive (attractive) for positive (negative) $a$, and it can be
dynamically modified for example in alkali atoms just by varying
an external magnetic field near a Feshbach resonance.

\subsubsection{The Gross-Pitaevskii equation}{--- }
The quantum state of a gas of $N$ particles is described by the
many-body wavefunction $\Psi(\myvec{r}_{\rm 1},\myvec{r}_{\rm
2},...,\myvec{r}_{\rm N})$, and the time evolution of the system
is determined by the Schr\"odinger equation. In a BEC, one can
describe the dynamics of the condensate just through the
Gross-Pitaevskii (GP) equation \cite{BB:Pitaevskii,BB:Pethick}
given by
\begin{equation}
\label{EQ:GP}
i\hbar \frac{\partial}{\partial t} \Psi_{\rm 0} (\myvec{r},t) = \left(-\frac{\hbar^2\myvec{\nabla}^2}{2m} + V_{\rm ext} (\myvec{r})
+ g |\Psi_{\rm 0} (\myvec{r},t)|^2\right) \Psi_{\rm 0} (\myvec{r},t),
\end{equation}
where $\Psi_{\rm 0} (\myvec{r},t)$ is the BEC wavefunction, also called the order parameter. The interaction between particles has been taken into account in a mean-field approximation by the term $g |\Psi_{\rm 0} (\myvec{r},t)|^2$, where $g$ is given in Eq. \ref{EQ:g}. $V_{\rm ext} (\myvec{r})$ is an external trapping potential, and
the order parameter is normalized to the total number of particles, i.e. $N = \int \ud^3r |\Psi_{\rm 0} (\myvec{r},t)|^2$.
Equation (\ref{EQ:GP}) was independently derived by Gross and Pitaevskii in 1961, it is one of the main theoretical tools for investigating dilute weakly interacting Bose gases at low temperatures, and it has the typical form of a mean field equation where the order
parameter must be calculated in a self-consistent way.
The GP equation has proven to be a very useful tool to describe the physics of weakly
interacting Bose-Einstein atomic condensates in the early ages of this field. With this
formalisms, and its extension to include small fluctuations given by Bogoliubov theory,
one can describe accurately, among others, the collective excitations of the systems, the
response to rotations including the formation of vortices, the propagation of sound, the
presence of dynamical instabilities. Generally speaking, the GP treatment is well suited
in the regime of full coherence, when a single macroscopically occupied matterwave
correctly describes the system. At the end of the '90, few years after the creation of
the first alkali BECs in the lab, the need of "going beyond GP" started to be very
strongly felt, due to the theoretical interest and experimental possibility of going into
the strongly correlated regime. In fact, the presence of strong interactions, strong
rotations and/or special trapping potentials can limit the validity of the GP equation.
For
instance a strong confinement in one or two dimensions can reduce the system to an
effectively 2D or 1D one. A strong rotation combined with interactions can lead to
quantum Hall physics. Also the presence of a deep optical lattices,
when the combined effect of interactions and trapping potential leads to a
"fragmentation" of the condensate, requires more sophisticated descriptions.

In this tutorial we are interested in describing the physics of
Bosons trapped in a periodic optical potential ($V_{\rm opt}$) and
eventually also confined in a magnetic harmonic trap ($V_{\rm
ho}$), the total external field being given by the sum
\begin{eqnarray}
V_{\rm ext} (\myvec{r}) &=& V_{\rm opt} (\myvec{r}) + V_{\rm ho} (\myvec{r}) \nonumber \\
                        &=& \sum_{i=x,y,z} V_{0,i} \cos^2 (k_{i} r_{i}) + \frac{1}{2}m \sum_{i=x,y,z} \omega_{\rm i}^2 r_{i}^2,\label{EQ:TrapPotential}
\end{eqnarray}
where $(V_{0,x},V_{0,y},V_{0,z})$ is the depth of the optical lattice in the three spatial directions and $(\omega_{\rm x},\omega_{\rm y},\omega_{\rm z})$
the frequencies of the harmonic trap. In order to describe the physics of Bosons trapped in the potential (\ref{EQ:TrapPotential}), we need to
"go beyond" the GP equation, and we will devote the following sections to this purpose.

\subsubsection{Bose-Hubbard model}{--- }
\label{SUBSEC:BHM}
The starting point of our discussion is Hamiltonian (\ref{EQ:Hamiltomian2NDQ}), written in the second quantization formalism in terms of the creation
and annihilation operators for Bosons, $\hat{\psi}^\dag(\myvec{r})$ and $\hat{\psi}(\myvec{r})$ respectively, and given by the expression
\begin{equation}
\label{EQ:Hamiltomian2NDQ}
\hat{H} = \int \ud^3r \hat{\psi}^\dag(\myvec{r}) \left[-\frac{\hbar^2 \nabla^2}{2m} + V_{\rm ext}(\myvec{r})
+ \frac{g}{2}\hat{\psi}^\dag(\myvec{r})\hat{\psi}(\myvec{r})  - \mu \right] \hat{\psi}(\myvec{r}),
\end{equation}
where the first term in square brackets is the kinetic energy, $V_{\rm ext}(\myvec{r}) = V_{\rm opt} (\myvec{r}) + V_{\rm ho} (\myvec{r})$ is the external trapping potential (\ref{EQ:TrapPotential}) and we have used the simplified contact interaction (\ref{EQ:ContactI}). We work in the grand canonical ensemble and the chemical
potential $\mu$ fixes the total number of particles. Additionally, we assume the harmonic confinement to change on a scale larger than the one of the optical lattice, such that we can consider the effect of the magnetic trapping to be constant over a single site of the lattice.

In this formalism, the field operators can be written in the basis of single-particle wave functions $\{\Phi_n(\myvec{r})\}_{n}$,
where $n$ is a complete set of single particle quantum numbers
\begin{equation}
\label{EQ:Fields}
\eqalign{
\hat{\psi}(\myvec{r})         &= \sum_{n} \Phi_n(\myvec{r}) \hat{a}_n \\
\hat{\psi}^\dag(\myvec{r}) &= \sum_{n} \Phi^*_n(\myvec{r}) \hat{a}^\dag_n,
}
\end{equation}
with $\hat{a}^\dag_n$ and $\hat{a}_n$ being the creation and annihilation operators for the mode $n$, i.e. $\hat{a}^\dag_n \ket{n} = \sqrt{n+1}\ket{n+1}$ and $\hat{a}_n \ket{n} = \sqrt{n}\ket{n-1}$. Also, the field operators satisfy the usual commutation relations for Bosons
\begin{equation}
\eqalign{
[\hat{\psi}(\myvec{r}),\hat{\psi}^\dag(\myvec{r^\prime})]
&= \sum_{n=0}^\infty\Phi_n(\myvec{r})\Phi^*_n(\myvec{r^\prime}) = \delta^{3}(\myvec{r} - \myvec{r^\prime}), \\
[\hat{\psi}(\myvec{r}),\hat{\psi}(\myvec{r^\prime})] &=  [\hat{\psi}^\dag(\myvec{r}),\hat{\psi}^\dag(\myvec{r^\prime})]  = 0.
}
\end{equation}

It is well known \cite{BB:Auerbach}, that the spectrum of a single particle in a periodic potential is
characterized by bands of allowed energies and energy gaps, and the single particle wave functions
are described by Bloch functions $\Phi_{\alpha\myvec{k}}(\myvec{r})$ with band index $\alpha$ and quasi-momentum $\hbar\myvec{k}$.
Alternatively, there exists a complementary single-particle basis given by the Wannier functions \cite{BB:Auerbach,BB:Korepin}
$w_\alpha(\myvec{r} - \myvec{R}_i)$, where $\myvec{R}_i$ is a lattice vector pointing at site $i$ and $w_\alpha(\myvec{r})$ are defined as
the Fourier transform of Bloch functions
\begin{equation}
w_\alpha(\myvec{r}) = \frac{1}{\sqrt{N_S}}\sum_{\myvec{k}}e^{-i\myvec{k}\cdot\myvec{r}}\Phi_{\alpha\myvec{k}}(\myvec{r}),
\end{equation}
where $N_S$, is the total number of sites in the lattice. The Wannier functions form a complete orthonormal set, so one may write the field operators
(\ref{EQ:Fields}) as
\begin{equation}
\label{EQ:OPWannier}
\eqalign{
\hat{\psi}(\myvec{r}) &= \sum_{\alpha\myvec{k}} \Phi_{\alpha\myvec{k}}(\myvec{r}) \hat{a}_{\alpha\myvec{k}} = \sum_{\alpha,i} w_\alpha(\myvec{r} - \myvec{R}_i) \hat{a}_{\alpha,i} \\
\hat{\psi}^\dag(\myvec{r}) &= \sum_{\alpha\myvec{k}} \Phi^*_{\alpha\myvec{k}}(\myvec{r}) \hat{a}^\dag_{\alpha\myvec{k}} = \sum_{\alpha,i} w^*_\alpha(\myvec{r} - \myvec{R}_i) \hat{a}^\dag_{\alpha,i}.
}
\end{equation}
Wannier functions are useful in the case of deep optical lattices where tight binding approximation apply. The big advantage of using Wannier functions
$w_\alpha(\myvec{r} - \myvec{R}_i)$ is that they are localized and centered around the lattice site pointed by $\myvec{R}_i$.

\begin{center}
\begin{figure}[h]
\begin{center}
\includegraphics[width=1\linewidth]{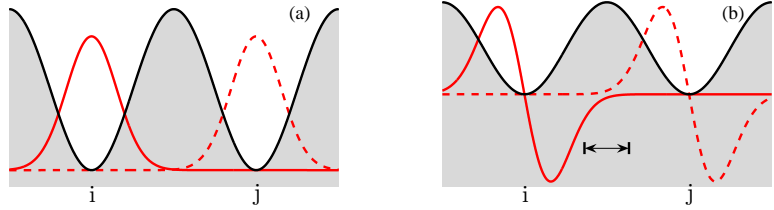}
\end{center}
\caption{(a) Two Gaussians localized on neighboring sites $i$ and $j$ of an optical lattice having negligible overlap. (b) The first derivatives of the
Gaussian functions instead, show a negative overlap in the region indicated by the arrow, which leads to a positive matrix element $J_{ij} > 0$.}
\label{FIG:Wannier}
\end{figure}
\end{center}

If the temperature of the system is low enough, and the interactions between the particles is not sufficient to induce transitions between the bands, one may restrict only to the first Bloch band because the particles have  insufficient energy to overcome the gap that separates the first band from the others. This amounts to keep
in (\ref{EQ:OPWannier}) only the lowest of the $\alpha$ indices, which we omit for simplicity of notation in the following. Therefore the Hamiltonian (\ref{EQ:Hamiltomian2NDQ}) becomes
\begin{equation}
\hat{H} = -\sum_{i,j} J_{ij} \; \hat{a}^\dag_i \hat{a}_j + \sum_{i,j,k,l}\frac{U_{ijkl}}{2} \; \hat{a}^\dag_i \hat{a}^\dag_j \hat{a}_k \hat{a}_l - \sum_{i,j} \mu_{ij} \; \hat{a}^\dag_i \hat{a}_j,
\end{equation}
where the quantities in the sums are given by
\begin{eqnarray}
J_{ij} &=& -\int \ud^3r w^*(\myvec{r} - \myvec{R}_i) \left[-\frac{\hbar^2 \nabla^2}{2m} + V_{\rm opt}(\myvec{r})\right]w(\myvec{r} - \myvec{R}_j) \label{EQ:Coeff01}\\
U_{ijkl} &=& g \int \ud^3r w^*(\myvec{r} - \myvec{R}_i)w^*(\myvec{r} - \myvec{R}_j)w(\myvec{r} - \myvec{R}_k)w(\myvec{r} - \myvec{R}_l)  \label{EQ:Coeff02} \\
\mu_{ij} &=&  \int \ud^3x w^*(\myvec{r} - \myvec{R}_i)\left[\mu - V_{\rm ho}(\myvec{r})\right]w(\myvec{r} - \myvec{R}_j).\label{EQ:Coeff03}
\end{eqnarray}
The Wannier functions are localized on the lattice sites, the deeper the lattice the more localized they are. For a sufficiently deep optical potential, in Eq. (\ref{EQ:Coeff02}) and (\ref{EQ:Coeff03}) the dominant contributions are given by $U_{iiii}$ and $\mu_{ii}$.
For the kinetic part (\ref{EQ:Coeff01}), there is a constant contribution given by $J_{ii}$ and due to the presence of the derivative in the integration, there is also a positive matrix element for nearest neighboring sites $J_{ij} > 0$. The two situations are qualitatively shown in Fig. (\ref{FIG:Wannier}) where we have approximated
the Wannier functions with two Gaussians respectively localized at site $i$ and $j$ of the lattice. However, we stress that the picture provided by Gaussian functions is only
qualitative. In fact, in order to be quantitatively correct, one needs to calculate the proper matrix elements with Wannier functions.
With the above considerations, we can now write the celebrated Bose-Hubbard Hamiltonian in the form
\begin{equation}
\label{EQ:HamiltonianBH}
\hat{H}_{\rm BH} = - J \sum_{\langle ij \rangle} \; \hat{a}^\dag_i \hat{a}_j + \frac{U}{2}\sum_i \hat{n}_i(\hat{n}_i-1)  - \sum_i \mu_i \hat{n}_i,
\end{equation}
where $\langle ij \rangle$ indicates sum over nearest neighbors, the tunneling coefficient $J = J_{ij} = J_{ji}$ for hermiticity, the on-site interaction
$U = g\int \ud^3r \mymod{w(\myvec{r})}^4$, $\hat{n}_i = \hat{a}^\dag_i \hat{a}_i$ is the number operator at site $i$, and we have neglected $J_{ii}$ since
it gives a constant contribution for each site.
The harmonic confinement, since it is assumed to be constant across one lattice site, has been taken into account in the chemical potential as
\begin{equation}
\label{EQ:LocalMu}
\mu_i = \mu - \frac{1}{2}m\vec{\omega}^{\;2}\cdot(\myvec{R}_i - \myvec{R_0})^2,
\end{equation}
where $\myvec{R_0}$ is the center of the harmonic trap with frequencies given by $\vec{\omega} = (\omega_{\rm x},\omega_{\rm y},\omega_{\rm z})$ in the three directions. The second
term on the right hand side of Eq. (\ref{EQ:LocalMu}) is practically a chemical potential that differs from site to site and it is often called the local chemical potential.

\begin{center}
\begin{figure}[h]
\begin{center}
\includegraphics[width=0.6\linewidth]{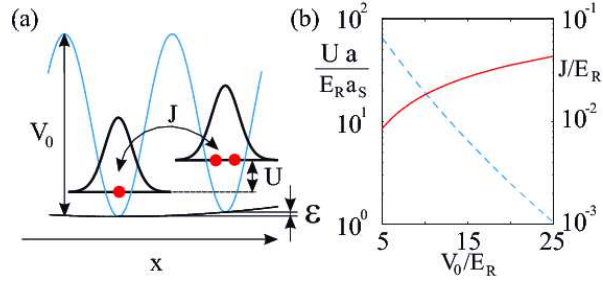}
\end{center}
\caption{(a) Schematic representation of a 1D optical lattice, where $\varepsilon$ is a local chemical potential; (b) scaled on-site $U$ (solid line) and tunneling coefficient $J$ (dashed line)
dependence on the optical potential depth $V_0$. The on-site interaction is multiplied by $a/a_s (\gg 1)$, where $a = \lambda/2$ is the lattice period and $a_s$ is the
s-wave scattering length for atoms of equal mass $m$. Figure courtesy of D. Jaksch \cite{BB:Jacksch}.}
\label{FIG:Jaksch}
\end{figure}
\end{center}

For a one dimensional optical lattice $V_{\rm opt}(x) = V_0\sin^2(kx)$ with wavevector $k = 2\pi/\lambda$, Fig. \ref{FIG:Jaksch} shows
both the on-site interaction $U$ (solid line) and the tunneling coefficient $J$ (dashed line) as a function of the optical lattice depth $V_0$, where all the
quantities are measured in terms of the recoil energy $E_R = \hbar^2k^2/2m$, that is the energy acquired by the atom after absorbing a photon with
momentum $\hbar k$. The lattice parameters $U$ and $J$ were calculated numerically in e.g. \cite{BB:Jacksch} for different values of $V_0$.
From Fig. \ref{FIG:Jaksch} (b), it is clear that it is possible to change the tunneling coefficient $J$ over a wide range, going from a situation of
practically isolated lattice sites at $V_0 = 25 E_R$ up to a regime in which particles can tunnel from site to site at $V_0 = 5 E_R$, only by changing
the optical potential depth by a few tens of recoil energies, and leaving the on-site interaction $U$ practically unchanged.

\subsection{Dipolar Bose gas}
\label{SEC:DipolarBoseGas}

\subsubsection{Properties of the dipole-dipole interaction}
\begin{center}
\begin{figure}[ht]
\begin{center}
\includegraphics[width=0.6\linewidth]{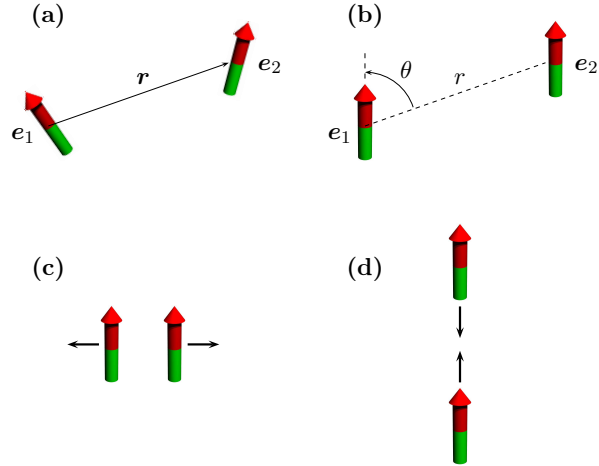}
\end{center}
\caption{(a) Two dipoles, $1$ and $2$, directed along unit vectors $\myvec{e}_1$ and $\myvec{e}_2$ and separated by a distance $r$.
(b) Polarized dipoles, for which the interaction depends on the angle $\theta$ between the direction of the dipoles and the interparticle separation
$r$. This results in a repulsive interaction for $\theta = \pi/2$ (c), and attractive for $\theta = 0, \pi$ (d).}
\label{FIG:Thierry}
\end{figure}
\end{center}

{--- }
Two particles $1$ and $2$ in a three dimensional space, at relative distance $\myvec{r}$ and with dipole moments
along the unit vectors $\myvec{e_1}$ and $\myvec{e_2}$ as in Fig. \ref{FIG:Thierry} (a), interact through the dipole-dipole interaction such that
their interaction energy is given by
\begin{equation}
\label{EQ:Dipole}
U_{\rm dd} (\myvec{r}) = \frac{C_{\rm dd}}{4\pi} \frac{(\myvec{e}_1 \cdot \myvec{e}_2)r^2 - 3(\myvec{e}_1 \cdot \myvec{r})(\myvec{e}_2 \cdot \myvec{r})}{r^5},
\end{equation}
where $r = |\myvec{r}|$, and $U_{\rm dd} (\myvec{r}) = U_{\rm dd} (\myvec{-r})$.
The dipolar coupling constant $C_{\rm dd}$ is different for particles having a permanent magnetic dipole
moment $\mu$, and for particles having a permanent electric dipole moment $d$, and is respectively given by
\begin{equation}
\label{EQ:Cdd}
C_{\rm dd} = \left\{
\begin{tabular}{ll}
$\mu_0 \mu^2$ & magnetic \\
$d^2/\varepsilon_0$ & electric,
\end{tabular}
\right.
\end{equation}
where $\mu_0$ is the vacuum permeability, and $\varepsilon_0$ is the vacuum permittivity.

The dipole-dipole interaction (\ref{EQ:Dipole}) has a {\it long-range} character; this is because at large distances it decays as $U_{\rm dd} \sim 1/r^3$,
contrary to the typical van der Waals potential that behaves like $U_{\rm vdW} \sim -1/r^6$. Also, from (\ref{EQ:Dipole}) it is easy to see the
{\it anisotropic} property of this interaction; for polarized atoms, i.e. all dipoles pointing in the same direction, the interaction reduces to
\begin{equation}
\label{EQ:DipoleI}
U_{\rm dd} (\myvec{r}) = \frac{C_{\rm dd}}{4\pi} \frac{1-3\cos^2\theta}{r^3},
\end{equation}
where $\theta$ is the angle between the dipole and the relative distance of the particles, as in Fig \ref{FIG:Thierry} (b). The interaction is repulsive for
$\theta = \pi/2$ as the example of Fig \ref{FIG:Thierry} (c), and attractive for $\theta = 0$ as shown in Fig \ref{FIG:Thierry} (d). The situation is reversed
for anti-parallel dipoles, where a minus sign appears in front of Eq. (\ref{EQ:DipoleI}), and therefore the interaction is attractive for $\theta = \pi/2$
while $\theta = 0$ gives rise to repulsion.

The scattering properties of ultracold atoms, in the
simple case of isotropic van der Waals interactions, are entirely described by the s-wave scattering length and the potential can be replaced by
the effective contact interaction (\ref{EQ:ContactI}). In the presence of a dipolar interaction as (\ref{EQ:Dipole}), because
of its long range (decay as $1/r^3$) and anisotropic character (strong dependence on the relative angles between the dipoles), all partial waves
contribute to the scattering problem and also partial waves with different angular momenta couple with each other. While for Fermions,
replacing the real potential (\ref{EQ:Dipole}) with an effective dipolar interaction as (\ref{EQ:ContactI}) is reasonable \cite{BB:Baranov},
 for Bosons this is not obvious, and in recent years it has been the subject of intensive studies \cite{BB:Derevianko, BB:Derevianko01, BB:Derevianko02, BB:Greg}.
In the presence of an optical lattice, it has been recently
argued \cite{BB:Yuan} that in a 1D geometry, replacing the real dipolar potential with an effective interaction as (\ref{EQ:ContactI}) is reasonable
as long as the optical lattice is shallow enough. However, in the most general case it is necessary to account for the full expression of the
dipole-dipole interaction potential (\ref{EQ:Dipole}).

\subsubsection{Polarized dipoles in anisotropic harmonic traps}{--- }
\label{SEC:PolarizedDD}
We now move  to the description of a BEC of polarized dipoles, pointing along the $z$ axis.
For polarized dipolar BECs, due to the anisotropy of the dipolar interactions, the geometry of the trapping potential plays a fundamental role,
first in determining the spatial distribution of the density, and second in the stability of the gas.

\begin{center}
\begin{figure}[h]
\begin{center}
\includegraphics[width=0.85\linewidth]{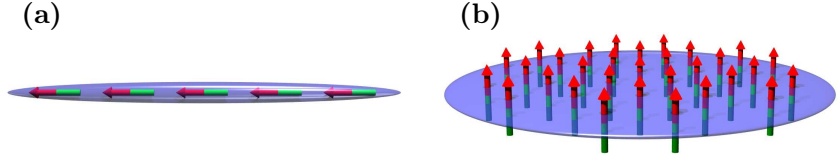}
\end{center}
\caption{Polarized dipoles in anisotropic harmonic potentials. (a) in a cigar shaped trap elongated in the direction of polarization, the resulting dipolar interaction is attractive, and (b) in a pancake trap with a strong confinement in the direction of polarization, the dipolar interactions are repulsive.}
\label{FIG:DipolarScenario}
\end{figure}
\end{center}

\begin{center}
\begin{figure}[h]
\begin{center}
\begin{tabular}{cc}
\includegraphics[width=0.5\linewidth]{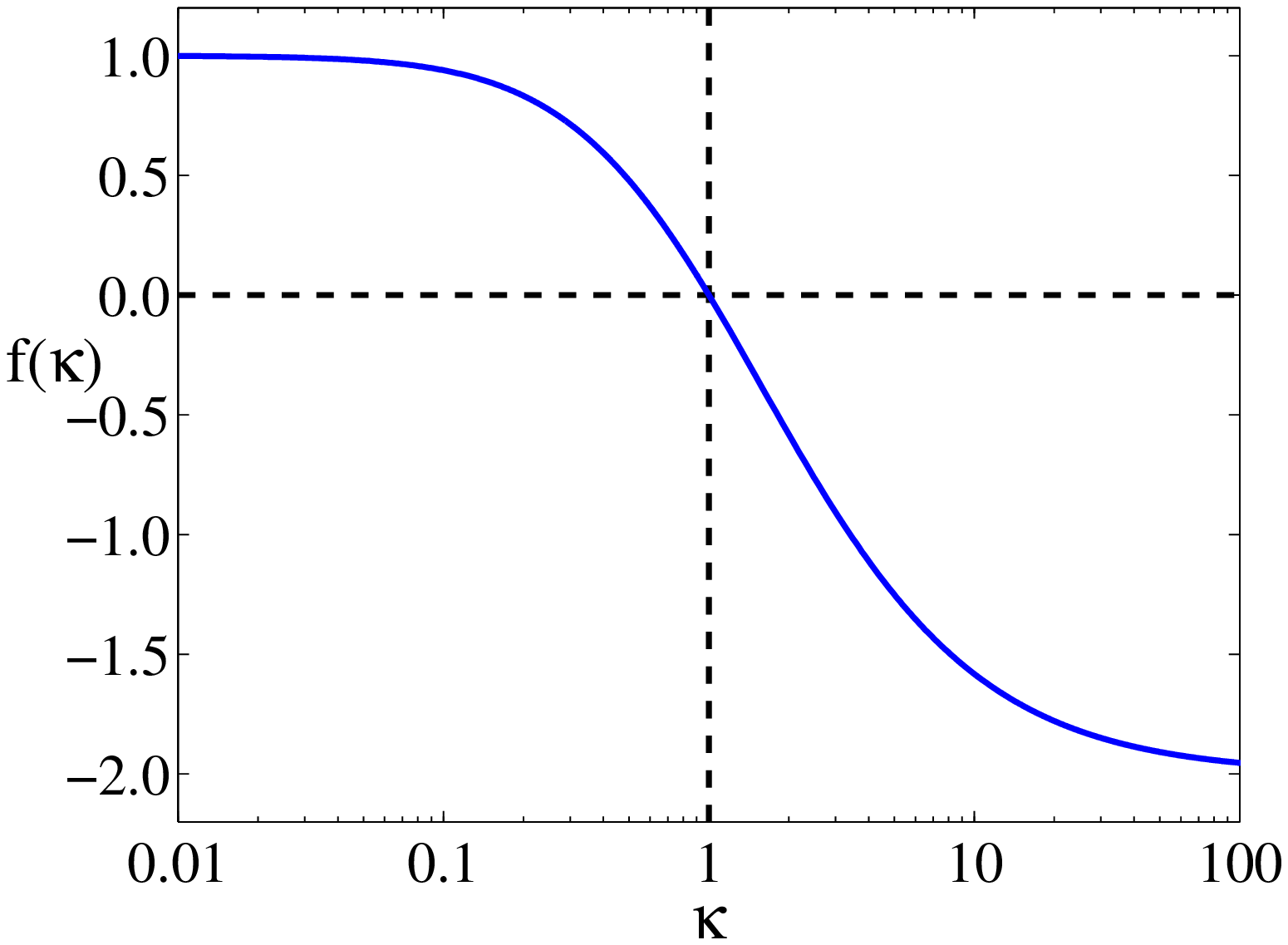} & \includegraphics[width=0.45\linewidth]{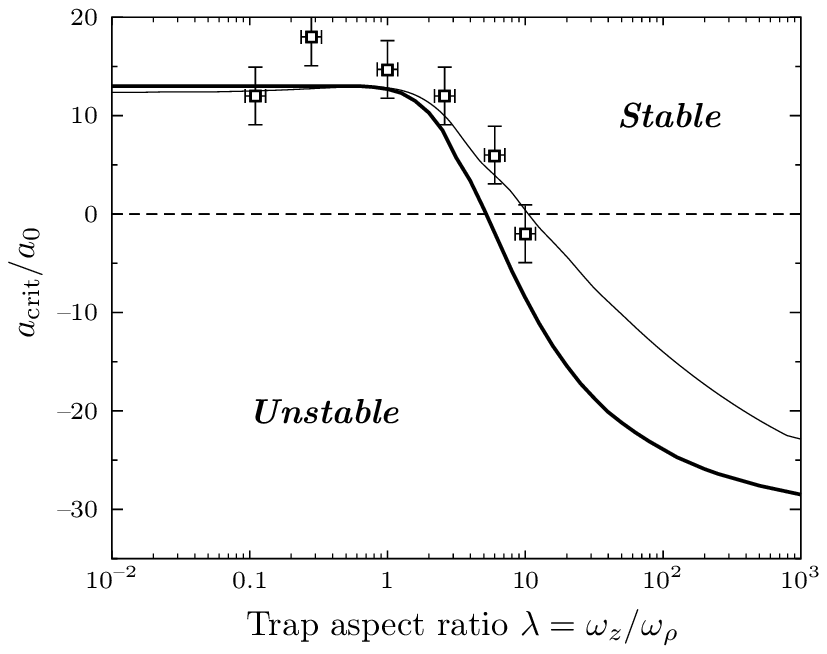}
\end{tabular}
\end{center}
\caption{(Left panel), $\kappa$ dependence of the $f(\kappa)$ function that appears in the mean-field dipolar interaction. (Right panel) Stability
diagram of a dipolar condensate: the thin line is the solution for $a_{\rm crit}(\lambda)/a_0$ calculated with the Gaussian Ansatz (\ref{EQ:Gaussian}),
where $a_0$ is the s-wave scattering length, while the thick line is
the numerical solution of the GP equation \cite{BB:BohnDipolar}. The dots with error bars are experimental data \cite{BB:Koch}.}
\label{FIG:FKappa}
\end{figure}
\end{center}

Qualitatively, there are two extreme scenarios depending on the shape of the confining potential, shown in Fig. \ref{FIG:DipolarScenario}:
(i) for a cigar-shaped trap elongated along the $z$ axis, i.e. with an aspect ratio between the axial $\omega_z$ and radial frequencies $\omega_\rho=\omega_x=\omega_y$ given by $\lambda = \omega_z/\omega_\rho \ll 1$, the density is mainly distributed along the polarization axis and the effect of dipole-dipole interaction is mostly attractive, which might lead to an instability of the gas even in the presence of a weak repulsive contact interaction; (ii), for a pancake-shaped trap, which is strongly confining along the $z$ axis, i.e. $\lambda \gg 1$, the dipolar interaction is mostly repulsive and the BEC is always stable for repulsive contact interactions and might be stable even for attractive contact interactions. In an intermediate situation in which the confining potential is perfectly spherical, the density distribution is then isotropic and the dipole-dipole interaction averages out to zero, which leads to a stable BEC for repulsive contact interactions. One can switch between one or the other scenario, just by adjusting the frequency of the confining potential along the $z$ axis with respect to the axial $x$ and $y$, and therefore it is natural to expect that
for any given $\lambda$ there is a critical value for the scattering length $a_{\rm crit}$ below which the BEC is unstable \cite{BB:Goral01}.
One can quantitatively describe the above scenarios starting from the Hamiltonian of the system, which in  the presence of the dipole-dipole interaction (\ref{EQ:DipoleI}) reads
\begin{eqnarray}
\hat{H} &=& \int \ud^3r \hat{\psi}^\dag(\myvec{r}) \left[-\frac{\hbar^2 \nabla^2}{2m} + V_{\rm ext}(\myvec{r})
+ \frac{g}{2}\hat{\psi}^\dag(\myvec{r})\hat{\psi}(\myvec{r})  - \mu \right] \hat{\psi}(\myvec{r}) \nonumber \\
&+&  \frac{1}{2}\int \ud^3r_1 \ud^3r_2 \hat{\psi}^\dag(\myvec{r}_1)\hat{\psi}^\dag(\myvec{r}_2) U_{\rm dd}(\myvec{r}_1 - \myvec{r}_2)
                                 \hat{\psi}(\myvec{r}_1)\hat{\psi}(\myvec{r}_2). \label{EQ:DipolarH2NDQ}
\end{eqnarray}
With the same approximations used to derive the Gross-Pitaevskii equation, one can write the Boson field operator
$\hat{\psi}(\myvec{r}) = \Psi_{\rm 0}(\myvec{r}) + \delta\hat{\psi}(\myvec{r})$ as a sum of a classical field $\Psi_{\rm 0}(\myvec{r})$, the condensate wave function,  plus the non condensate component $\delta\hat{\psi}(\myvec{r})$ \cite{BB:Pitaevskii}. By neglecting the fluctuations $\delta\hat{\psi}(\myvec{r})$, one can calculate the energy
of the BEC given by
\begin{eqnarray}
E\big[\Psi_{\rm 0} \big] &=& \int \Big[ -\frac{\hbar^2}{2m} |\myvec{\nabla}\Psi_{\rm 0}(\myvec{r})|^2 + V_{\rm ext}(\myvec{r}) |\Psi_{\rm 0}(\myvec{r})|^2 + \frac{g}{2} |\Psi_{\rm 0}(\myvec{r})|^4 \nonumber \\
              && + \frac{1}{2} |\Psi_{\rm 0}(\myvec{r})|^2 \int  U_{\rm dd}(\myvec{r} - \myvec{r}^\prime) |\Psi_{\rm 0}(\myvec{r}^\prime)|^2 \ud^3r^\prime \Big]\ud^3r,
\label{EQ:EFunctional}
\end{eqnarray}
where the macroscopic wavefunction $\Psi_{\rm 0}$ is normalized to the total number of particles $N$.
Within a variational Ansatz, we assume the condensate wave function to be a Gaussian of axial width $\sigma_z$ and radial width $\sigma_x = \sigma_y = \sigma_\rho$, normalized to the total number of particles $N$, namely
\begin{equation}
\label{EQ:Gaussian}
\Psi_{\rm 0}(z,\myvec{\rho}) = \sqrt{\frac{N}{\pi^{3/2}\sigma_\rho^2 \sigma_z a_{\rm ho}^3}}\exp \left[-\frac{1}{2a_{\rm ho}^2} \left( \frac{\rho^2}{\sigma_\rho^2} + \frac{z^2}{\sigma_z^2}\right)  \right],
\end{equation}
where $a_{\rm ho} = \sqrt{\hbar/(m\bar{\omega})}$ is the harmonic oscillator length with average trap frequency $\bar{\omega} = (\omega_\rho^2\omega_z)^{1/3}$.
Therefore, inserting Ansatz (\ref{EQ:Gaussian}) into the energy functional Eq. (\ref{EQ:EFunctional}), after integration we find the energy of the BEC to be a function of the widths of the Gaussians, namely
\begin{equation}
\label{EQ:EnergyDD}
E_{0}(\sigma_z,\sigma_\rho) = E_{\rm kin} + E_{\rm trap} + E_{\rm contact} + E_{\rm dd},
\end{equation}
with the kinetic energy
\begin{equation}
\label{EQ:EKinDD}
E_{\rm kin} = \frac{N \hbar \bar{\omega}}{4} \left(\frac{2}{\sigma_\rho^2} + \frac{1}{\sigma_z^2}\right),
\end{equation}
the potential energy due to the trap
\begin{equation}
\label{EQ:ETrapDD}
E_{\rm trap} = \frac{N \hbar \bar{\omega}}{4\lambda^{2/3}} \left(2\sigma_\rho^2 + \lambda^2\sigma_z^2\right),
\end{equation}
the contact interaction energy given by
\begin{equation}
\label{EQ:EContactDD}
E_{\rm contact} = \frac{\hbar \bar{\omega}}{\sqrt{2\pi} a_{\rm ho}}\frac{1}{\sigma_\rho^2\sigma_z} a_s,
\end{equation}
and the contribution coming from the dipolar term
\begin{equation}
\label{EQ:EddDD}
E_{\rm dd} = -\frac{\hbar \bar{\omega} a_{\rm dd}}{\sqrt{2\pi} a_{\rm ho}}\frac{1}{\sigma_\rho^2\sigma_z} f(\kappa).
\end{equation}
We have introduced the {\it dipolar length} $a_{\rm dd} = \frac{C_{\rm dd}m}{12\pi\hbar^2}$, with $C_{\rm dd}$ given in Eq. (\ref{EQ:Cdd}), which measures the absolute strength of the dipolar interaction, $\kappa = \sigma_\rho/\sigma_z$ is the {\it aspect ratio} of the density distribution, and the function $f$ is given by
\begin{equation}
f(\kappa) = \frac{1+2\kappa^2}{1-\kappa^2} - \frac{3\kappa^2\textrm{artanh}{\sqrt{1-\kappa^2}}}{(1-\kappa^2)^{3/2}}.
\end{equation}
While the integrals needed to obtain (\ref{EQ:EKinDD},\ref{EQ:ETrapDD},\ref{EQ:EContactDD}) are easy to calculate since they contain only Gaussian functions and their derivatives, the integral to get (\ref{EQ:EddDD}) is not straightforward due to the presence of the dipolar potential $U_{\rm dd}(\myvec{r}_1 - \myvec{r}_2)$. See
 section \ref{SUBSEC:DIESpherical} for more details.
In the left panel of Fig. \ref{FIG:FKappa}, we show the behavior of the function $f(\kappa)$ as $\kappa$ is continuously varied from $\kappa = 10^{-2}$ to
$\kappa = 10^{2}$. The function takes the asymptotic values of $f(0)=1$, $f(\infty)=-2$, and it vanishes for $\kappa = 1$, which implies that for a spherical
density distribution the dipole-dipole mean-field interaction (\ref{EQ:EddDD}) averages out to zero. Therefore
we notice that it is possible to control the strength and the sign of the mean-field dipolar interaction just by adjusting the aspect ratio $\lambda$ between the axial and
the radial frequencies of the confining trap. The total interaction energy is provided by the sum of the contact (\ref{EQ:EContactDD}) plus the dipolar interaction energy (\ref{EQ:EddDD}), given by
\begin{equation}
\label{EQ:OnSiteUGaussian}
E_{\rm int} = \frac{\hbar \bar{\omega} a_{\rm dd}}{\sqrt{2\pi} a_{\rm ho}}\frac{1}{\sigma_\rho^2\sigma_z} \left( \frac{a_s}{a_{\rm dd}} - f(\kappa)\right).
\end{equation}
The stability of the gas requires a repulsive interaction $E_{\rm int} > 0$, which
leads to the condition
\begin{equation}
\frac{a_s}{a_{\rm dd}} - f(\kappa) > 0,
\end{equation}
and can be adjusted ad-hoc by changing the frequencies of the trap in the three directions.
To determine the stability threshold $a_{\rm crit}(\lambda)$, one needs to minimize the energy (\ref{EQ:EnergyDD}) with
respect to the variational parameters $\sigma_\rho$ and $\sigma_z$ for fixed values of $N$, $\lambda$ and $\bar{\omega}$. The results are
summarized in the right panel of Fig. \ref{FIG:FKappa} as a thin line, while the thick line represents more accurate results calculated from
solving numerically the Gross-Pitaevskii equation \cite{BB:BohnDipolar}. The dots with error bars correspond to experimental data \cite{BB:Koch}.

\subsubsection{Mean-field dipolar interaction in a spherical trap}{--- }
\label{SUBSEC:DIESpherical}
In order to calculate the mean-field dipolar interaction energy (\ref{EQ:EddDD}), we insert the Gaussian Ansatz (\ref{EQ:Gaussian}) into the
the second of the integrals (\ref{EQ:EFunctional}), and we get to the expression
\begin{equation}
E_{\rm dd} = \frac{1}{2}\int \ud^3r_1 \ud^3r_2 \rho(\myvec{r}_1)  \rho(\myvec{r}_2) U_{\rm dd}(\myvec{r}_1 - \myvec{r}_2),
\end{equation}
with $\rho(\myvec{r}) = |\Psi_{\rm 0}(\myvec{r})|^2$ being the condensate density at $\myvec{r}$.
The last integral can be simplified by means of the convolution theorem \cite{BB:Goral01, BB:Goral} which states
\begin{equation}
\int \ud^3r_2 U_{\rm dd}(\myvec{r}_1 - \myvec{r}_2) \rho(\myvec{r}_2) = \mathcal{F}^{-1}\left\{ \widetilde{U_{\rm dd}}(\myvec{k}) \; \widetilde{\rho}(\myvec{k})\right\},
\end{equation}
where $\widetilde{U_{\rm dd}}(\myvec{k})$ and $\widetilde{\rho}(\myvec{k})$ are the Fourier transform respectively of the dipole-dipole potential
and the density. $\mathcal{F}^{-1}$ indicates the inverse Fourier transform, and using its definition we can write
\begin{eqnarray}
E_{\rm dd} &=& \frac{1}{2}\int \ud^3r_1 \rho(\myvec{r}_1) \frac{1}{(2\pi)^3}\int \ud^3k \, \widetilde{U_{\rm dd}}(\myvec{k}) \;\widetilde{\rho}(\myvec{k}) e^{i\myvec{k}\cdot \myvec{r}_1} \nonumber \\
            &=& \frac{1}{2(2\pi)^3} \int \ud^3k \, \widetilde{U_{\rm dd}}(\myvec{k}) \;\widetilde{\rho}^{\;2}(\myvec{k}), \label{EQ:Edd}
\end{eqnarray}
where in the last step we have used the relation $\widetilde{\rho}(\myvec{k}) = \widetilde{\rho}(\myvec{-k})$.

The Fourier transform of the dipole-dipole interaction (\ref{EQ:DipoleI}) is given by
\begin{equation}
\widetilde{U_{\rm dd}}(\myvec{k}) = \int \ud^3r U_{\rm dd}(\myvec{r}) e^{-i\myvec{k}\myvec{r}}  = C_{\rm dd} (\cos^2\gamma - 1/3),
\end{equation}
where $\gamma$ is the angle between $\myvec{k}$ and the polarization direction, and $C_{\rm dd}$ is given by expression (\ref{EQ:Cdd}) \cite{BB:Goral01}.
In order to evaluate the integral of Eq. (\ref{EQ:Edd}) we need to insert the condensate wave function, which
in the simple case of isotropic potential ($\sigma_z = \sigma_\rho$) becomes a product of three Gaussian distributions with
equal widths $\sigma$. Therefore the Fourier transform of the condensate density is readily calculated as
\begin{equation}
\widetilde{\rho}(\myvec{k}) = \frac{N}{(\sqrt{\pi}\sigma a_{\rm ho})^3} \int \ud^3r\, e^{-i\myvec{k}\cdot\myvec{r}}e^{-\frac{\myvec{r}^2}{\sigma^2 a_{\rm ho}^2}} = \exp \left[- \frac{\sigma^2 a_{\rm ho}^2}{4} \myvec{k}^2\right].
\end{equation}
This expression has to be inserted into the integral (\ref{EQ:Edd}), which can be easily evaluated in polar $(r,\gamma,\varphi)$ coordinates \footnote{remember, $\gamma$ is the angle between the polarization axis $z$ and $\myvec{k}$.}, giving
\begin{eqnarray}
E_{\rm dd} &=& \frac{NC_{\rm dd}}{2}\int \sin\gamma \ud\gamma \ud\varphi k^2\ud k (\cos^2\gamma - 1/3)
\exp \left[- \frac{\sigma^2 a_{\rm ho}^2}{2} \myvec{k}^2\right] \nonumber \\
&=&  NC_{\rm dd} 2\pi \int \ud k k^2 \exp \left[- \frac{\sigma^2 a_{\rm ho}^2}{2} \myvec{k}^2\right] \int_{-1}^{+1} \ud x (x^2 - 1/3) \nonumber\\
&=& 0,
\end{eqnarray}
where we have performed the change of variable $x = \cos\gamma$. The generalization to anisotropic density distributions is mathematically
more demanding but in principle straightforward, and leads to Eq. (\ref{EQ:EddDD}).

\subsubsection{Extended Bose-Hubbard model}{--- }
\label{SEC:EBHM}
As in section \ref{SUBSEC:BHM}, we expand the field operators
in the basis of Wannier functions (\ref{EQ:OPWannier}), and we keep only the lowest index corresponding to the first Bloch band.
Within this approximation the first line of Eq. (\ref{EQ:DipolarH2NDQ}) leads to the Bose-Hubbard Hamiltonian (\ref{EQ:HamiltonianBH}).
Instead the dipolar term gives rise to a further contribution
\begin{equation}
\label{EQ:HDip2Q}
\hat{H}_{\rm dd} =  \frac{1}{2}\int \ud^3r_1 \ud^3r_2 \hat{\psi}^\dag(\myvec{r}_1)\hat{\psi}^\dag(\myvec{r}_2) U_{\rm dd}(\myvec{r}_1 - \myvec{r}_2)
                                 \hat{\psi}(\myvec{r}_1)\hat{\psi}(\myvec{r}_2),
\end{equation}
which after expansion of the field operators in the basis of Wannier functions, becomes
\begin{equation}
\hat{H}_{\rm dd} = \sum_{i,j,k,l} \frac{V_{ijkl}}{2}\; \hat{a}^\dag_i \hat{a}^\dag_j \hat{a}_k \hat{a}_l,
\end{equation}
and the matrix elements $V_{ijkl}$ are given by the integral
\begin{equation}
\label{EQ:VIntegral}
\fl
 V_{ijkl} = \int \ud^3r_1 \ud^3r_2 w^*(\myvec{r}_1 - \myvec{R}_i)w^*(\myvec{r}_2 - \myvec{R}_j)U_{\rm dd}(\myvec{r}_1 - \myvec{r}_2)
 w(\myvec{r}_1 - \myvec{R}_k)w(\myvec{r}_2 - \myvec{R}_l).
\end{equation}
The Wannier functions are centered at the bottom of the optical lattice wells with a spatial localization that we assume to be $\sigma$. For deep enough optical potentials we can assume $\sigma$ to be much smaller than the optical lattice spacing $d$, i.e. $\sigma \ll d$. In this limit, each function $w(\myvec{r} - \myvec{R}_i)$ is
 significantly non-zero for $\myvec{r} \sim \myvec{R}_i$, and the integral (\ref{EQ:VIntegral})
is significantly non-zero for the indices $i=k$ and $j=l$. Therefore there are two main contributions to the integral (\ref{EQ:VIntegral}):
the {\it off-site} matrix element $V_{ijij}$ corresponding to $k=i\neq j=l$, and the {\it on-site} $V_{iiii}$ when all the indices are equal.
Below we will explain the physical meaning of these two contributions.
\paragraph{Off-site ---}The dipolar potential $U_{\rm dd}(\myvec{r}_1 - \myvec{r}_2)$ changes slowly on the scale of $\sigma$, therefore one may approximate it with the constant $U_{\rm dd}(\myvec{R}_i - \myvec{R}_j)$ and take it out of the integration. Then the integral reduces to
\begin{equation}
V_{ijij} \simeq U_{\rm dd}(\myvec{R}_i - \myvec{R}_j) \int \ud^3r_1 \mymod{w(\myvec{r}_1 - \myvec{R}_i)}^2
\int \ud^3r_2 \mymod{w(\myvec{r}_2 - \myvec{R}_j)}^2,
\end{equation}
which leads to the off-site Hamiltonian
\begin{equation}
\hat{H}_{\rm dd}^{\rm off-site} = \sum_{i \neq j}\frac{V_{ij}}{2}\hat{n}_i \hat{n}_j.
\end{equation}
In the last expression $V_{ij} = U_{\rm dd}(\myvec{R}_i - \myvec{R}_j)$, $\hat{n}_i = \hat{a}^\dag_i \hat{a}_i$ is the bosonic number operator at site $i$, and the sum runs over all different sites of the lattice.
\paragraph{On-site ---}At the same lattice site $i$, where $|\myvec{r}_1 - \myvec{r}_2| \sim \sigma$, the dipolar potential changes very rapidly and
diverges for $|\myvec{r}_1 - \myvec{r}_2| \rightarrow 0$. Therefore the above approximation is not valid any more and the integral
\begin{equation}
\label{EQ:VInt}
V_{iiii} = \int \ud^3r_1 \ud^3r_2 \rho(\myvec{r}_1) U_{\rm dd}(\myvec{r}_1 - \myvec{r}_2) \rho(\myvec{r}_2),
\end{equation}
with $\rho(\myvec{r}) = \mymod{w(\myvec{r})}^2$ being the single particle density, has to be calculated taking into account the atomic spatial distribution
at the lattice site, similarly to what has been described in Sec. \ref{SEC:PolarizedDD} \footnote{Since $\myvec{R}_i$ is
a constant, we have renamed the variables as $\myvec{r}_u - \myvec{R}_i = \myvec{r}_u$ for $u = 1,2$.}.
We have already encountered this kind of integral in Sec. \ref{SEC:PolarizedDD}, and we have seen that, a part from a factor of $2$, the solution can be found by Fourier transforming, i.e.
\begin{eqnarray}
V_{iiii}  = \frac{1}{(2\pi)^3} \int \ud^3k \, \widetilde{U_{\rm dd}}(\myvec{k}) \;\widetilde{\rho}^{\;2}(\myvec{k}).
\end{eqnarray}
Which leads to an on-site dipolar contribution to the Hamiltonian of the type
\begin{equation}
\label{EQ:HddContact}
\hat{H}_{\rm dd}^{\rm on-site} = \sum_{i} \frac{V_{iiii}}{2} \hat{n}_i(\hat{n}_i-1).
\end{equation}
The extended Bose-Hubbard Hamiltonian is given by the sum
of the Bose-Hubbard (\ref{EQ:HamiltonianBH}) and the dipolar Hamiltonians calculated above, leading to the expression
\begin{equation}
\label{EQ:eBHH}
\fl
\hat{H}_{\rm eBH} = - J \sum_{\langle ij \rangle} \; \hat{a}^\dag_i \hat{a}_j + \frac{U}{2}\sum_i \hat{n}_i(\hat{n}_i-1)  - \sum_i \mu_i \hat{n}_i + \sum_{i \neq j} \frac{V_{ij}}{2} \hat{n}_i\,\hat{n}_j,
\end{equation}
where $U$ is now taken into account as an effective on-site interaction
\begin{equation}
\label{EQ:OnSiteU}
U = g\int \ud^3r \mymod{w(\myvec{r})}^4  + \frac{1}{(2\pi)^3} \int \ud^3k \, \widetilde{U_{\rm dd}}(\myvec{k}) \;\widetilde{\rho}^{\;2}(\myvec{k}),
\end{equation}
which contains the contribution of the contact potential, with $g$ given in Eq. (\ref{EQ:g}), plus the dipolar contribution coming from (\ref{EQ:HddContact}).
Approximating each lattice site with a tiny harmonic trap, and approximating the atomic density distribution with Gaussians, $U$ looks like
Eq. (\ref{EQ:OnSiteUGaussian}), and one can see that the resulting on-site interaction can be increased or decreased by changing the lattice confinement.

\section{Hubbard models: theoretical methods}
\label{CH:Methods}

\subsection{Superfluid$-$Mott insulator quantum phase transition in the Bose-Hubbard model}
\label{SEC:SF_MI}
Consider the Bose-Hubbard Hamiltonian as derived in Sec. (\ref{SUBSEC:BHM}),
\begin{equation}
\label{EQ:BH2}
\hat{H}_{\rm BH} = - J \sum_{\langle ij \rangle} \; \hat{a}^\dag_i \hat{a}_j + \frac{U}{2}\sum_i \hat{n}_i(\hat{n}_i-1)  - \mu \sum_i  \hat{n}_i,
\end{equation}
with a uniform chemical potential $\mu$, and a total number of Bosons given by the expectation value of the operator $\hat{N} = \sum_i \hat{n}_i$.  There are three parameters in this Hamiltonian, namely $J$, $U$ and $\mu$, but it is a convention to reduce the analysis of the phase diagram of $\hat{H}_{\rm BH}$ to the ratio of two of them over the third one, e.g. $J/U$ and $\mu/U$.

\begin{center}
\begin{figure}[h]
\begin{center}
\includegraphics[width=0.7\linewidth]{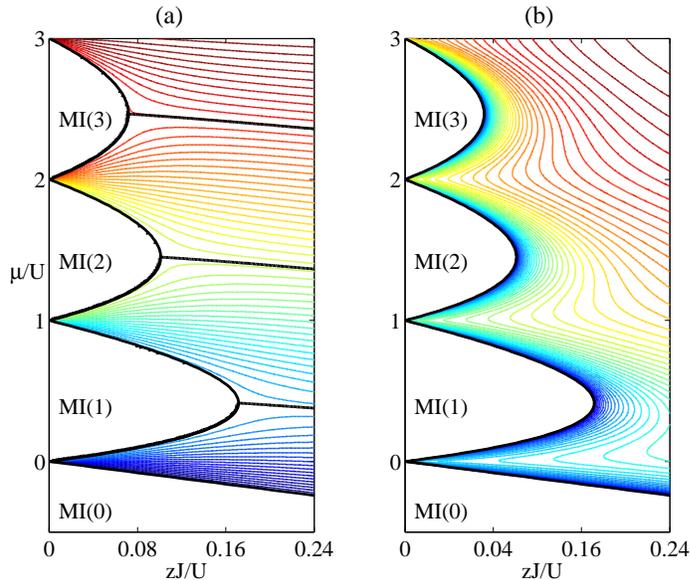}
\end{center}
\caption{Mean-field phase diagram of the Bose Hubbard Hamiltonian. (a) Contour plot of the density per site; (b) contour plot of the order parameter (see Sec. \ref{SEC:GMF}). MI($\bar{n}$) indicate a Mott insulating phase with fixed $\bar{n}$ atoms per site.}
\label{FIG:MILobes}
\end{figure}
\end{center}

The ground state of Hamiltonian (\ref{EQ:BH2}) is easily understood for two opposite regimes of parameters: (i) for {\it shallow lattices}, i.e. $U/J \ll 1$, the system is in a gapless superfluid phase (SF) characterized by on-site density fluctuations and the particles delocalized over the whole lattice; (ii) for {\it deep lattices}, i.e $J/U \ll 1$, and commensurate filling, on-site density fluctuations are completely suppressed, each site is occupied by an integer number of atoms $\bar{n}$, and the ground state is a product of single-site Fock states
\begin{equation}
\label{EQ:GSFock}
\ket{GS} = \ket{\bar{n},\bar{n},\cdots}.
\end{equation}
This filling is energetically favorable in the range of chemical potential $\bar{n} - 1 < \mu/U < \bar{n}$. The system is gapped and incompressible, as beautifully explained in the famous paper of Fisher {\it et. al.} \cite{BB:Fisher}, and it is called a Mott insulator MI($\bar{n}$). For small values of $J/U$ the MI($\bar{n}$) phase persists in a closed and finite area of the $J$ vs. $\mu$ plane \cite{BB:Fisher}, which is called the Mott lobe for MI($\bar{n}$). The larger $J/U$ value of the lobe is called the tip of the lobe or also critical point $(J/U)_{\rm c}$.
The critical point changes with the dimensionality and geometry of the system. In Fig. \ref{FIG:MILobes} we plot the first $\bar{n}=0,1,2,3$ insulating lobes, calculated for an infinite optical lattice within the mean-field approximation, which will be discussed in Sec. \ref{SEC:Semi}. The thick black lines enclose the lobes and mark the boundaries between the MI and SF phases. Outside the insulating lobes, the phase is SF. The colored lines of Fig. \ref{FIG:MILobes}(a) indicate
a contour plot of constant fractional density, while the thick black lines departing from the tip of the lobes and extending into the SF region, correspond to an integer value $\bar{n}$ of the density.

We will derive the mean-field Mott insulating lobes of Fig. \ref{FIG:MILobes} in a more rigorous way in Sec. \ref{SEC:Semi}, but for the moment we just list
the critical points $(J/U)_{\rm c}$, for the $\bar{n} = 1$ lobe, that have been estimated with different methods and for different dimensions of the lattice.
In one dimension, the critical point has been estimated to be $(J/U)_{\rm c} \simeq 0.29$ \cite{BB:Monien1} using Density Matrix Renormalization Group calculations (DMRG). In two dimensions, with quantum Monte Carlo calculations, the critical point has been estimated to be $(J/U)_{\rm c} \simeq 0.061$
\cite{BB:Wessel}, while in the three dimensional model the location of the critical point has been estimated with perturbative expansions \cite{BB:Monien2}, and quantum Monte Carlo simulations \cite{BB:Svistunov3D} to be at $(J/U)_{\rm c} \simeq 0.034$.
In the next section we will derive the mean-field lobes.

\subsection{The Gutzwiller mean-field approach}
\label{SEC:GMF}

The Gutzwiller mean-field approach provides an approximated many-body wave function for Hubbard-type Hamiltonians, of the form
\begin{equation}
\label{EQ:GW}
\ket{\Psi} = \prod_i \sum_{n=0}^{n_{\rm max}} f_n^{(i)} \ket{n}_i,
\end{equation}
where $\ket{n}_i$ represents the Fock state of $n$ atoms occupying the site $i$, $n_{\rm max}$ is a cut off in the maximum number of atoms per site, and $f_n^{(i)}$
is the probability amplitude of having the site $i$ occupied by $n$ atoms. The probability amplitudes are normalized to unity $\sum_n |f_n^{(i)}|^2 = 1$.
The Gutzwiller wave function (\ref{EQ:GW}) has been extensively used in the literature \cite{BB:Jacksch,BB:Krauth,BB:Rokhsar,BB:Jaksch02}, it predicts that there exists a critical value $(J/U)_{\rm mf}$ in a given range of $\mu$, below which the ground state is a product of single Fock states $f_n^{(i)} = \delta_{n,\bar{n}}$ with exactly $\bar{n}$ particles per site, as (\ref{EQ:GSFock}). Moreover, for $J/U > (J/U)_{\rm mf}$ the Gutwiller Ansatz predicts a superfluid ground state with fluctuating on-site particle number.
The Gutzwiller critical point, for $\bar{n} = 1$, is found to be $(J/U)_{\rm mf} = 1/5.8z$ \cite{BB:Zwerger}, where $z=\sum_{\langle j\rangle_i} 1$ is the number of nearest neighbor connections at each site of the lattice.
In table \ref{TAB:CriticalXXYY} we show the comparison of the critical points predicted by the Gutzwiller Ansatz for different dimensions of the system, with the more
precise ones discussed in Sec. (\ref{SEC:SF_MI}).
\begin{table}[htdp]
\caption{Comparison of the Gutzwiller critical points $(J/U)_{\rm mf}$ with the more precise, up to now, critical points $(J/U)_{\rm c}$, for different dimensions D of the system. }
\begin{center}
\begin{tabular}{ccccccc}
\br
D & &  $z$ & &  $(J/U)_{\rm mf}$ & &  $(J/U)_{\rm c}$ \\
\mr
1 & & 2 & &  0.0862 & &  0.29 \\
2 & &  4 & &  0.0431 & &  0.061 \\
3 & &  6 & &  0.0287 & &  0.034 \\
\br
\end{tabular}
\end{center}
\label{TAB:CriticalXXYY}

\end{table}
From the comparison, one can deduce that the Guzwiller is unsatisfactory for 1D systems ($z=2$) while it is satisfactory for a 3D one ($z = 6$).
Also, in the limit of $J/U \rightarrow \infty$ the difference between the Gutzwiller predictions and the exact results are negligible \cite{BB:Zwerger}. Summarizing,
the Gutzwiller predictions are exact in the two limiting cases of $J/U \rightarrow 0$, and $J/U \rightarrow \infty$, while for intermediate cases the performance of the Gutzwiller approach strongly depends on the dimensionality of the lattice, since it does not correctly account for the quantum fluctuations at the phase transition.

\begin{center}
\begin{figure}[h]
\begin{center}
\includegraphics[width=1\linewidth]{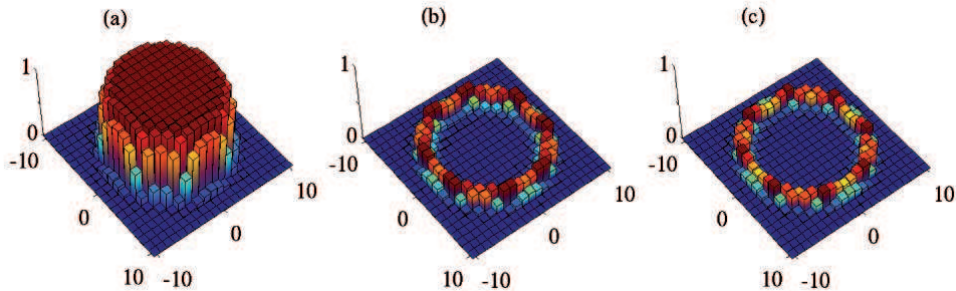}
\end{center}
\caption{Mean-field ground state of a 2D optical lattice, calculated for $J=0.025U$, and $\mu_i = 0.45U - \Omega i^2$, where $\Omega = 8 \times 10^{-3} U/\hbar$ is the frequency of the harmonic oscillator confinement. (a) the vertical axis shows the value of the density at each site of the lattice, (b) the corresponding absolute value of the order parameter, and (c) on-site density fluctuations.}
\label{FIG:BHTrap}
\end{figure}
\end{center}

Two important quantities are the {\it order parameter} $\varphi_i = \bra{\Psi} \hat{a}_i \ket{\Psi}$, namely the expectation value of the Bosonic annihilation
operator at the $i$-th site of the lattice, and the {\it density fluctuations} at the site $i$ given by $\delta n_i = \bra{\Psi} \hat{n}_i^2 \ket{\Psi} - \bra{\Psi} \hat{n}_i \ket{\Psi}^2$. By using the Gutzwiller wavefunction (\ref{EQ:GW}) one gets the following expressions for the order parameter
\begin{equation}
\varphi_i = \sum_n \sqrt{n+1} f_n^{*(i)} f_{n+1}^{(i)},
\end{equation}
and for the density fluctuations
 \begin{equation}
\delta n_i = \sum_n n^2 |f_n^{(i)}|^2 - \Big[ \sum_n n |f_n^{(i)}|^2 \Big]^2.
\end{equation}
The order parameter $\varphi_i$ together with the density fluctuations $\delta n_i$ describe the phase of the system at the site $i$ of the lattice:
in the Mott phase $\varphi_i = 0$, and density fluctuations are suppressed $\delta n_i = 0$, while the superfluid phase is characterized by $\varphi_i \neq 0$ and presence of density fluctuations $\delta n_i \neq 0$. In the uniform system, the lattice is translationally invariant and therefore all sites are self-similar, which means that only one site is sufficient to determine the phase of the whole system. In Fig. \ref{FIG:MILobes}(b) we plot the absolute value of the order parameter $\varphi_i$ for such a system, in the $J/U$ vs. $\mu/U$ plane. The colored lines outside the insulating lobes correspond to a contour plot of constant non-zero value of $\varphi_i$ typical of the SF phase. Instead, in a non-uniform system as it is in the presence of an external confining harmonic potential, different phases can coexist. As an example, in Fig. \ref{FIG:BHTrap} we plot the density of the ground state (a), the order parameter at each site (b), and the density fluctuations (c) of a 2D lattice in the presence of a confining harmonic potential. Notice that the MI phase at the center of the harmonic trap $(0,0)$ is surrounded by a ring of SF phase, in a wedding-cake like structure, as first discussed in \cite{BB:Jacksch}. The ground state of Fig. \ref{FIG:BHTrap} was obtained within the mean-field approximation through the imaginary-time evolution technique, that will be discussed in Sec. \ref{SEC:DGA}.
In the presence of dipolar interactions, as we shall see later on, it is also necessary to account for non-uniform quantum phases, because even in the uniform system the presence of dipolar interactions may lead to spontaneous symmetry breaking of translational invariance on a scale larger than the lattice constant.

\subsubsection{Dynamical Gutzwiller approach}{--- }
\label{SEC:DGA}
The time dependent version of the Gutzwiller wavefunction (\ref{EQ:GW}) is obtained by allowing the Gutzwiller amplitudes to depend on time $f_n^{(i)} (t)$ \cite{BB:Jaksch02}.
Then the equations of motion for the amplitudes are readily obtained  by minimizing the action of the system, given by $S = \int \ud t \mathcal{L}$, with respect to the variational parameters $f_n^{(i)}(t)$ and their complex conjugates $f_{n}^{*(i)}(t)$. The Lagrangian of the system in the quantum state $\ket{\Psi}$,
is given by  \cite{BB:Perez}
\begin{equation}
\mathcal{L} = i\hbar\frac{\braket{\Psi}{\dot{\Psi}} - \braket{\dot{\Psi}}{\Psi}}{2} - \bra{\Psi} \hat{H}\ket{\Psi},
\end{equation}
where $\ket{\dot{\Psi}}$ is the time derivative of the wave function (\ref{EQ:GW}).
By equating to zero the variation of the action with respect to $f_{n}^{*(i)}$, one gets the equations
\begin{eqnarray}
i\hbar \frac{\ud}{\ud t} f_n^{(i)} &=& -J \Big[\bar{\varphi}_i \sqrt{n} f_{n-1}^{(i)} + \bar{\varphi}_i^* \sqrt{n+1} f_{n+1}^{(i)} \Big] \nonumber \\
                                           & & + \Big[\frac{U}{2}n(n-1) + n \sum_{j\neq i} V_{ij} \langle \hat{n}_j\rangle - \mu_i n\Big] f_n^{(i)}, \label{EQ:FDynamics}
\end{eqnarray}
where $\bar{\varphi}_i = \sum_{\langle j \rangle_i} \varphi_j$, the sum runs over all nearest neighbors $j$ of site $i$, $\langle \hat{n}_j\rangle = \bra{\Psi} \hat{a}_j^\dag \hat{a}_j \ket{\Psi}$ is the average particle number at site $j$, and the total number of particles is given by
$N = \sum_i \langle \hat{n}_i\rangle$. It is not difficult to verify the commutation relation $[\hat{N},\hat{H}_{\rm BH}] = 0$,
which implies that the total number of Bosons is a conserved quantity for the dynamics in real time \cite{BB:Sachdev}.
These equations are of mean-field type, because for each site $i$ the influence of the neighboring sites is taken into account in a mean way
into the "field" $\bar{\varphi}_i$ together with $\sum_{j\neq i} V_{ij} \langle \hat{n}_j\rangle$, which have to be determined self-consistently. Eqs. (\ref{EQ:FDynamics}) are a set of coupled equations, the coupling arising from the tunneling part, and can be written in the matrix form
\begin{equation}
\label{EQ:FDinamics2}
i\hbar \frac{\ud}{\ud t} \vec{f} = \mathcal{M}[\vec{f},\mu,U,J] \cdot \vec{f},
\end{equation}
where $\vec{f} = \left[ f_0^{(1)}, f_1^{(1)}, \cdots, f_n^{(i)}, \cdots f_{n_{\rm max}}^{(N_S)} \right]^{\rm T}$, is the vector of the Gutzwiller amplitudes ordered from site $1$ to site $N_S$, the latter being the total number of sites. It is worth noticing that the matrix $\mathcal{M}[\vec{f},\mu,U,J]$ is itself a functional of the coefficients $\vec{f}$ through the fields $\bar{\varphi}_i$ and $\sum_{j\neq i} V_{ij} \langle \hat{n}_j\rangle$, which have to be calculated in a self-consistent way.
Let us clarify this point with an example. Suppose we want to solve Eq. (\ref{EQ:FDinamics2})
between an initial time $t_i=0$ and a final time $t_f$, with a given initial condition $\vec{f}(0)$. We discretize the time interval in $N$ steps of size $\Delta t$, with $N$ finite, and define $t_s = s \Delta t$ such that $t_{s=0} \equiv 0$ and $t_{s=N} \equiv t_f$. Therefore, to calculate the solution at a certain point in time $\vec{f}(t_{s+1})$
we need to know the solution right at the preceding time $\vec{f}(t_s)$, with which we can compute the fields that in turn determine $\mathcal{M}[\vec{f}(t_s),\mu,U,J]$,
and the solution is readily found to be
\begin{equation}
\label{EQ:SolTimeStep}
\vec{f}(t_{s+1}) = e^{-i \mathcal{M}[\vec{f}(t_s),\mu,U,J] \Delta t/\hbar} \vec{f}(t_s).
\end{equation}
Starting from $s=0$, in $N+1$ steps we have determined the solution at the desired time $t_f$. At the computational level, this is the
simplest procedure one can implement to calculate the dynamics of the system. However, one needs to be careful in the choice of the time step $\Delta t$, especially
for fast-oscillating dynamics. In such cases, a Runge-Kutta with adaptive stepsize control has proven to be more efficient. Instead in the case of imaginary time evolution, which requires solving stiff dynamics, the simple procedure described above has shown to be enough accurate and faster.

Equations (\ref{EQ:FDinamics2}) can be solved in real time $t$ or also imaginary time $\tau = it$. The imaginary time evolution is a standard technique that has been thoroughly used, because due to dissipation is supposed to converge to the ground state of the system.
Two things are worth to be noticed. First, because the imaginary time evolution is not unitary, it does not conserve the norm of the Gutzwiller wavefunction, which has to be renormalized after each time step. Second, the total number of particles is not a conserved quantity anymore.
For dipolar Hamiltonians the imaginary time evolution does not always converge to the true ground state and it gets blocked in configurations which are a local minimum of the energy. On the one hand this makes it a difficult task to identify the ground state of such systems, and on the other hand it is a signature of the existence of metastable states as we will discuss in details in Sec. \ref{CH:MS:DipolarBosons}.

\subsubsection{Perturbative mean-field approach}{--- }
\label{SEC:Semi}
A more convenient method to determine the insulating phases of a dipolar Hamiltonian is to use a mean-field approach perturbative in $\varphi_i$.
From statistical mechanics, the expectation value of the annihilation operator at the $i$-th site is given \cite{BB:Tannoudji01} by the trace
\begin{equation}
\label{EQ:Distribution}
\varphi_i  = \langle \hat{a}_i \rangle = {\rm Tr} (\hat{a}_i\hat{\rho}),
\end{equation}
where $\hat{\rho} = Z^{-1}e^{-\beta \hat{H}}$ is the density matrix operator, $Z = {\rm Tr} (e^{-\beta \hat{H}})$ its normalization, and  $\beta = 1/K_B T$ is the inverse temperature of the system. We write Hamiltonian (\ref{EQ:eBHH}) in the form $\hat{H} = \hat{H}_0 + \hat{H}_1$ where
\begin{eqnarray}
\hat{H}_0 &=&  \frac{U}{2}\sum_i \hat{n}_i(\hat{n}_i-1)  - \mu \sum_i \hat{n}_i + \sum_{i \neq j} \frac{V_{ij}}{2} \hat{n}_i\,\hat{n}_j \\
\hat{H}_1 &=& - J \sum_{\langle ij \rangle} \; \hat{a}^\dag_i \hat{a}_j \label{EQ:H1},
\end{eqnarray}
and we assume a uniform chemical potential $\mu$. The generalization to a site-dependent chemical potential is straightforward. Furthermore, we assume low temperatures $\beta \rightarrow \infty$, and the tunneling coefficient to be the smallest energy in the system, i.e $J \ll U,\mu,V_{ij} $ such that we can treat $\hat{H}_1$ as a small perturbation on $\hat{H}_0$, and use the Dyson expansion at the first order in $\hat{H}_1$ for all the exponential operators, so that one obtains
\begin{equation}
e^{-\beta(\hat{H}_0 + \hat{H}_1)} \simeq e^{-\beta \hat{H}_0}\Big[ \hat{\openone} - \int_0^\beta e^{\tau \hat{H}_0} \hat{H}_1 e^{-\tau \hat{H}_0} \ud \tau \Big].
\end{equation}
We now write Hamiltonian (\ref{EQ:H1}) as a sum of single site Hamiltonians. Writing the annihilation operator as
$\hat{a}_i = \hat{A}_i + \varphi_i$, we can perform the mean field decoupling
on the tunneling term
\begin{eqnarray}
\hat{a}_i^\dag \hat{a}_j &=& \hat{A}_i^\dag \varphi_j + \hat{A}_j \varphi_i + \varphi_i \varphi_j + \hat{A}_i^\dag \hat{A}_j \nonumber \\
          &\simeq& \hat{a}_i^\dag \varphi_j + \hat{a}_j \varphi_i - \varphi_i \varphi_j ,
\end{eqnarray}
where in the last step we have assumed small fluctuations,
characteristic of the Mott, or the deep superfluid
states, and replaced $\hat{A}_i^\dag \hat{A}_j \simeq 0$. In Hamiltonian (\ref{EQ:H1}) we
now replace $\hat{a}_i^\dag \hat{a}_j$ with the expression calculated above,
we neglect terms of the order of $\varphi^2$ and find the mean field tunneling Hamiltonian
\begin{equation}
\hat{H}_1^{\scriptscriptstyle \mathrm{MF}} = -J\sum_i \left( \hat{a}_i^\dag \bar{\varphi}_i + \bar{\varphi}_i^*\hat{a}_i \right).
\end{equation}
Given a classical distribution of atoms in a lattice such as
\begin{equation}
\ket{\Phi} = \prod_i\ket{n_i}_i,
\end{equation}
satisfying $\hat{H}_0 \ket{\Phi} = E_\Phi \ket{\Phi}$, let us suppose that this configuration is a local minimum of the
energy, it can be the ground state, namely the absolute minimum, or another local minimum.
We will be more
rigorous at the end of this section regarding the meaning of local minimum of energy but for the moment let us refer to the common picture of a local minimum.
In the basis of the eigenfunctions of $\hat{H}_0$, satisfying the relation $\hat{H}_0 \ket{\Upsilon} = E_\Upsilon \ket{\Upsilon}$ the partition function then takes the simple form
\begin{equation}
\label{EQ:PartitionF}
Z \simeq {\rm Tr} (e^{-\beta \hat{H}_0}) = \sum_{\ket{\Upsilon}} \bra{\Upsilon} e^{-\beta \hat{H}_0} \ket{\Upsilon} \;
\stackrel{\scriptscriptstyle{\beta \mapsto \infty}}{\longrightarrow } \; e^{-\beta E_\Phi},
\end{equation}
where the last limit holds because we do not trace over all the states of the basis but only around the state $\ket{\Phi}$ , which is assumed to be a local minimum of the energy.
Using again a Dyson expansion of the exponential of the density operator, we obtain the order parameter as
\begin{eqnarray}
\varphi_i &\simeq& - e^{\beta E_\Phi}\int_0^\beta {\rm Tr} \left[ \hat{a}_i \; e^{-(\beta - \tau) \hat{H}_0} \; \hat{H}_1^{\scriptscriptstyle \mathrm{MF}} \; e^{-\tau \hat{H}_0} \right] \ud \tau \nonumber \\
             &=& J \bar{\varphi}_i e^{\beta E_\Phi} \int_0^\beta \sum_{\ket{\Upsilon}} \bra{\Upsilon} \hat{a}_i \; e^{-(\beta - \tau) \hat{H}_0} \; \hat{a}_i^\dag \; e^{-\tau \hat{H}_0} \ket{\Upsilon}, \label{EQ:IntPhi}
\end{eqnarray}
which is easy to calculate. The trace is then non trivial only for
$\ket{\Upsilon} = \ket{\Phi}$ and $\ket{\Upsilon} = \frac{\hat{a}_i}{\sqrt{n_i}}\ket{\Phi}$, where $n_i$ is integer on $\ket{\Phi}$, and after the integration in the $\beta \mapsto \infty$ limit we are left with the result
\begin{equation}
\label{EQ:OrderParameter}
\varphi_i = J\bar{\varphi}_i \left[ \frac{n_i + 1}{E_{\scriptscriptstyle \mathrm{P}}^i} + \frac{n_i}{E_{\scriptscriptstyle \mathrm{H}}^i}\right],
\end{equation}
where the quantities $E_{\scriptscriptstyle \mathrm{P}}^i$, $E_{\scriptscriptstyle \mathrm{H}}^i$ are defined as
\begin{eqnarray}
\eqalign{
\label{EQ:PHExcitations}
E_{\scriptscriptstyle \mathrm{P}}^i &= - \mu + Un_i + V_{\rm dip}^{1,i} \\
E_{\scriptscriptstyle \mathrm{H}}^i &= \mu - U(n_i-1) - V_{\rm dip}^{1,i},
}
\end{eqnarray}
and are respectively the energy cost for a particle (P) and hole (H) excitation on top of the $\ket{\Phi}$ configuration. In the previous expressions $V_{\rm dip}^{1,i} = \sum_{j\neq i} V_{ij} n_j$
is the dipole-dipole interaction that one atom placed at site $i$ feels with the rest of the atoms in the lattice.
We performed the integral (\ref{EQ:IntPhi}) in the limit of $\beta \mapsto \infty$, and in such a limit one finds that the integral converges only for positive
values of the particle and hole excitation energies, namely
\begin{equation}
\label{EQ:Conditions01}
U(n_i-1) + V_{\rm dip}^{1,i} < \mu < Un_i + V_{\rm dip}^{1,i}.
\end{equation}
This requirements have to be fulfilled at every site $i$ of the lattice and they simply state that the configuration $\ket{\Phi}$ is a local minimum with respect to adding and removing particles at any site. In the light of this statement, the restriction on the trace of Eq. (\ref{EQ:PartitionF}) is now rigorous, and is in perfect agreement
with the treatment done in \cite{BB:Fisher}.
Notice that if $\ket{\Phi}$ is not a local minimum, then one finds that conditions (\ref{EQ:Conditions01}) are never satisfied and the integral (\ref{EQ:IntPhi}) indeed diverges.
This treatment is of course also valid for $\ket{\Phi}$, being in particular the ground state of the system.

One finds such an equation (\ref{EQ:OrderParameter}), and conditions (\ref{EQ:Conditions01}) for every site $i$ of the lattice.
The convergence conditions are simple and among them one has to choose the most stringent to find the boundary of the lobe at J = 0.
Instead the equations for the order parameters are coupled due to the $\bar{\varphi}_i$ term, they can be written in a matrix
form $\mathcal{M}(\mu,U, J) \cdot \vec{\varphi} = 0$, with $ \vec{\varphi} \equiv (\cdots \varphi_i \cdots )$, and have
a non trivial solution. For every $\mu$, the smallest $J$ for which ${\rm det} [\mathcal{M}(\mu,U, J)] = 0$ gives the lobe of the $\ket{\Phi}$ configuration in the
$J$ vs. $\mu$ plane.

\subsubsection{Perturbative mean-field vs. dynamical Gutzwiller approach}{--- }
\label{SS:Compare}
The predictions of the perturbative mean-field treatment are in perfect agreement with the results of the dynamical Gutzwiller approach,
 since they both rely on the same mean-field approximation. The first looks at the stability of a given density distribution
$\ket{\Phi} = \prod_i\ket{n_i}_i$ of integer $n_i$ atoms per site, with respect to particle and hole excitations, while the latter minimizes the energy of a random initial configuration with respect to particle and hole excitations leading to the distribution $\ket{\Phi}$ if the initial condition is sufficiently close. However, the first method can only identify the phase boundaries of the insulating lobes without providing any further information on the SF phases outside the lobes, which can instead be explored with the imaginary time evolution. Nevertheless for dipolar Hamiltonians, due to the presence of many local minima of the energy, as we will see in the next part \cite{BB:MenottiPRL,BB:Trefzger01}, it is very difficult to identify the ground state with the dynamical Gutzwiller approach. This can be achieved more
efficiently through the perturbative mean-field approach. Therefore the two methods complement each other.
As an example, in Fig \ref{FIG:MILobes} (a,b) the black lines are calculated with the perturbative method ($V_{ij} = 0$) while the SF region outside the lobes is explored using imaginary time evolution showing perfect agreement between the two approaches.

\section{Dipolar Bosons in a 2D optical lattice}
\label{CH:MS:DipolarBosons}

\subsection{The model}
\label{SEC:TheModel01}

\begin{center}
\begin{figure}[h]
\begin{center}
\includegraphics[width=0.5\linewidth]{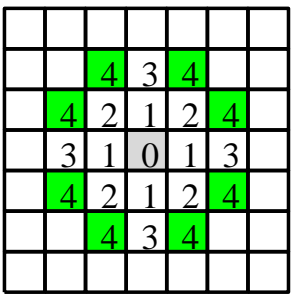}
\end{center}
\caption{Representation of the first four nearest neighbors of the site labeled as $0$ in the 2D lattice.}
\label{FIG:Truncation}
\end{figure}
\end{center}

In \cite{BB:MenottiPRL,BB:Trefzger01}, we have studied the properties of dipolar Bosons in an infinite 2D optical lattice,
mimicked by an elementary cell of finite dimensions $L\times L$ ($N_S = L^2$ sites) satisfying periodic boundary conditions.
The dipoles are aligned and point perpendicularly out of the plane so that the dipole-dipole interaction (\ref{EQ:DipoleI}) between two particles at relative distance
$\myvec{r}$  becomes $U_{\rm dd} (\myvec{r}) = C_{\rm dd}/(4\pi r^3)$ repulsive and isotropic in the 2D plane of the lattice, where $C_{\rm dd}$ is given by
Eq. (\ref{EQ:Cdd}).
Furthermore for computational simplicity we truncate the range of the off-site interactions at a finite number of nearest neighbors, as shown in Fig. \ref{FIG:Truncation} up to
the range number four.
We have studied the phase diagram of the system described by the Hamiltonian
\begin{equation}
\label{EQ:EBH2D}
\fl
\hat{H} = - J \sum_{\langle i j\rangle} \hat{a}^{\dag}_i \hat{a}_j -  \sum_i \mu \hat{n}_i  +\sum_i \frac{U}{2}  \; \hat{n}_i(\hat{n}_i-1) +
\frac{U_{\rm NN}}{2} \sum_{\vec \ell} \sum_{\langle \langle ij \rangle \rangle_{\vec \ell}} \frac{1}{{|\;\vec{\ell}\phantom{1}| }^{3}} \; \hat{n}_i \hat{n}_j ,
\end{equation}
where $U_{\rm NN} = C_{\rm dd}/(4\pi d_{\scriptscriptstyle \rm 2D}^3)$ is the dipole-dipole interaction between nearest neighboring sites, $d_{\scriptscriptstyle \rm 2D}$ is the lattice period, and $\langle \langle ij \rangle \rangle_{\vec \ell}$ represents neighbors at relative distance $\vec{\ell}$ which is measured in units of $d_{\rm 2D}$.
All other quantities were introduced previously.

\subsection{Metastability}
\label{SEC:Metastability}
Let us start the discussion by introducing our definition of stability of a given classical distribution of atoms in the lattice, i.e. a product over single-site Fock states
\begin{equation}
\label{EQ:Metastable01}
\ket{\Phi} = \prod_i \ket{n_i}_i.
\end{equation}
At $J=0$, we define the state (\ref{EQ:Metastable01}) to be stable if there exists a finite interval $\Delta\mu = \mu_{\rm max} - \mu_{\rm min} > 0$
in the $\mu$ domain, in which the particle (P) and hole (H) excitations at each site
$i$ of the lattice are positive, and the system is gapped.
Using the dipolar Hamiltonian (\ref{EQ:EBH2D}), in Sec. \ref{SEC:Semi} we have calculated the particle and hole excitation energies of $\ket{\Phi}$ to be
\begin{eqnarray}
\eqalign{
\label{EQ:PHExcitations02}
E_{\scriptscriptstyle \mathrm{P}}^i &= - \mu + Un_i + V_{\rm dip}^{1,i} \\
E_{\scriptscriptstyle \mathrm{H}}^i &= \mu - U(n_i-1) - V_{\rm dip}^{1,i},
}
\end{eqnarray}
where we recall $V_{\rm dip}^{1,i}\geq 0$ to be the dipolar interaction experienced by one atom sitting at the site $i$ of the lattice.
From Eqs. (\ref{EQ:PHExcitations02}) it is then straightforward to find $\Delta\mu$, if it exists, given by the set of inequalities
\begin{equation}
\label{EQ:StabilityConditions}
U(n_i-1) + V_{\rm dip}^{1,i} < \mu < Un_i + V_{\rm dip}^{1,i}.
\end{equation}
This is consistent with the
stability conditions discussed in the seminal paper of Fisher {\it et. al.} \cite{BB:Fisher}. Indeed, in the absence of dipolar interactions $U_{\rm NN}=0$
into Eqs. (\ref{EQ:StabilityConditions}), one recovers the well known conditions $U(n_i-1) < \mu < Un_i$ for the stability of the MI($n_i$), with $n_i$ particles per site.
One can extend the stability analysis to small values of $J$, and for a given stable state calculate its insulating lobe with the perturbative
mean-field approach we have developed in Sec. \ref{SEC:Semi}. In this context, we therefore define a state like (\ref{EQ:Metastable01}) to be {\it metastable} if it satisfies two conditions: the first is that the state must have an insulating lobe inside which it is gapped, and the second is that the energy of the state must be higher than the ground state energy. In other words a metastable state is a local minimum of the energy.

In the absence of dipolar interactions $U_{\rm NN}=0$, no metastable states are found. In the low tunneling region the ground state of the system consists of Mott insulating lobes with integer filling factors $\nu = N_{\rm a}/N_{\rm S}$ (number of atoms$/$number of sites), while for large values of $J$ the system is superfluid.
In our treatment metastable states appear as soon as one introduces at least one nearest neighbor of the dipolar interaction. In fact, the imaginary time evolution, which for Bose Hubbard Hamiltonians with only on-site interactions converges unambiguously to the ground state, for the dipolar Hamiltonian (\ref{EQ:EBH2D}) often converges to different
metastable configurations depending on the exact initial condition. Moreover, in the real time evolution, their stability manifests as typical small oscillations of frequency $\omega_0$ around the local minimum of the energy.

\begin{center}
\begin{figure}[h!]
\begin{center}
\includegraphics[width=0.9\linewidth]{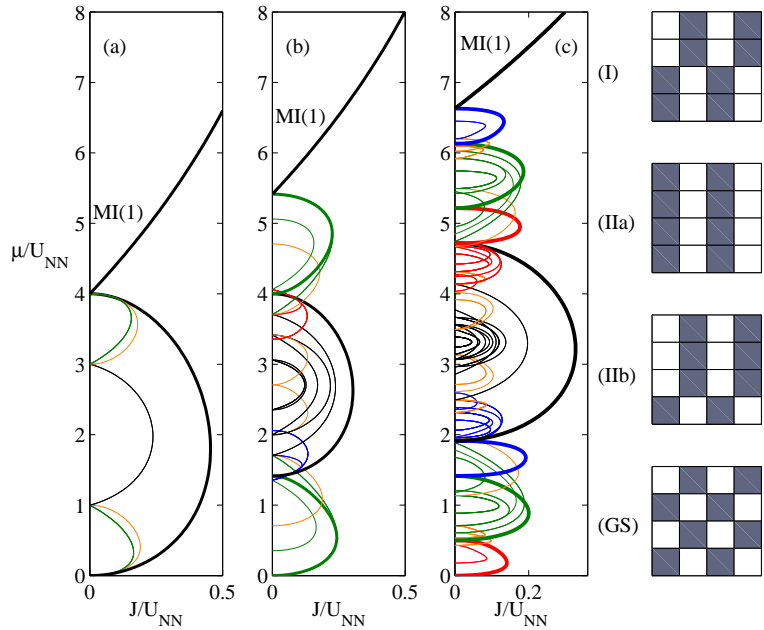}
\end{center}
\caption{(a), (b), (c) Phase diagram with a range of
the dipole-dipole interaction cut at the first, second, and fourth nearest
neighbor, respectively. The thick line is the ground state and the
other lobes correspond to the metastable states, the same color corresponding
to the same filling factor. In (c) filling factors range
from $\nu=1/8$ to $\nu=1$. In the right column we present metastable
configurations for $\nu = 1/2$ appearing at the first nearest neighbor (I), and second
(IIa, IIb), and the corresponding ground state (GS); those metastable
states remain stable for all larger ranges of the dipole-dipole interaction. White sites are empty, while gray sites are occupied by one atom.}
\label{FIG:Lobes}
\end{figure}
\end{center}

Our main results are summarized in Figs. \ref{FIG:Lobes} (a,b,c), where we plot the phase diagram of the Hamiltonian (\ref{EQ:EBH2D}) for a $L=4$ elementary cell satisfying periodic boundary conditions, for different values of the cut off range of the dipolar interactions respectively at one (a), two (b), and four nearest neighbors. The on-site interaction is given by $U/U_{\rm NN} = 20$ and $U_{\rm NN}=1$ is the unit of energy.
The thick lines correspond to the ground state while the thin lines correspond to metastable insulating states, with the color identifying the same filling
factor $\nu$. The difference with the Bose Hubbard phase diagram of Fig. \ref{FIG:MILobes},
where only integer filling factors $\nu = \bar{n}$ are present, is evident already with one nearest neighbor of the dipolar interaction (a). In fact the MI(1) lobe
undergoes a global shift of $zU_{\rm NN}$ ($z = 4$ in the figure) towards higher values of $\mu$, and the new fractional filling factor $\nu = 1/2$ appears with a ground state density distribution modulated in a checkerboard pattern, shown in Fig. \ref{FIG:Lobes} (GS).
Instead, the density distribution shown in Fig. \ref{FIG:Lobes} (I) is metastable with $\nu=1/2$, and its insulating lobe is given by the thin line extending from $1 < \mu/U_{\rm NN} < 3$.
Remarkably, in Fig. \ref{FIG:Lobes}(a) the two lobes extending from $0 < \mu/U_{\rm NN} < 1$ correspond to metastable configurations at filling factors $\nu = 1/4, 5/16$, while the two lobes between $3 < \mu/U_{\rm NN} < 4$ correspond to metastable states at filling factors $\nu = 3/4, 11/16$, but no ground state is found for these fillings. In the region immediately outside the ground state lobes, we found evidences of supersolid (SS) phases, where the order parameters $\varphi_i$ are different than zero and are spatially modulated, e.g. in a CB structure.
Before our work, studies of BH models with extended interactions have pointed out the existence of novel quantum phases, like the SS and checkerboard phases, but not the existence of the metastable states.

Increasing further the range of the dipolar interactions leads to the
appearance of more metastable states, as (IIa) and (IIb) found at $\nu=1/2$ for two nearest neighbors in the range of dipolar interactions. The MI(1) undergoes
a larger shift which is accompanied by
the emergence of other insulating fractional filling factors, as shown in Fig. \ref{FIG:Lobes} (c), where the dipole-dipole interaction is cut at the fourth nearest neighbor and the ground state is a series of lobes with
$\nu$ multiple of $\nu = 1/8$. The number of metastable states varies depending on the parameters of the Hamiltonian and the filling factor; it is found to be up to 400 for
$U/U_{\rm NN} = 20$ at filling $\nu=1/2$, and up to 1500 for $U/U_{\rm NN} = 2$ at unit filling \cite{BB:MenottiPRL}. With this picture in mind, it is now clear why the imaginary time evolution,
which often converges to different metastable configurations, is very inefficient both to find the ground state of the system and to compute the lobe boundaries of a given
metastable state. Instead, the mean-field perturbative approach we have derived in Sec. \ref{SEC:Semi} has proven to be satisfactory for this purpose but it also has
some limitations. In fact, all possible values of $\nu$ and the corresponding configurations which are detectable with this method is limited by the size of the elementary cell. Evidently the possible filling factors of an elementary cell of size $L$ are given by multiples of $\nu = 1/L^2$. It is worth to notice, that despite the inefficiency
of the imaginary time evolution in finding the insulating lobes of the system, the corresponding equations in real time turn out to be very useful, for example, to compute the excitation spectrum $\omega({\bf k})$ of the system. For a given metastable configuration, one can calculate $\omega({\bf k})$ from the small fluctuations $\delta f_{n}^{(i)}(t)$ around the unperturbed metastable state coefficients $\overline{f}_{n}^{(i)}$. Writing $f_{n}^{(i)} = \overline{f}_{n}^{(i)} + \delta f_{n}^{(i)}(t)$ into Eq. (\ref{EQ:FDynamics}), and taking into account only linear terms in the fluctuations, one obtains a set of coupled equations for $\delta f_{n}^{(i)}(t)$, which can be easily solved in the Fourier domain as explained in \cite{BB:Trefzger01}.

Finally, we find that there is usually a gap between the ground state and the lowest metastable state, which might allow to reach
the ground state by ramping up the optical lattice under some adiabaticity condition. However, this feature is strongly reduced in the case of larger elementary cells because the number of metastable configurations and the variety of their patterns increase very rapidly with the size of
the elementary cell $L$. Indeed, we have found that there exist many metastable configurations that differ from the ground state only by small localized defects, and the energy of these reduces the size of the gap.

\subsection{The lifetime}
\label{SEC:Lifetime}
\begin{center}
\begin{figure}[h]
\begin{center}
\includegraphics[width=0.9\linewidth]{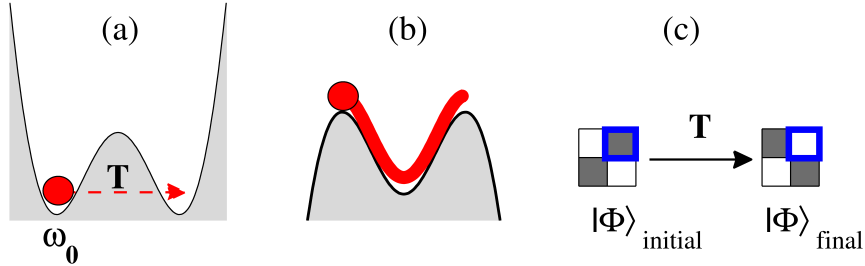}
\end{center}
\caption{(a) Particle in a minimum of a potential barrier, the particle oscillates with frequency $\omega_0$ around the local minimum and tunnels
into the right well in a time $T$; (b) the instanton; and (c) the process for which a checkerboard state tunnels into the anti-checkerboard that
shown complete exchange of particle with holes and vice versa. The process happens in a time $T$ in analogy with (a).}
\label{FIG:Connection}
\end{figure}
\end{center}

We have studied the stability of the metastable states with a path integral formulation in imaginary time and a generalization of the instanton theory \cite{BB:Wen}. For any given initial metastable configuration $\ket{\Phi}_{\rm initial}$, we are able to estimate the time $T$ in which  $\ket{\Phi}_{\rm initial}$ has tunneled completely into a different metastable state $\ket{\Phi}_{\rm final}$. We do this in analogy with the case of a classical particle tunneling through a potential barrier shown in Fig. \ref{FIG:Connection} (a), with the difference that we do not have any information a priory on the characteristics of the potential barrier separating initial and final state.
Nevertheless we can estimate the barrier and the time $T$ in three steps: (i) first we construct the imaginary time Lagrangian
of the system described by a quantum state $\ket{\Phi}$, (ii) we make use of a variational method on $\ket{\Phi}$ with only one variational parameter $q$, and its conjugate momenta $P$, that interpolate continuously between $\ket{\Phi}_{\rm initial}$ and $\ket{\Phi}_{\rm final}$, and (iii) through the variation of $q$ we calculate the minimal action $S_0$, with the imaginary time Lagrangian, along the stationary path starting at $\ket{\Phi}_{\rm initial}$; this path is called an {\it instanton path}, in short instanton.
It connects $\ket{\Phi}_{\rm initial}$ and $\ket{\Phi}_{\rm final}$ only if the two states are degenerate, otherwise the stationary path connects $\ket{\Phi}_{\rm initial}$ with
an intermediate state called the {bouncing point} $\ket{\Phi}_{\rm bounce}$. We get an estimate of the energy barrier separating the two states by evaluating the Lagrangian from $\ket{\Phi}_{\rm initial}$ to $\ket{\Phi}_{\rm final}$ and imposing zero "momentum" $P=0$, as one would do in the Lagrangian of a classical particle in a potential.
Once the minimal action $S_0$ is known, the tunneling time $T$ is readily calculated \cite{BB:Wen} as
\begin{equation}
\omega_0 T = \frac{\pi}{2}e^{S_0},
\end{equation}
where $\omega_0$ is of the order of the frequency of the typical small oscillations of $\ket{\Phi}_{\rm initial}$ around the local minimum of the energy. In analogy to a classical particle tunneling through a barrier, the instanton has the nice interpretation of the stationary path connecting the two local minima in the inverted potential, as schematically represented in Fig. \ref{FIG:Connection} (b).

In units of $\hbar=1$, the imaginary time Lagrangian of a system \cite{BB:Perez} described by a quantum state $\ket{\Phi}$, is given by
\begin{equation}
\label{EQ:Lagrangian}
\mathcal{L} = \frac{\braket{\Phi}{\dot{\Phi}} - \braket{\dot{\Phi}}{\Phi}}{2} + \bra{\Phi}\hat{H}\ket{\Phi},
\end{equation}
with $\ket{\dot{\Phi}}$ indicating the time-derivative, and $\hat{H}$ being the Hamiltonian of the system. In the approximation where $\ket{\Phi}$ is the Gutzwiller wave function of a given metastable state, we write its amplitudes as
\begin{equation}
\label{EQ:GWAmplitudesStability}
f_{\rm n}^{(i)} = \frac{1}{\sqrt{2}} \left( x_{\rm n}^{(i)} + ip_{\rm n}^{(i)}\right),
\end{equation}
where $x_{\rm n}^{(i)}$, and $p_{\rm n}^{(i)}$ are real numbers which are going to be
related to the variational parameters and their conjugate momenta in the following.
For simplicity, we consider states with a maximum occupation number of
$n_{\rm max} = 1$ (i.e. $n=0,1$), and therefore we have a total of $4N_S$ parameters, $N_S$ being the total number of sites.
The Lagrangian (\ref{EQ:Lagrangian}) as well as the expectation value of the Hamiltonian $\bra{\Phi}\hat{H}\ket{\Phi}$, become functional of the $4N_S$ parameters, namely
\begin{equation}
\label{EQ:Lagrangian01}
\mathcal{L}[x_{\rm n}^{(i)},p_{\rm n}^{(i)}] = -i\sum_{i,n=0}^1 p_{\rm n}^{(i)} \dot{x}_{\rm n}^{(i)} + \bra{\Phi}\hat{H}\ket{\Phi}.
\end{equation}
After introducing the new coordinates $q_{\rm n}^{(i)} = x_{\rm n}^{(i)}$, and their conjugate momenta $P_{\rm n}^{(i)} = \partial \mathcal{L}/\partial \dot{q}_{\rm n}^{(i)} = -i p_{\rm n}^{(i)}$, we can put the Lagrangian (\ref{EQ:Lagrangian01}) in its canonical form
\begin{equation}
\label{EQ:Lagrangian02}
\mathcal{L}[q_{\rm n}^{(i)},P_{\rm n}^{(i)}] = \sum_{i,n=0}^1 P_{\rm n}^{(i)} \dot{q}_{\rm n}^{(i)} - \mathcal{H}[q_{\rm n}^{(i)},P_{\rm n}^{(i)}],
\end{equation}
where $\mathcal{H}[q_{\rm n}^{(i)},P_{\rm n}^{(i)}] = - \bra{\Phi}\hat{H}\ket{\Phi}$ is a constant of the motion \footnote{note that in the analogy of a classical particle in a potential $V(x)$, the conserved quantity in the imaginary time would be $\mathcal{H} = \frac{P^2}{2m} - V(x)$, which describes the particle's motion in the inverted potential.}. We now want to reduce the dynamic described by the Lagrangian (\ref{EQ:Lagrangian02}) to a one dimensional problem, described only by one variable $q$ and its conjugate momentum $P$. Through the variation of $(q,P)$ we want to describe the interchange between the state $\ket{\Phi}_{\rm initial}$ and $\ket{\Phi}_{\rm final}$, as for example the one represented in Fig. \ref{FIG:Connection}(c).
In \ref{SEC:Parametrization}, we show how to reduce the number of variational parameters to one, $q$, and its conjugate momentum $P$,
by making use of a variational Ansatz as well as the normalization condition on the coefficients (\ref{EQ:GWAmplitudesStability}) and the conservation of the total number of particles.
These conditions are enforced through Lagrange multipliers $\lambda_c$. Consequently the equations of motion
given by $\dot{q} = \partial \mathsf{H}/ \partial P$, and $\dot{P} = -\partial \mathsf{H}/ \partial q$
are governed by an Hamiltonian which also includes the constraints as follows
\begin{equation}
\label{EQ:HamiltonianStationary}
\mathsf{H} = \mathcal{H}[q,P] + \sum_c \lambda_c \mathcal{C}_c,
\end{equation}
where an explicit expression for the conditions $\mathcal{C}_c$ is given in \ref{SEC:Parametrization}.
The action is then readily calculated along the stationary path of Eq. (\ref{EQ:HamiltonianStationary}) as follows
\begin{equation}
S_0 = \int \mathcal{L}[q,P] \ud \tau = \int_{\rm path} \mathcal{L}[q,P] \frac{\ud q}{\dot{q}},
\end{equation}
with $\dot{q} = \partial \mathsf{H}/ \partial P$ from Eq. (\ref{EQ:HamiltonianStationary}).

\subsubsection{Action and tunneling time}
\begin{center}
\begin{figure}[h!t]
\begin{center}
\includegraphics[width=0.7\linewidth]{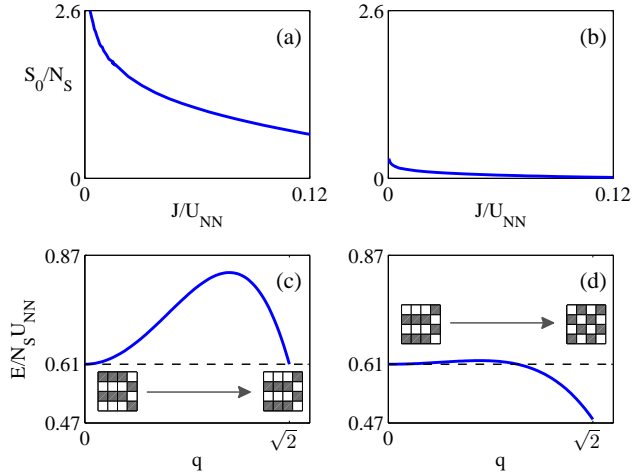}
\end{center}
\caption{(a,b) Action per site and (c,d) energy barrier for the process sketched in panels (c,d). In both cases the initial state is configuration (IIb) of Fig. \ref{FIG:Lobes} and the value $J = 0.12 U_{\rm NN}$ corresponds to the tip of its insulating lobe. The first one (a,c) is for degenerate initial
and final configurations while for the second one (b,d) the final configuration is energetically deeper. The difference in the two processes manifests also
in the height of the barrier which is smaller for the second case, leading to a smaller action and consequently a smaller life-time.}
\label{FIG:Actions}
\end{figure}
\end{center}

{--- }
In Fig. \ref{FIG:Actions} (a,b) we plot the minimal action
divided by the total number of sites $N_S$ of the cell, as a function of the tunneling coefficient $J$, for two different processes. The first one (a,c), in which initial and final state are degenerate,
shows the exchange of particles with holes in the whole lattice, and is sketched in the lower part of Fig. \ref{FIG:Actions} (c). There we also plot
the potential barrier between initial and final state calculated as $-\mathsf{H}(q,P=0)$. Instead, in the second process (b,d) the final state is the ground state, i.e. deeper in energy with respect to the initial state, and only a few sites of the lattice exchange particles with holes during the process,
as sketched in the upper part of  Fig. \ref{FIG:Actions} (d) along with the potential barrier.
A side remark, the point where the thick line of the barrier encounters the dashed line is called {\it bouncing point} $\ket{\Phi}_{\rm bounce}$.
The action in general diverges for $J \rightarrow 0$ indicating a divergent tunneling time $T$, and then decreases monotonically up to a minimum value
in correspondence of the tip of the lobe, $J = 0.12 U_{\rm NN}$ here, signaling a minimum life-time at the tip of the lobe, as expected. In between these
two extreme behaviors, the action increases monotonically with the number of sites involved in the exchange of particles with holes; the more sites involved as in
the case of Fig. \ref{FIG:Actions} (a,c), the bigger the action is.
Summarizing, from the figures above, we conclude that small energy differences between the initial $\ket{\Phi}_{\rm initial}$
and the final states $\ket{\Phi}_{\rm final}$ and large regions of the lattice undergoing
particle-hole exchange in the tunneling process contribute to
large barriers, i.e. long life times $T$. On the contrary, for big energy differences and small regions
of the lattice undergoing particle-hole exchange, the barrier is small.
Hence, in general it is more likely for a
given state to tunnel into a state deeper in energy, e.g., the
ground state, than into its complementary, which implies the
exchange of particles with holes in the whole lattice.

\section{Multiple layers and up-down mixtures}

\subsection{Dipolar Bosons in a bilayer optical lattice}
\label{SEC:TheModel}

\begin{center}
\begin{figure}[h!]
\begin{center}
\includegraphics[width=0.6\linewidth]{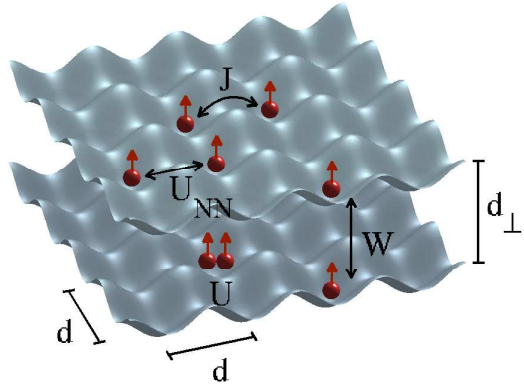}
\end{center}
\caption{Schematic representation of two 2D
optical lattice layers populated with dipolar bosons polarized
perpendicularly to the lattice plane. The particles feel repulsive
on site $U$ and nearest-neighbor $U_{\rm NN}$ interactions. Interlayer
tunneling is completely suppressed, while a nearest-neighbor
interlayer attractive interaction $W$ is present.}
\label{FIG:Layers}
\end{figure}
\end{center}

In \cite{BB:Trefzger02} we consider polarized dipolar particles in two decoupled
2D optical lattice layers (see Fig. \ref{FIG:Layers}), where the potential barrier between the two layers is large enough to prevent any inter-layer hopping.
This is the simplest multi-layer structure and can be
obtained by using anisotropic optical lattices or superlattices,
which can exponentially suppress tunneling in one
direction.

The in-plane dipolar interaction is isotropic
and repulsive. The interlayer interaction depends on the
relative position between the two dipoles, but is dominated
by the nearest-neighbor attractive interaction $W < 0$ between
two atoms at the same lattice site in different layers.
We include only nearest-neighbor (NN)
in-plane ($U_{\rm NN}$) and out-of-plane ($W$) dipolar interactions.
Since tunneling is suppressed between the layers particles
belonging to the different layers cannot mix and behave in
practice like two different species \footnote{Because of this analogy we will often refer
to the two layers as the two species and vice versa.}. The problem is analogous to that of two bosonic species on a 2D optical lattice
with an inter-species attraction $W < 0$ at the same lattice site, and intra-species repulsion $U_{\rm NN}$.
The relative strength between $U_{\rm NN}$ and $W$ can be tuned by
changing the spacing $d_\perp$ between the two layers, relative to
the 2D optical lattice spacing $d$. Because of the dependence
of the dipole-dipole interaction like the inverse cubic
power of the distance, the ratio $|W|/U_{NN}$ can be tuned
over a wide range. While it can be negligible for $d_\perp \gg d$
making the system asymptotically similar to a single 2D
lattice layer, it can also become relevant and give rise to
interesting physics, not existing in the single layer model as pointed out in \cite{BB:Arguelles,BB:Chen,BB:Wang,BB:Wang01,BB:Yi,BB:Wang02,BB:Klawunn,BB:Klawunn01}.

\newpage
The system is described by the Hamiltonian $\hat{H} = \hat{H}_0 + \hat{H}_1$, with
\begin{eqnarray}
\fl
\hat{H}_0 &=& \sum_{i,\sigma} \Big[ \frac{U}{2} \hat{n}_i^\sigma (\hat{n}_i^\sigma - 1) + \sum_{\langle j \rangle_i}\frac{U_{\rm NN}}{2} \hat{n}_i^\sigma \hat{n}_j^\sigma  - \mu \hat{n}_i^\sigma \Big] + W \sum_i \hat{n}_i^a \hat{n}_i^b \label{EQ:H3Da}~\\
\fl
\hat{H}_1 &=&  - J\sum_{\langle ij \rangle} [\hat{a}_i^\dag \hat{a}_j + \hat{b}_i^\dag \hat{b}_j ] \label{EQ:H3Db},
\end{eqnarray}
where  $\sigma = \big(a, b\big)$ indicates the two species (which in the
specific case considered here are atoms in the lower and
upper 2D optical lattice layer, respectively), $U$ is the on-site
energy, $U_{\rm NN}$ the intralayer nearest neighbors repulsion,
$W$ the interlayer attraction, $J$ the intralayer tunneling parameter,
and $\mu$ the chemical potential, as schematically represented in Fig. \ref{FIG:Layers}.
The parameters $U$
and $J$ are equal for the upper and lower layers and the
chemical potentials $\mu$ are the same, since equal densities in
the two layers are assumed. Notice that since $W<0$, it is necessary to have $U+W > 0$ to avoid collapse. The symbols $\langle ij \rangle$ and $\langle j \rangle_i$ indicate nearest neighbors.

We focus on the physical situation in which the two layers are very close to one another ($d_\perp \ll d$) such that $(U+W) \ll U$, and small in-plane dipolar interactions $U_{\rm NN} \ll U$, because in these limits particles at the same lattice site $i$ of different layers pair into composites.
The composites localize in a MI state for small values of the tunneling
coefficient, while for larger values of $J$ the pairs hop around in the optical lattice forming a pair-superfluid (PSF) phase \cite{BB:Arguelles}.
Furthermore, the presence of the in-plane long-range interactions leads to the formation of a novel pair-supersolid phase (PSS), namely, a supersolid of composites \cite{BB:Mathey}. Finally, it is useful to introduce the operator of the sum of the two
species number operators at each site of the lattice, namely
\begin{equation}
\label{EQ:ms}
\hat{m}_i = \frac{\hat{n}_i^a + \hat{n}_i^b}{2},
\end{equation}
which is diagonal on a given Fock state $\ket{m_i}_i$, with $m_i = (n_i^a + n_i^b)/2$ being the average occupation per layer at the site $i$.

\subsubsection{Low-energy subspace and effective Hamiltonian}{--- }
\label{SEC:LES}
In the limit where $(U+W),U_{\rm NN},J\ll U$, there exist a low-energy subspace spanned by the states
\begin{equation}
\label{EQ:Alpha}
\ket{\alpha} = \prod_i \ket{n_i,n_i}_i,
\end{equation}
with equal occupation of the two species $a$ and $b$ at each site. These states can be equivalently written in terms of the pair occupations as
$\ket{\alpha} = \prod_i \ket{m_i}_i$, where $m_i$ is the number of pairs at site $i$, and is given by the eigenvalue of operator (\ref{EQ:ms}).

\begin{center}
\begin{figure}[h!]
\begin{center}
\includegraphics[width=0.85\linewidth]{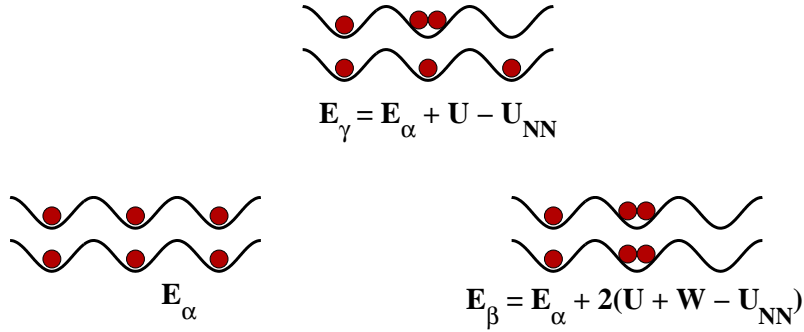}
\end{center}
\caption{Schematic representation of the low-energy subspace. The state $\ket{\alpha}$ is uniformly occupied by one particle per site, and its energy is given by $E_\alpha$. A second order process in the tunneling connects the two states $\ket{\alpha}$, and $\ket{\beta}$, through the state $\ket{\gamma}$, which belongs to the virtual subspace. It is
straightforward to notice that in the limit of $(U + W), U_{\rm NN}, J \ll U$, the energies above satisfy the necessary condition $E_\alpha,E_\beta \ll E_\gamma$ for the existence of the subspace.}
\label{FIG:Subspace}
\end{figure}
\end{center}

Single particle-hole excitations necessarily break one pair, and the energy cost of these excitations is
of the order of $U$. Therefore, in the limit where $(U+W),U_{\rm NN},J\ll U$ breaking one pair is
energetically very costly. In the limit $(U + W)/U \rightarrow 0$, asymptotically all states
$\ket{\alpha}$ become stable with respect to single-particle-hole excitations
and develop an insulating lobe at finite $J$, which tend to overlap as shown in Fig \ref{FIG:CB_2CB} (left) by the thin blue lobes.
However in this system, and for the above choice of parameters, single-particle-hole excitations are not the lowest lying ones. One
should also consider two-particles-two-holes excitations, which can be accounted for within a low-energy theory described by
an effective Hamiltonian acting on the low-energy subspace spanned by the states $\ket{\alpha}$-s. The validity of the effective
Hamiltonian relies on the existence of a low-energy subspace
well separated in energy from the subspace of virtual
excitations, to which it is coupled through the tunneling Hamiltonian $\hat{H}_1$ (\ref{EQ:H3Db}) via single-particle hopping.
The relevant virtual subspace is obtained from the states $\ket{\alpha}$ by breaking one composite, namely
\begin{equation}
\label{EQ:Subspace}
\eqalign{
\ket{\gamma_{ij}^{(a)}} &= \frac{\hat{a}_i^\dag \hat{a}_j }{\sqrt{n_j^{a}(n_i^{a} + 1 )}} \ket{\alpha} \\
\ket{\gamma_{ij}^{(b)}} &= \frac{\hat{b}_i^\dag \hat{b}_j }{\sqrt{n_j^{b}(n_i^{b} + 1 )}} \ket{\alpha},
}
\end{equation}
as schematically represented in Fig. \ref{FIG:Subspace}. All other states are not coupled to $|\alpha\rangle$ via single particle hopping and hence do not contribute.

The energy difference between the virtual states $\ket{\gamma}$, and the states $\ket{\alpha}$, is given by the sum of single-particle plus single hole-excitation energies of the states  $\ket{\alpha}$, which are of the order of $U$ at $J=0$, and are minimized by the width of the lobes $\ket{\alpha}$ at finite $J$ (see, e.g., Fig. \ref{FIG:CB_2CB} (left)).

Slow processes drive the system through different states of the low energy subspace via second order tunneling; this happens through a fast coupling with the virtual subspace. Since we are interested in the long time physics of the system, we have to average out all the fast processes and therefore we write
an effective Hamiltonian in the subspace of pairs, and include tunneling through second order perturbation
theory \cite{BB:Tannoudji02,BB:Kuklov}. In the pair-state basis, the matrix elements of such a Hamiltonian in second order perturbation theory are given
by
\begin{eqnarray}
\bra{\alpha}\hat{H}_{\rm eff}\ket{\beta} &=& \bra{\alpha}\hat{H}_0 \ket{\beta} - \frac{1}{2} \sum_\gamma \bra{\alpha}\hat{H}_1 \ket{\gamma} \bra{\gamma}\hat{H}_1 \ket{\alpha}\nonumber \\
& & \times \left[ \frac{1}{E_\gamma - E_\alpha} + \frac{1}{E_\gamma - E_\beta} \right], \label{EQ:DefHeff}
\end{eqnarray}
where $\hat{H}_0$, given by the interaction terms (\ref{EQ:H3Da}), is diagonal on the states $\ket{\alpha}$, and the
single-particle tunneling term $\hat{H}_1$ in Eq. (\ref{EQ:H3Db}) is treated at second order.
For a given state $|\alpha\rangle$,
\begin{eqnarray}
E_{\gamma_{ij}}-E_\alpha=U+(U+W)(m_i-m_j)+ U_{\rm NN} \Delta
m^{ij}_{\rm NN}, \label{den}
\end{eqnarray}
with $\Delta m^{ij}_{\rm NN}=\sum_{\langle k\rangle_i}m_{k}-\sum_{\langle k \rangle_j} m_{k}-1$, where $m_i$
indicates the pair occupation number at site $i$ as defined in Eq. (\ref{EQ:ms}). For $(U+W),U_{\rm NN}\ll U$, the denominators
$E_{\gamma_{ij}}-E_\alpha$ are all of order $U$, which leads to
\begin{eqnarray}
\label{heff0} &&\hat{H}_{\rm eff}^{(0)} = \hat{H}_0 - \frac{2J^2}{U} \sum_{\langle ij \rangle} \left[ \hat{m}_i (\hat{m}_j+1) + \hat{c}_i^\dag \hat{c}_j \right],
\end{eqnarray}
where $\hat{c}_i$ and $\hat{c}_i^\dag$ are the pair destruction and creation operators such that
\begin{equation}
\eqalign{
\hat{c}_i |m_i\rangle &= m_i|m_i-1\rangle \\
\hat{c}_i^\dag|m_i\rangle &= (m_i+1)|m_i + 1\rangle.
}
\end{equation}
One can easily obtain corrections to $\hat{H}_{\rm eff}^{(0)}$  by expanding (\ref{den}) at higher orders in $(U+W)/U$ and $U_{\rm NN}/U$ but, as we will see,
the zeroth order is already quite accurate to describe the physics of the system for the range of parameters we consider.

\subsubsection{Insulating lobes}{--- }
We now make use of the effective Hamiltonian $\hat{H}_{\rm eff}^{(0)}$ derived above to study the ground state phase diagram of the system, starting from the insulating states.
For every classical distribution of pairs in the lattice we can calculate the pair order parameters $\psi_i = \langle \hat{c}_i \rangle$ with the perturbative mean-field method derived in Sec. \ref{SEC:Semi}, and get the expression
\begin{equation}
\label{Eqn:MF}
 \psi_i = \frac{2J^2}{U}\left[ \frac{(m_i+1)^2}{E_{\scriptscriptstyle \mathrm{2P}}^i(J)}
 + \frac{m_i^2}{E_{\scriptscriptstyle \mathrm{2H}}^i(J)} \right]\bar{\psi_i},
\end{equation}
where $\bar{\psi_i} = \sum_{\langle j \rangle_i} \psi_j$, and the energy costs of adding a pair (2P) and removing a pair (2H) can be calculated from the diagonal part of Eq. (\ref{heff0}), and are respectively given by
\begin{equation}
\fl
\eqalign{
\label{EQ:PHE_2}
E_{\scriptscriptstyle \mathrm{2P}}^i(J) &= -2\mu + 2U m_i + (2m_i + 1)W + 2 V_{\rm dip}^{1,i} - \frac{2J^2}{U}\sum_{\langle k \rangle_i}(2m_k + 1)\\
E_{\scriptscriptstyle \mathrm{2H}}^i(J) &= 2\mu - 2U(m_i-1) - (2m_i - 1)W - 2 V_{\rm dip}^{1,i} + \frac{2J^2}{U}\sum_{\langle k \rangle_i}(2m_k + 1),
}
\end{equation}
with $V_{\rm dip}^{1,i} = U_{\rm NN} \sum_{\langle j \rangle_i} m_j$ being the in-plane dipolar interaction, i.e. the dipole-dipole interaction that one atom positioned
at site $i$ of the lattice, experiences with the rest of the particles belonging to the same plane.

Using Eq. (\ref{Eqn:MF}), one can calculate the
mean-field lobes for any given configuration of pairs in
the lattice. The ground state lobes for the checkerboard and doubly
occupied checkerboard are shown in Fig. \ref{FIG:CB_2CB} (right) for the 0th
(full lines) and 1st order (dashed lines) effective
Hamiltonians. The comparison between the two shows
that, for the parameters considered here, the 0th order
already captures the physics accurately. It is worth noticing that the $J^2$ dependence
of the energy of the elementary excitations is at the origin
of the reentrant behavior of the lobes, which was predicted by
exact matrix-product-state calculations for the 1D geometry
in \cite{BB:Arguelles}

\subsubsection{Pair superfluid/supersolid}{--- }
\label{SEC:BilayerSFSS}
While the MI phases are predictable through the perturbative mean-field approach of Eqs. (\ref{Eqn:MF}) for the pair order parameters, to identify
the SF phases, both PSF and PSS outside of the lobes, it is necessary to make use of the imaginary time evolution introduced in Sec. \ref{SEC:DGA} based on a Gutzwiller Ansatz for the pair wave function. Therefore, we need to calculate the dynamic equations equivalent to (\ref{EQ:FDynamics}) in the low energy subspace described by the effective Hamiltonian $\hat{H}_{\rm eff}^{(0)}$ of Eq. (\ref{heff0}).

\begin{figure}[h!]
\begin{center}
\begin{tabular}{cc}
\includegraphics[width=0.475\linewidth]{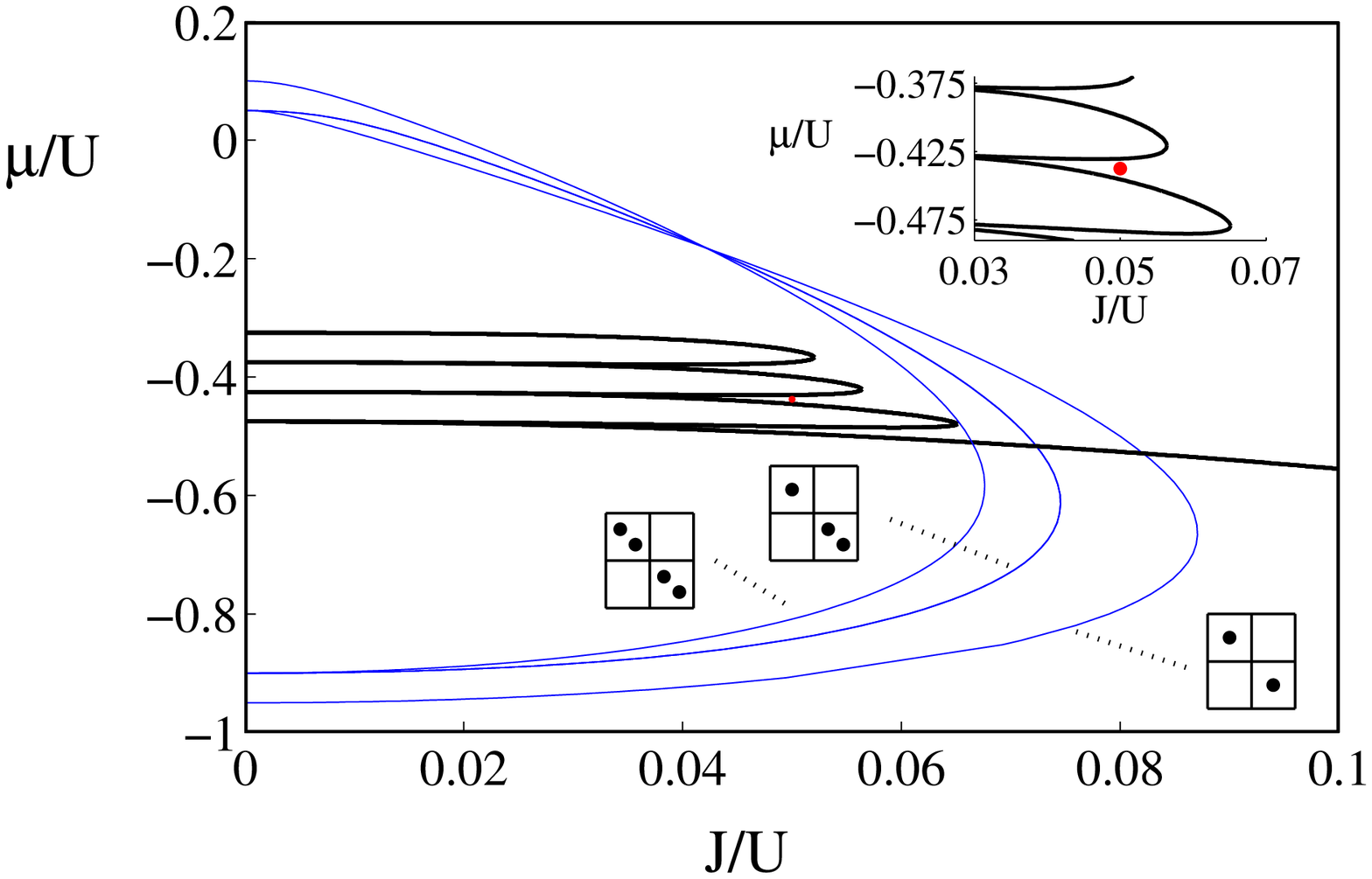} & \includegraphics[width=0.475\linewidth]{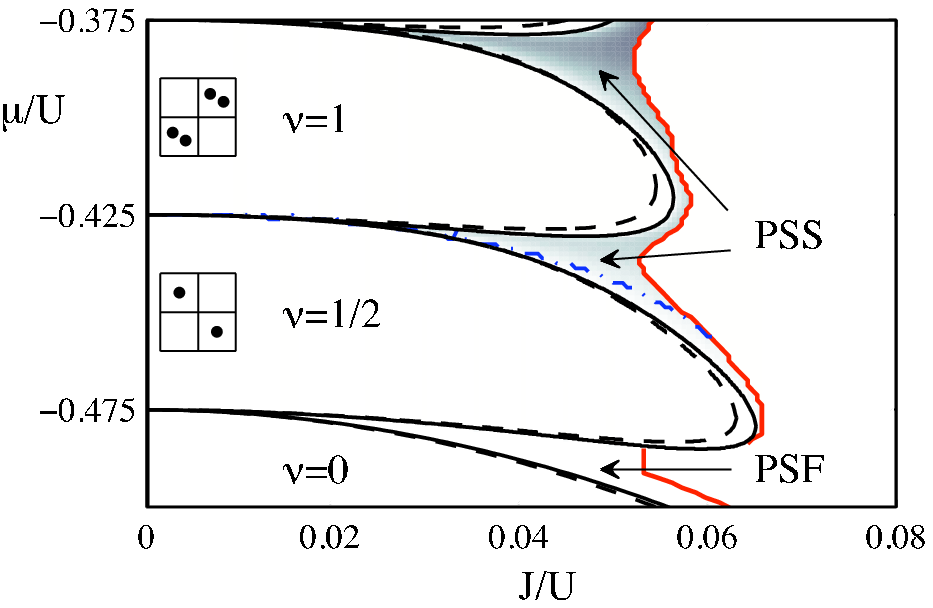}
\end{tabular}
\end{center}
\caption{(left) Pair insulating lobes for $\nu=0,1/2,1,3/2$ (thick
lines); Lobes with respect to single particle-hole excitations
(thin blue lines) for the dominant configurations in the ground
state at $J=0.05U$ and $\mu=-0.4375U$, namely $m_i=0$ and $m_j=1,2$
(for $i,j$ nearest neighboring sites). The inset shows a zoom of
the pair phase diagram. (right) Phase diagram of the effective Hamiltonian.
The black full lines are the semi-analytic
solution of Eq. (\ref{Eqn:MF}) indicating the boundaries of the
insulating lobes for the checkerboard ($\nu=1/2$) and the doubly
occupied checkerboard ($\nu=1$). The black dashed lines are the
boundaries of the insulating lobes for 1st order expansion of
$\hat{H}_{\rm eff}$. The shaded area is the PSS phase predicted by the
Gutzwiller approach. The red line indicates the estimated limit of
validity of $\hat{H}_{\rm eff}^{(0)}$ (see text). The lattice parameters are $U_{\rm NN}=0.025U$ and $W = -0.95 U$, which can be obtained for
$d_{\perp}=0.37d$. }
\label{FIG:CB_2CB}
\end{figure}

The time dependent Gutzwiller Ansatz for the pairs, is given by
\begin{equation}
\label{EQ:GWPairs}
\ket{\Phi} = \prod_i \sum_m f_{\rm m}^{(i)} \ket{m}_i,
\end{equation}
where we allow the Gutzwiller amplitudes $f_{\rm m}^{(i)} (t)$ to depend on time. The order parameter is readily obtained from (\ref{EQ:GWPairs}) as
\begin{equation}
\psi_i = \bra{\Phi} \hat{c}_i \ket{\Phi} = \sum_m (m+1) f_{\rm m}^{*(i)} f_{\rm m+1}^{(i)}.
\end{equation}
As in Sec.  \ref{SEC:DGA}, equating to zero the variation of the action with respect to $f_{\rm m}^{*(i)}$ leads to the equations
\begin{eqnarray}
\fl
i\hbar \frac{\ud}{\ud t} f_{\rm m}^{(i)} &=&
\Big[Um(m-1) -2\mu m  +  2\Big(U_{NN} - \frac{2J^2}{U}\Big) m \sum_{\langle j \rangle_i} \langle \hat{m}_j \rangle
- \frac{2J^2}{U}z m  \Big] f_{\rm m}^{(i)} \nonumber \\
\fl
&-&\frac{2J^2}{U} \Big[\bar{\psi}_i m f_{\rm m-1}^{(i)} + \bar{\psi}_i^* (m+1) f_{\rm m+1}^{(i)} \Big], \label{EQ:FDynamicsEff2}
\end{eqnarray}
where $\langle \hat{m}_i \rangle= \sum_m m|f_{\rm m}^{(i)}|^2$, the fields $\bar{\psi}_i = \sum_{\langle j \rangle_i} \psi_j$ and
$\sum_{\langle j \rangle_i}  \langle \hat{m}_j\rangle$, have to be calculated self-consistently, and $z = \sum_{\langle j \rangle_i} 1$ is the coordination number in each lattice layer (here $z=4$). The solution of Eq. (\ref{EQ:FDynamicsEff2}) can be
easily obtained numerically, and by making use of the imaginary time evolution we get the ground state phase diagram of Fig. \ref{FIG:CB_2CB} (right).

To get reliable results, one should combine the Gutzwiller
predictions with an estimate of the limits of validity of $\hat{H}_{\rm
eff}^{(0)}$, beyond which the subspace of pairs looses its
meaning. Before starting the discussion on the validity
of the subspace, let us explain how we define the
dominant classical configurations of a given state $\ket{\Phi}$.
It is not difficult to see that Eq. (\ref{EQ:GWPairs}) can be equivalently written as
\begin{equation}
\label{EQ:GWPairs02}
\ket{\Phi} = \sum_{\{\vec{m}\}} g_{\vec{m}}\prod_i \ket{m_i}_i,
\end{equation}
where $\vec{m} = (m_1,...,m_i,..., m_{\scriptscriptstyle \mathrm{N_S}})$ is a collection of the indices $m$ at
each site, and we have introduced the notation such that
$g_{\vec{m}} = \prod_i f_{\rm m_i}^{(i)}$. The advantage of writing
the Gutzwiller state $\ket{\Phi}$ in the form (\ref{EQ:GWPairs02}), lays in the product over single-site Fock
states $\ket{\alpha} = \prod_i \ket{m_i}_i$, which is nothing but a classical distribution of atoms in the lattice.
Therefore we can rewrite Eq. (\ref{EQ:GWPairs02}) as
\begin{equation}
\label{EQ:GWAlphas}
 \ket{\Phi} = \sum_{\{\alpha\}} g_{\rm \alpha}\ket{\alpha}.
\end{equation}
For each point of the phase diagram, from the ground
state Gutzwiller wavefunction, we define the dominant classical
configurations with the criteria
$|g_{\vec{m}}| = |\prod_i f_{\rm m_i}^{(i)}|>0.02$, and $|f_{\rm m_i}^{(i)}|^2>0.05$,
implying that each of the contributing $f_{\rm m_i}^{(i)}$
should also be sufficiently large.
For each of these configurations, we calculate the lobe with respect to single
particle-hole excitations \footnote{We have checked that the validity region is not
strongly modified upon small changes in these conditions.}. If the system at this given point of
the phase diagram turns out to be stable against all dominant
single particle-hole excitations (in other words, if this point is
inside all selected single particle-hole lobes), $\hat{H}_{\rm
eff}^{(0)}$ is considered valid at that point. This procedure is
shown for $J=0.05U$ and $\mu=-0.4375U$ in Fig.~\ref{FIG:CB_2CB} (left),
and gives the red line of Fig. \ref{FIG:CB_2CB} (right). On the right hand side of
this red line, the low energy subspace is not well defined and therefore the effective
Hamiltonian looses its meaning, leaving the description of the system to the
domain of single-particle single-hole excitation theory that predicts
SF and SS phases for each component separately.

To summarize, in the ground state phase diagram of Fig. \ref{FIG:CB_2CB} (right) we identify four different types of phases characterized by different values of the order parameters,
the single-particle order parameters for each species $\varphi_i^{a} = \langle \hat{a}_i \rangle$, $\varphi_i^{b} = \langle \hat{b}_i \rangle$,
and the pair order parameter $\psi_i = \langle \hat{c}_i \rangle$.
\begin{enumerate}
\item The Mott insulating checkerboard phases (MI) characterized by fractional filling factors $\nu$ and vanishing order parameters
$\varphi_i^{a} = \varphi_i^{b} = \psi_i = 0$ inside the lobes.
\item The pair-superfluid phase (PSF) in which the single particle order parameters are zero $\varphi_i^{a} = \varphi_i^{b} = 0$, while the pair order parameter is uniformly non-zero $\psi_i \neq 0$, signaling a finite fraction of superfluid density of the pairs.
\item The pair-supersolid (PSS) characterized by vanishing single-particle order parameters $\varphi_i^{a} = \varphi_i^{b} = 0$, and non vanishing pair order parameter $\psi_i \neq 0$, coexisting with broken translational symmetry, namely, a modulation of both density and order parameter on a scale larger than the one of the lattice
spacing, analogously to the supersolid phase.
\item We infer superfluid and supersolid phases of both components SF$_a$ and SF$_b$ (SS$_a$ and SS$_b$ respectively) with non vanishing single species order parameters $\varphi_i^{a} \neq 0$ and $\varphi_i^{b} \neq 0$, as well as non vanishing $\psi_i \neq 0$, on the right hand side of the red line of Fig. \ref{FIG:CB_2CB} (right), which we estimate to be the limit of validity of the low energy subspace for the parameters considered here.
\end{enumerate}

\subsection{Up-down mixture in a 2D optical lattice}
In most cases, if not all cases considered so far, dipolar gases are polarized, i.e. magnetic or electric dipoles point in the same direction.
This, however, does not have to be always the case. A prominent example concerns molecules that follow the Hund's rule (a) \cite{BB:Hund}. For such molecules the electric dipole is either parallel or anti-parallel to the direction of the magnetic moment. Thus, if one takes a sample of such molecules and polarizes it magnetically, say in the up direction, one will obtain in this way a dipolar gas with certain fraction of electric dipoles pointing up, and the remaining fraction pointing down. In fact there was recently a spectacular progress in cooling and trapping of magnetically confined neutral OH molecules. This progress opens important perspectives to study novel quantum phases with ultracold dipoles \cite{BB:Lev01, BB:Lev02}.

\begin{center}
\begin{figure}[t]
\begin{center}
\includegraphics[width=0.6\linewidth]{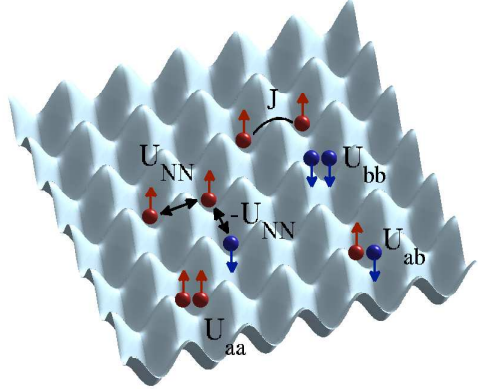}
\end{center}
\caption{Schematic representation of a 2D
optical lattice populated with dipolar Bosons polarized
in both directions perpendicular to the lattice plane. The particles feel repulsive
intra-species $U_{aa}$, $U_{bb}$, and inter-species $U_{ab}$ repulsive on-site energies.
The nearest-neighbor interaction is repulsive $U_{\rm NN}>0$ for aligned dipoles, while it is attractive $-U_{\rm NN}$ for anti-aligned particles. The hopping term $J$ is equal for both species.}
\label{FIG:UpDown}
\end{figure}
\end{center}

In \cite{BB:TrefzgerCSS} we consider a sample of dipoles in the presence of a 2D optical lattice, and an extra
confinement in the perpendicular direction. The dipoles are free to point in both directions normal to the lattice plane, which results in a nearest neighbor interaction either repulsive for aligned dipoles, or attractive for anti-aligned one, as shown in Fig. \ref{FIG:UpDown}.
The system is described by the Hamiltonian
\begin{eqnarray}
\hat{H} &=& \sum_{i,\sigma} \left[ \frac{U_\sigma}{2} \hat{n}_i^\sigma (\hat{n}_i^\sigma - 1) +  \frac{U_{\sigma\sigma^\prime}}{2}\hat{n}_i^\sigma \hat{n}_i^{\sigma^\prime} -  \mu_\sigma \hat{n}_i^\sigma \right]\nonumber \\
          &+& \frac{1}{2}\sum_{ i\neq j,\sigma}\frac{U_{NN}}{|r_{ij}|^3} \left[ \hat{n}_i^\sigma \hat{n}_j^\sigma - \hat{n}_i^\sigma \hat{n}_j^{\sigma^\prime}\right]  - J\sum_{\langle ij \rangle} [\hat{a}_i^\dag \hat{a}_j + \hat{b}_i^\dag \hat{b}_j ],\nonumber\\
          \label{EQ:H3D}
\end{eqnarray}
where  $\sigma = \big(a, b\big)$ indicates the two species, i.e. dipoles pointing in the up and
down direction perpendicular to the 2D plane of the lattice, respectively.
$U_{aa}$ and $U_{bb}$ are the on-site
energies for particles of the same species, while $U_{ab}$ is the on-site energy for different species.
The long-range dipolar interaction potential decays as the inverse cubic power of the relative distance $r_{ij}$, which we
express it in units of the lattice spacing. For computational reasons, in most theoretical approaches the range is cutoff at a certain neighbors.
In the present work, we consider a range of interaction up to the 4th nearest neighbor.
The first nearest neighbor dipolar interaction is attractive (repulsive) for particles of the same
(different) species, with strength $U_{NN}>0$ (or respectively $-U_{NN}$),
and we consider an equal tunneling coefficient $(J)$ for both species, while
the densities of the dipoles pointing upwards and downwards are fixed by the corresponding chemical potentials $\mu_\sigma$.
The on-site interactions have two contributions (see Eq. \ref{EQ:OnSiteU}): one is arising from the s-wave scattering length, and the second one is due to the on-site dipole-dipole interaction. We consider that the s-wave scattering length is independent on the orientation of the dipoles. Instead, the on-site
dipolar contribution $U_{\rm dd}$ depends both on the orientation of the dipoles and on the geometry of the trapping potential, and it can be varied by changing the
ratio between the vertical to the axial confinement.
For simplicity we will focus on the specific case of a spherically symmetric confinement, where the on-site dipolar interactions
average out to zero $U_{\rm dd} = 0$, and the resulting on-site interactions are all equal to $U$.
We consider the case of dipole-dipole interactions to be 600 times weaker with respect to the on-site
interaction, i.e. $U_{\rm NN} = U/600$ \footnote{In standard experiments with $^{52}$Cr, which features a magnetic moment of $\mu = 6\mu_{\scriptscriptstyle \mathrm{B}}$,
this ratio is given by $U_{\rm NN} \simeq U/400$, for an optical lattice depth of $20E_{\rm R}$, where $E_{\rm R}$ is the recoil
energy at $\lambda = 500$nm.}.

The properties of the system are conveniently extracted using the operators given by the sum (filling factor) and by the difference (imbalance) of the two species number operators at each site of the lattice, namely by
\begin{equation}
\label{EQ:PMOperators}
\hat{\nu}_i = \frac{\hat{n}_i^a + \hat{n}_i^b}{2}, \;\;\;  \hat{m}_i = \frac{\hat{n}_i^a - \hat{n}_i^b}{2},
\end{equation}
which are simultaneously diagonal on a given Fock state $\ket{\nu,m}_i$.
Notice that the eigenvalues of these two operators are not independent. In fact, by fixing $\nu$ the eigenvalues of $\hat{m}_i$ can only assume
$2\nu + 1$ values given by $m = \{-\nu,  -\nu+1,..., +\nu\}$, in complete analogy with the angular momentum operator $\hat{S}_i^2$ and its projection along the
$z$ axis $\hat{S}_i^z$. It is also useful to introduce the average magnetization of the system, defined as
\begin{equation}
M = \frac{1}{N_{\scriptscriptstyle \mathrm{S}}}\sum_i m_i,
\end{equation}
where $N_{\scriptscriptstyle \mathrm{S}}$ is the total number of lattice sites.

Substituting Eqs. (\ref{EQ:PMOperators}) into Eq. (\ref{EQ:H3D}) allows us to express the system Hamiltonian as $\hat{H} = \hat{H}_{\rm 0}^\nu + \hat{H}_{\rm 0}^m + \hat{H}_{\rm 1}^{\nu m}$,
where
\begin{eqnarray}
\hat{H}_{\rm 0}^\nu       &=& \sum_i \Big[-2\mu_{\scriptscriptstyle \mathrm{+}} \hat{\nu}_i + 2U\hat{\nu}_i\Big(\hat{\nu}_i - \frac{1}{2}\Big)\Big]\label{EQ:HNUM01} \\
\hat{H}_{\rm 0}^m         &=&\sum_i \Big[-2\mu_{\scriptscriptstyle \mathrm{-}} \hat{m}_i + 2U_{NN}\sum_{j\neq i}\frac{\hat{m}_i\hat{m}_j}{|r_{ij}|^3} \Big]\label{EQ:HNUM02} \\
\hat{H}_{\rm 1}^{\nu m} &=& - J\sum_{\langle ij \rangle} [\hat{a}_i^\dag \hat{a}_j + \hat{b}_i^\dag \hat{b}_j ]. \label{EQ:HNUM03}
\end{eqnarray}
In Eqs. (\ref{EQ:HNUM01},\ref{EQ:HNUM02}), we have introduced the chemical potentials
\begin{equation}
\mu_\pm = \frac{\mu_a \pm \mu_b}{2},
\end{equation}
which respectively fix the eigenvalues of the filling factor ($+$), and the imbalance operators ($-$) in Eq. (\ref{EQ:PMOperators}).
In the following we consider $\hat{H}_{\rm 1}^{\nu m}$ to be a small perturbation on the interaction terms (\ref{EQ:HNUM01}) and (\ref{EQ:HNUM02}).

\subsubsection{Low-energy subspace and effective Hamiltonian}{--- }
In the limit where $U \gg U_{NN}$, there exist a low-energy subspace spanned by the states
\begin{equation}
\label{EQ:Alphas01}
\ket{\alpha} = \prod_i \ket{\nu,m_i}_i,
\end{equation}
with uniform total on-site occupation $2\nu$. Single particle hopping changes the total on-site population, the energy cost of these excitations is of the order of the on-site interaction energy $U$, and becomes very large in the limit where $U \gg \big(U_{NN},J\big)$.
Thus, a successful description of such a system is obtained through an effective Hamiltonian $\hat{H}_{\rm eff}$ restricted to the low-energy subspace, where single-particle hopping is suppressed and tunneling is included at second order in perturbation theory.

\begin{center}
\begin{figure}[h!]
\begin{center}
\includegraphics[width=0.6\linewidth]{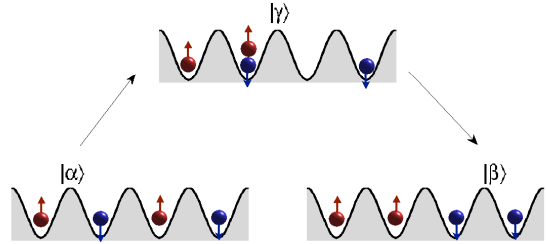}
\end{center}
\caption{Schematic representation of a two-particle hopping between the states $\ket{\alpha}$ and $\ket{\beta}$, belonging to the low-energy subspace at
$\nu = 1/2$. These states are coupled through virtual excitations to the states $\ket{\gamma}$ by single-particle hopping.}
\label{FIG:SubspacePH}
\end{figure}
\end{center}

This situation is in fact similar to the one discussed in Sec. \ref{SEC:TheModel} for a bilayer optical lattice, and therefore we apply
the same technique to compute $\hat{H}_{\rm eff}$. In the basis of constant total on-site population $2\nu$, using Eq. (\ref{EQ:DefHeff}) we calculate the matrix elements of such a Hamiltonian at second order in perturbation theory, where the subspace of virtual excitations $\ket{\gamma}$ is obtained from the states $\ket{\alpha}$ via single particle hopping, as schematically represented in Fig. \ref{FIG:SubspacePH}.
For a given state $|\alpha\rangle$,
\begin{eqnarray}
\label{EQ:Den02}
E_{\gamma_{ij}}-E_\alpha = U + U_{NN} \Delta m^{ij}_{\rm NN},
\end{eqnarray}
with $\Delta m^{ij}_{\rm NN}=\sum_{k\neq i}2m_k/|r_{ik}|^3 - \sum_{k\neq j} 2m_{k}/|r_{jk}|^3 - 1$, where $m_i$
indicates the population imbalance at site $i$ of Eq. (\ref{EQ:PMOperators}). For $U \gg U_{NN}$, the denominators
$E_{\gamma_{ij}}-E_\alpha$ are all of order $U$, which leads to
\begin{eqnarray}
\hat{H}_{\rm eff}^{(0)}  &=&   \hat{H}_{\rm 0}^\nu - \frac{2J^2}{U} \sum_{\langle ij \rangle} \hat{\nu}_i(\hat{\nu}_j+1) \nonumber \\
&+& \hat{H}_{\rm 0}^m - \frac{2J^2}{U} \sum_{\langle ij \rangle} \left[ \hat{m}_i \hat{m}_j + \hat{c}_i^\dag \hat{c}_j \right], \label{Heff0_num}
\end{eqnarray}
where $\hat{c}_i=\hat{a}_i \hat{b}_i^\dag$ and $\hat{c}_i^\dag = \hat{a}_i^\dag \hat{b}_i$ are composite operators, corresponding to the creation of a particle of one species and a hole of the other species, such that
\begin{equation}
\label{EQ:CreationAnn}
\eqalign{
\hat{c}_i |\nu,m_i\rangle &= \sqrt{\nu(\nu+1) - m_i(m_i-1)}|\nu,m_i - 1\rangle\\
\hat{c}_i^\dag|\nu,m_i\rangle &= \sqrt{\nu(\nu+1) - m_i(m_i+1)}|\nu,m_i + 1\rangle.
}
\end{equation}

\subsubsection{Insulating lobes}{--- }
In the limit where $U \gg U_{\rm NN}$, asymptotically all classical states
$\ket{\alpha}$ become stable with respect to single-particle-hole excitations
and develop an insulating lobe at finite $J$. As single particle hopping changes the total on-site population, it breaks the translational invariance of the ground state with respect to the total on-site occupation $2\nu$.
The energy cost of these excitations is of the order of the on-site interaction energy $U$, and is therefore very large in the limit where $U \gg \big(U_{NN},J\big)$.

Instead, the lowest-lying excitations within the subspace are obtained by adding (PH) or removing (HP) one composite, made of a particle of
the upper-polarized dipoles (species $a$) and a hole of the lower-polarized dipoles (species $b$), at the $i$-th site of the lattice.
For any given configuration $\ket{\alpha}$, one can calculate the corresponding energy costs from the diagonal terms of the effective Hamiltonian (\ref{Heff0_num}), which are respectively given by
\begin{equation}
\eqalign{
\label{EQ:PHE_CSS}
E_{\scriptscriptstyle \mathrm{PH}}^i(J) &=-2\mu_{\scriptscriptstyle \mathrm{-}} + 4U_{\rm NN}\sum_{k\neq i}\frac{m_k}{|r_{ik}|^3} - \frac{4J^2}{U}\sum_{\langle k \rangle_i}m_k,\\
E_{\scriptscriptstyle \mathrm{HP}}^i(J) &= 2\mu_{\scriptscriptstyle \mathrm{-}} - 4U_{\rm NN}\sum_{k\neq i}\frac{m_k}{|r_{ik}|^3} + \frac{4J^2}{U}\sum_{\langle k \rangle_i}m_k.
}
\end{equation}
The order parameters $\psi_i=\langle \hat{c}_i \rangle$
for $\ket{\alpha}$, are readily calculated as in Sec. \ref{SEC:Semi}, and satisfy the equations
\begin{eqnarray}
\fl
 \psi_i = \frac{2J^2}{U}\Big[ \frac{\nu(\nu+1) - m_i(m_i+1)}{E_{\scriptscriptstyle \mathrm{PH}}^i(J)}
  \; + \; \frac{\nu(\nu+1) - m_i(m_i-1)}{E_{\scriptscriptstyle \mathrm{HP}}^i(J)} \Big]\bar{\psi_i},\label{Eqn:MFCSS}
\end{eqnarray}
where $\bar{\psi_i} = \sum_{\langle j \rangle_i} \psi_j$. With Eqs. (\ref{Eqn:MFCSS}) one can calculate
the mean-field lobes of any distribution of atoms in the lattice $\ket{\alpha}$.
\begin{center}
\begin{figure}[h!]
\begin{center}
\includegraphics[width=0.6\linewidth]{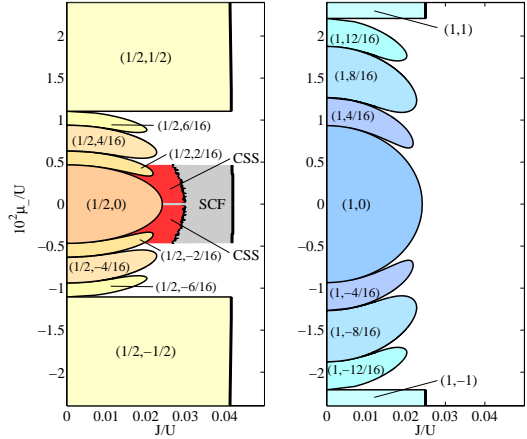}
\end{center}
\caption{Ground state of a $4\times4$ square lattice satisfying periodic boundary conditions, for $\nu=1/2$ at
$\mu_{\scriptscriptstyle \mathrm{+}} = 0.5U$ (left), and $\nu=1$ at $\mu_{\scriptscriptstyle \mathrm{+}} = 1.4U$ (right),
and $U_{\rm NN} = U/600$. The text in parentheses $(\nu,M)$ indicate the filling factor $\nu$, and the average magnetization $M$, respectively.}
\label{FIG:GroundStateCSS}
\end{figure}
\end{center}
In Fig. \ref{FIG:GroundStateCSS} we plot the ground state insulating lobes calculated in this way for $\nu=1/2$ (left) and $\nu=1$ (right).
For all filling factors $\nu$, we find an anti-ferromagnetic (AM) ground state $(\nu,M=0)$. The AM insulating lobes are symmetric with respect to the $\mu_{\scriptscriptstyle \mathrm{-}} = 0$ axis, and present a spatial distribution of alternating
sites occupied by particles of species $a$ and $b$ resembling a checkerboard structure. Remarkably, the larger the $\nu$, the more stable is
the AM ordering with respect to flipping the direction of a dipole. By increasing the absolute value of $\mu_{\scriptscriptstyle \mathrm{-}}$ we find RM ground states with rational values of the average magnetization, corresponding to $M = (\pm 2\nu, \; \pm 4\nu, \; \pm 6\nu)/8$. The exact fractional values of $M$ in the ground state, depend on the cutoff range in the dipolar interactions, and on the size of the lattice. We have used a $4\times 4$ elementary cell with periodic boundary conditions, and dipolar interaction range cut at the 4th nearest neighbor. By considering more neighbors in the interactions, and larger lattices, we expect to find RM states appearing at all rational $M$, asymptotically approaching a Devil's staircase as recently shown in \cite{BB:Barbara2, BB:Cooper}. Finally we find a FM ground state $(\nu,M=\pm \nu)$, in which only particles of one type are present.

\begin{center}
\begin{figure}[h!]
\begin{center}
\includegraphics[width=0.6\linewidth]{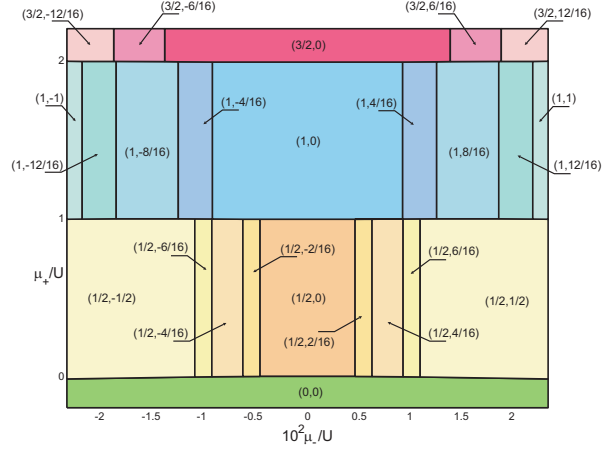}
\end{center}
\caption{Ground state of the system at $J=0$, calculated for a $4\times 4$ elementary cell satisfying periodic boundary conditions, and
$U_{\rm NN} = U/600$. The text in parentheses $(\nu,M)$ indicates the filling factor $\nu$ and the average magnetization $M$.}
\label{FIG:GSJ0}
\end{figure}
\end{center}

It is worth noticing that the insulating lobes calculated in this way, do not contain any dependence on $\mu_{\scriptscriptstyle \mathrm{+}}$,
which does not enter into Eqs. (\ref{Eqn:MFCSS}).
Therefore, for any given value of $\mu_{\scriptscriptstyle \mathrm{+}}$, in order to obtain the ground state phase diagram one has to compare the energies of the ground state configurations at different $\nu$. Using the effective Hamiltonian (\ref{Heff0_num}), for any value of $\mu_{\scriptscriptstyle \mathrm{+}}$, $J$, and $\mu_{\scriptscriptstyle \mathrm{-}}$, we compare the energies of the ground state configurations for different $\nu$, and select the state with the smaller energy. In this way we have obtained the phase diagram at $J=0$ shown in Fig. \ref{FIG:GSJ0}.

\subsubsection{Counterflow superfluid/supersolid}{--- }
In the low-energy subspace at constant $\nu$, the Gutzwiller Ansatz on the
wave function of the system reads
\begin{equation}
\label{EQ:NuConstant}
\ket{\Phi} = \prod_i \sum_{\rm m=-\nu}^{\rm \nu} f_{\rm \nu,m}^{(i)} \ket{\nu,m}_i,
\end{equation}
where we allow the Gutzwiller amplitudes $f_{\rm \nu,m}^{(i)}(t)$ to depend on time.
We obtain the equations of motion for the amplitudes by minimizing the action of the system with respect to the variational parameters $f_{\rm \nu,m}^{*(i)}(t)$,
\begin{eqnarray}
i\hbar \frac{\ud}{\ud t} f_{\rm \nu,m}^{(i)} &=&
\Big[-2\mu_{\scriptscriptstyle \mathrm{-}} - \frac{4J^2}{U}\sum_{\langle j \rangle_i} \langle \hat{m}_j\rangle \nonumber \\&+& 4U_{NN}\sum_{j\neq i}  \frac{\langle\hat{m}_j \rangle}{|r_{ij}|^3} \Big] m f_{\rm \nu,m}^{(i)} \nonumber \\
&-& \frac{2J^2}{U} \Big[\bar{\psi}_i \sqrt{\nu(\nu+1) - m(m-1)} \; f_{\rm \nu,m-1}^{(i)} \nonumber \\
&+& \bar{\psi}_i^*\sqrt{\nu(\nu+1) - m(m+1)} \; f_{\rm \nu,m+1}^{(i)} \Big]  \label{EQ:FDynamicsEffCSS},
\end{eqnarray}
where $\langle \hat{m}_i \rangle= \sum_{\rm m=-\nu}^{\rm \nu} m |f_{\rm \nu,m}^{(i)}|^2$, the fields $\bar{\psi}_i = \sum_{\langle j \rangle_i} \psi_j$,
$\sum_{\langle j \rangle_i}  \langle \hat{m}_j\rangle$, and $\sum_{j\neq i}  \langle \hat{m}_j\rangle/|r_{ij}|^3$ have to be calculated in a self consistent way, and the
order parameter $\psi_i = \bra{\Phi} \hat{c}_i \ket{\Phi}$ is given by
\begin{equation}
\psi_i = \sum_{\rm m=-\nu}^{\rm \nu} \sqrt{\nu(\nu+1) - m(m+1)} \; f_{\rm \nu,m}^{*(i)} f_{\rm \nu,m+1}^{(i)}.
\end{equation}
We solve Eqs. (\ref{EQ:FDynamicsEffCSS}) in imaginary time $\tau=it$, and in Fig. \ref{FIG:GroundStateCSS} we show the ground state phase diagram of the system for $\nu = 1/2$ (left) computed in this way.

For $\nu=1/2$, in the region immediately outside the insulating AM lobe, depending on the values of $J$
and $\mu_{\scriptscriptstyle \mathrm{-}}$ we find either super-counter-fluid SCF or counterflow-supersolid CSS.
In the SCF phase, while the single-particle order parameters vanish $\langle \hat{a}_i \rangle = \langle \hat{b}_i \rangle = 0$,
the composite order parameters are non-zero $\langle \hat{c}_i \rangle \neq 0$, indicating the presence of counterflow \cite{BB:Kuklov}.
The CSS is characterized by vanishing single-particle order parameters $\langle \hat{a}_i \rangle = \langle \hat{b}_i \rangle = 0$,
and non-vanishing composite order parameters $\langle \hat{c}_i \rangle \neq 0$, coexisting with broken translational
symmetry, namely, a modulation of both $m_i$, and $\langle \hat{c}_i \rangle$ on a scale larger than the one of the lattice
spacing, analogously to the supersolid phase. Note that we don't find any evidence of CSS at
$\mu_{\scriptscriptstyle \mathrm{-}} = 0$, which indicates that a finite imbalance between the two
species is a necessary condition for the system in order to sustain CSS.
Finally, with a similar method described in Sec. \ref{SEC:BilayerSFSS}, we estimate the limits of validity of the effective Hamiltonian to be given by the vertical thick lines of Fig. \ref{FIG:GroundStateCSS}.

\section{Path Integral Monte Carlo and the Worm algorithm}
Any physical system consisting of $N$ non-relativistic particles can be in principle described by the many-body Schr{\"o}dinger equation. In three dimensions (3D),
the number of degrees of freedom in the Schr{\"o}dinger equation becomes 3 times $N$. For typical physical systems such as electrons in conducting materials or BEC that have a large number of constituents $N$, the Schr{\"o}dinger equation becomes difficult to solve exactly in a reasonable amount of time even for parallel computing. Monte Carlo methods overcome this problem, they allow for a description of the many-body system relying on repeated random sampling, at the cost of statistical uncertainty which can be reduced with more simulation time. The typical basic steps of a Monte Carlo algorithm can be summarized as follows
\begin{enumerate}
\item Define the {\it configuration space} (here by configuration we mean a collection of indices, see below).
\item Generate configurations randomly and accept them with a certain probability which depends on the specific problem. This is called
the {\it updating procedure}.
\item Perform a computation, i.e. calculate quantities of interest, based on the randomly generated configurations.
\item Add the result of the computation to the final result.

\end{enumerate}

In the spirit of the Metropolis-Hastings algorithm \cite{BB:Metropolis,BB:Hastings}, two requirements must be satisfied: a) ergodicity, that is, given an initial configuration it has to be possible to reach any other allowed configuration via the updating procedure; b) the probability of having a certain configuration appearing in sums which calculate quantities of interest has to be proportional to its Boltzmann weight.

There is a large class of Quantum Monte Carlo methods that can simulate quantum many-body systems, like for example the Variational Monte Carlo
\cite{BB:Polls, BB:Variational}, the Diffusion
Monte Carlo \cite{BB:Diffusion01, BB:Greg, BB:Greg01}, the Path Integral Monte Carlo \cite{BB:PathIN, BB:Worm,BB:Worm2}, auxiliary field Monte Carlo \cite{BB:Auxiliary01, BB:Auxiliary02}, etc. Most methods aim at computing the ground-state wavefunction of the system, with the exception of Path Integral Monte Carlo, and finite-temperature auxiliary field Monte Carlo, which calculate the density matrix. The results presented in this work are based on the Path Integral Monte Carlo (PIMC) and the Worm algorithm (WA), which was originally developed by Prokof'ev, Svistunov and Tupitsyn \cite{BB:Worm,BB:Worm2}.

\subsection{Path Integral Monte Carlo}
\label{SEC:PIMC}
Consider a system described by the Hamiltonian $\hat{H} = \hat{H}_0 + \hat{H}_1$, where $\hat{H}_0$ is diagonal in the basis
of eigenstates $\{\ket{\alpha}\}$ satisfying the eigenvalue equation
\begin{equation}
\hat{H}_0 \ket{\alpha} = E_\alpha  \ket{\alpha},
\end{equation}
and $\hat{H}_1$ is non-diagonal. The thermodynamic properties of the system at equilibrium, can be derived from the partition function which is given by the
trace of the density matrix operator $Z = {\rm Tr} \left[e^{-\beta \hat{H}} \right]$, where $\beta = 1/K_BT$ is the inverse temperature and $K_B$
the Boltzmann constant. In the interaction picture \cite{BB:Tannoudji01} one may write
\begin{equation}
\label{EQ:Z1}
Z = {\rm Tr} \left[ e^{-\beta(\hat{H}_0 + \hat{H}_1)}\right] = {\rm Tr} \left[ e^{-\beta \hat{H}_0} \hat{\mathcal{T}}_\tau e^{-\int_0^\beta \ud \tau \hat{H}_1(\tau)}  \right],
\end{equation}
where $\hat{\mathcal{T}}_\tau$ is the time-ordering operator, $\hat{H}_1(\tau) = e^{\tau \hat{H}_0} \hat{H}_1 e^{-\tau \hat{H}_0}$ is the non-diagonal part of the Hamiltonian expressed in the interaction picture, and the variable $\tau$ is usually called the {\it imaginary time} \footnote{This is because by replacing $\tau = it$, with $t$ being the real time, the operator $e^{-\tau\hat{H}}$ becomes the usual time-evolution operator in quantum mechanics.}.
One can write the partition function using the Feynman path integral formulation, and by Taylor expanding the second exponent in the right-hand-side of Eq. (\ref{EQ:Z1}) one gets
\begin{eqnarray}
Z &=& \sum_{\alpha} e^{-\beta E_\alpha}\bra{\alpha} \hat{\openone} - \int_0^\beta \ud \tau \hat{H}_1(\tau) + \nonumber \\
   &+& \sum_{m=2}^\infty (-1)^m \int_0^\beta\ud \tau_{\rm m}\;..\; \int_0^{\tau_2}\ud \tau_1 \hat{H}_1(\tau_{\rm m}) \;..\; \hat{H}_1(\tau_1)\ket{\alpha}, \label{EQ:Z2}
\end{eqnarray}
where the integrals are ordered in time and the sum over the states $\ket{\alpha}$ comes from the trace.
Now we explicitly make use of the completeness property of the $\{\ket{\alpha}\}$ basis, and insert $m-1$ identity operators
$\hat{\openone} = \sum_{\alpha} \ket{\alpha} \bra{\alpha}$ between the products of $\hat{H}_1(\tau_{\rm m})$ operators, therefore we can write
\begin{eqnarray}
\fl
\bra{\alpha} \hat{H}_1(\tau_m) \;..\; \hat{H}_1(\tau_1) \ket{\alpha} = \sum_{\alpha_1,..,\alpha_{\rm m-1}} H_1^{\alpha \alpha_{\rm m-1}} (\tau_{\rm m})\;..\;
H_1^{\alpha_2 \alpha_1}(\tau_2)H_1^{\alpha_1 \alpha}(\tau_1),
\end{eqnarray}
where the matrix elements
\begin{equation}
\label{EQ:MatrixElements}
H_1^{\alpha^\prime \alpha}(\tau) = e^{\tau E_{\alpha^\prime}} H_1^{\alpha^\prime \alpha} e^{-\tau E_{\alpha}} = \bra{\alpha^\prime} \hat{H}_1 \ket{\alpha}
e^{-\tau(E_{\alpha} - E_{\alpha^\prime})},
\end{equation}
contain both diagonal ($E_\alpha$) and off-diagonal ($H_1^{\alpha^\prime \alpha}$) matrix elements. We now insert the last equation into expression
(\ref{EQ:Z2}) and get the final expression for the partition function
\begin{eqnarray}
\fl
Z &=& \sum_\alpha e^{-\beta E_\alpha}\Big\{1 -  \int_0^\beta \ud \tau H_1^{\alpha\alpha} (\tau)  +   \label{EQ:Z}\\
\fl
   &+&   \sum_{\rm m=2}^\infty(-1)^m \int_0^\beta \ud \tau_{\rm m} \;..\;  \int_0^{\tau_2} \ud \tau_1 \sum_{\alpha_1,..,\alpha_{\rm m-1}}
     H_1^{\alpha\alpha_{\rm m-1}} (\tau_{\rm m}) \;..\;  H_1^{\alpha_1\alpha} (\tau_1) \Big\}, \nonumber
\end{eqnarray}
which contains only matrix elements of the operators $\hat{H}_0$ and $\hat{H}_1$. Therefore, by using this formalism of path integrals,
the calculation of the partition function reduces to a classical problem since only scalars enter into Eq. (\ref{EQ:Z}), but we have payed
the price of the extra dimension $\tau$. In other words, the original $d$-dimensional quantum system is equivalent to a ($d+1$)-dimensional
classical system.

It is worth noticing that since the partition function is a trace, periodic boundary conditions in the imaginary time $\tau$ must apply. This is
easily understood by looking at the $m$-th order term of $Z$, which contains the product of $m$ matrix elements
$H_1^{\alpha\alpha_{\rm m-1}} (\tau_{\rm m}) \;..\;  H_1^{\alpha_1\alpha} (\tau_1)$ that are ordered in time from the first at $\tau_1$, to the last
at $\tau_m$. Therefore, for any given $\alpha$ in the trace, the first matrix element brings $\alpha$ to some $\alpha_1$ at time $\tau_1 \geq 0$,
while the last matrix element brings $\alpha_{\rm m-1}$ back to $\alpha$ at time $\tau_{\rm m} \leq \beta$.
All the possible configurations which are periodic in imaginary time and that enter into the expression for the partition function Eq. (\ref{EQ:Z}), define the configuration space spanned by a PIMC algorithm.

\subsubsection{Path Integral Monte Carlo and the 2D extended Bose-Hubbard model}{--- }
\label{SEC:PIMC_BH}
We now consider a 2D system of $L\times L$ sites filled with polarized dipolar Bosons, we assume spatial periodic boundary conditions and the dipoles to be polarized  perpendicularly to the 2D plane as explained in Sec. \ref{CH:MS:DipolarBosons}. The system is therefore described by the extended Bose-Hubbard Hamiltonian (\ref{EQ:EBH2D}), which, to be consistent with the notations in our publication \cite{BB:Barbara2}, we rewrite in this form
\begin{equation}
\label{EQ:MCH}
\hat{H} = - J \sum_{\langle i j\rangle} \hat{b}^{\dag}_i \hat{b}_j +  \sum_i \left[\frac{U}{2}\hat{n}_i(\hat{n}_i-1) -\mu_i \hat{n}_i\right]   +
V\sum_{i<j} \frac{\hat{n}_i \hat{n}_j}{r_{ij}^3},
\end{equation}
where $\hat{b}^{\dag}_i$ ($\hat{b}_i$) is the boson creation (annihilation) operator at site $i$, $\hat{n}_i =  \hat{b}^{\dag}_i \hat{b}_i$ is the number operator, $V = D/a^3 >0$
is the dipole-dipole interaction strength $D$ divided by the lattice spacing $a$, $r_{ij} = |i-j|$ is the distance between two sites of the lattice, and $\mu_i = \mu - \Omega i^2$
contains the chemical potential $\mu$ which fixes the number of particles, and the curvature $\Omega$ of an external harmonic confinement.

We choose to work in the basis of the interaction term of the Hamiltonian (\ref{EQ:MCH}), i.e. Fock states $\ket{\alpha} = \prod_i^{L^2} \ket{n_i}_i$ of localized particles in the $L\times L$ square lattice, where $n_i$ is the occupation number at site $i$. Therefore in this basis, the diagonal matrix elements entering Eq. (\ref{EQ:MatrixElements}) take the form
 \begin{equation}
 \label{EQ:Diagonal}
 E_\alpha = \frac{U}{2}\sum_i n_i(n_i-1) - \sum_i \mu_in_i + V\sum_{i<j} \frac{n_i n_j}{r_{ij}^3}.
 \end{equation}
 The off-diagonal ones are given by the expression
 \begin{equation}
 \label{EQ:OffDiagonal}
 -H_1^{\alpha^\prime\alpha} = J\bra{\alpha^\prime} \hat{b}^{\dag}_i \hat{b}_j \ket{\alpha} = J \sqrt{(n_i^{\alpha}+1)n_j^{\alpha}},
 \end{equation}
 and they connect states $\ket{\alpha^\prime}$ and $\ket{\alpha}$ that differ only in the occupation number of the two nearest neighboring sites $i$ and $j$, namely
 $\ket{\alpha^\prime} \equiv \frac{\hat{b}^{\dag}_i \hat{b}_j}{\sqrt{(n_i^{\alpha}+1)n_j^{\alpha}}} \ket{\alpha} $ with $n_i^\alpha$ being the integer number of particles at the $i$-th site of the state $\ket{\alpha}$.

\begin{center}
\begin{figure}[h]
\begin{center}
\includegraphics[width=0.8\linewidth]{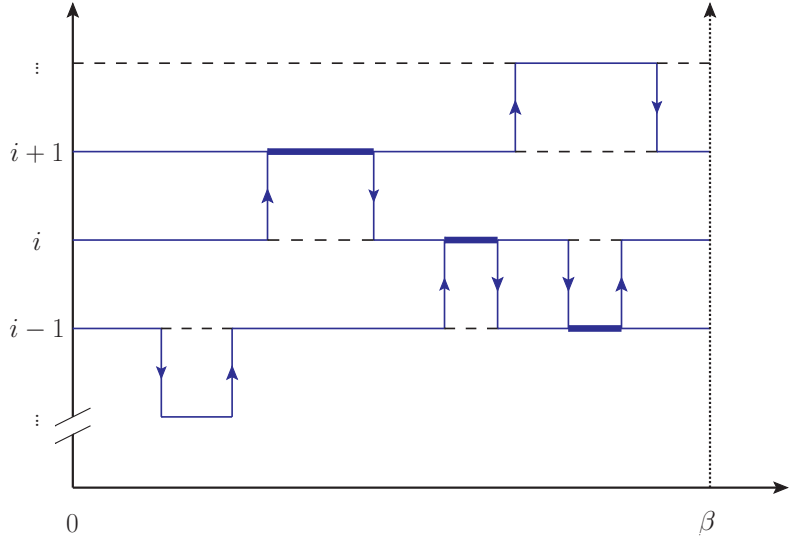}
\end{center}
\caption{Schematic representation of one configuration which enters the calculation of $Z_{\scriptscriptstyle \rm eBH}$. Each line is called a worldline and it represents a trajectory of a particle in imaginary time. Configurations have to fulfill periodic boundary conditions in the imaginary time $\tau$, owing to the definition of the partition function as a trace. Vertical arrows correspond to changes of the state of the system, and are called kinks. In the sketch, the thickness of intervals between the kinks shows the number of particles: the dashed black line is for $n$ particles while the solid and bold blue lines have occupation numbers equal to $n+1$ and $n+2$ respectively. }
\label{FIG:Configuration}
\end{figure}
\end{center}

To write the partition function for the 2D extended Bose-Hubbard model, we notice that the first order term vanishes since the matrix elements (\ref{EQ:OffDiagonal}) are off-diagonal, i.e. $H_1^{\alpha\alpha} =0$, and due to the geometry of the system (2D square lattice) it is not difficult to see that all the terms with an odd value of $m$
also vanish, owing to periodic boundary conditions in imaginary time. Therefore by rearranging the exponentials and renaming $\alpha \equiv \alpha_0$, we get to the expression
\begin{eqnarray}
\label{EQ:ZeBH0}
\fl
Z_{\scriptscriptstyle \rm eBH} &=& \sum_{\alpha_0} e^{-\beta E_{\alpha_0}} + \sum_{\rm m=2}^\infty
J^m A_{\rm m} \times \\
\fl
&\times& \int_0^\beta \ud \tau_{\rm m} \;..\;  \int_0^{\tau_2} \ud \tau_1 \sum_{\alpha_0,\alpha_1,..,\alpha_{\rm m-1}}
\exp \Big\{-\beta E_{\alpha_0} -\sum_{p=0}^{m-1} E_{\alpha_p}(\tau_{p+1} - \tau_p)\Big\}, \nonumber
\end{eqnarray}
where $A_{\rm m}$ is a product of $m$ square root factors coming from Eq. (\ref{EQ:OffDiagonal}) and we have introduced $\tau_0 = \tau_m$ to compact the notation.
We can compact further the notation by defining $A_{\rm m=0} = 1$, and keeping in mind that for $m=0$ there is no summation in the exponent of Eq. (\ref{EQ:ZeBH0})
we then write the partition function as follows
\begin{eqnarray}
\label{EQ:ZeBHCompact}
\fl
Z_{\scriptscriptstyle \rm eBH} &=& \sum_{\rm m=0}^\infty \; \sum_{\alpha_0,\alpha_1,..,\alpha_{\rm m-1}}  J^m A_{\rm m}  \times \\
\fl
&\times&\int_0^\beta \ud \tau_{\rm m} \;..\;  \int_0^{\tau_2} \ud \tau_1
\exp \Big\{-\beta E_{\alpha_0} -\sum_{p=0}^{m-1} E_{\alpha_p}(\tau_{p+1} - \tau_p)\Big\}. \nonumber
\end{eqnarray}
From the last expression one can formally write
\begin{equation}
Z_{\scriptscriptstyle \rm eBH} = \sum_{\rm \nu} W_{\rm \nu},
\end{equation}
with $W_{\rm \nu}$ being the weight of each configuration $\nu \equiv \left[m,\alpha_0(\tau_1),\alpha_1(\tau_2),..,\alpha_{\rm m-1}(\tau_m)\right]$. A configuration can be
pictorially represented as in Fig. \ref{FIG:Configuration}, with imaginary time $\tau$ on the horizontal axis, and lattice sites on the vertical axis. 
Each line represents a trajectory of a particle in imaginary time and is called a {\bf worldline}. The latter has to close on itself owing to the fact that the partition function is a trace. Moreover if one assumes spatial periodic boundary conditions one can imagine the configuration of Fig. \ref{FIG:Configuration} to be wrapped on a torus (in the case of one dimensional systems). We call the phase space of all possible configurations the {\it closed path configuration space} (CP).

If we cut one configuration at a certain instant in imaginary time, we get the system in a particular quantum state. The points in imaginary time where the system changes state are called {\bf kinks}, which in Fig. \ref{FIG:Configuration} are represented by vertical arrows. A configuration with a number of kinks equal to $m$,
contributes to the $m$-th order term of the partition function Eq. (\ref{EQ:ZeBHCompact}), and it is straightforward to see that there exist an infinite number of different configurations with the same number of kinks, the difference being the time at which the kinks take place and/or the different states they connect. The updating procedure of a PIMC algorithm therefore consists of changing the number of kinks and/or their position in imaginary time. We will discuss the updating procedure specifically
for the Worm algorithm in the next section.

\subsection{The Worm algorithm}
\label{SEC:WAGF}
\begin{center}
\begin{figure}[h]
\begin{center}
\includegraphics[width=0.8\linewidth]{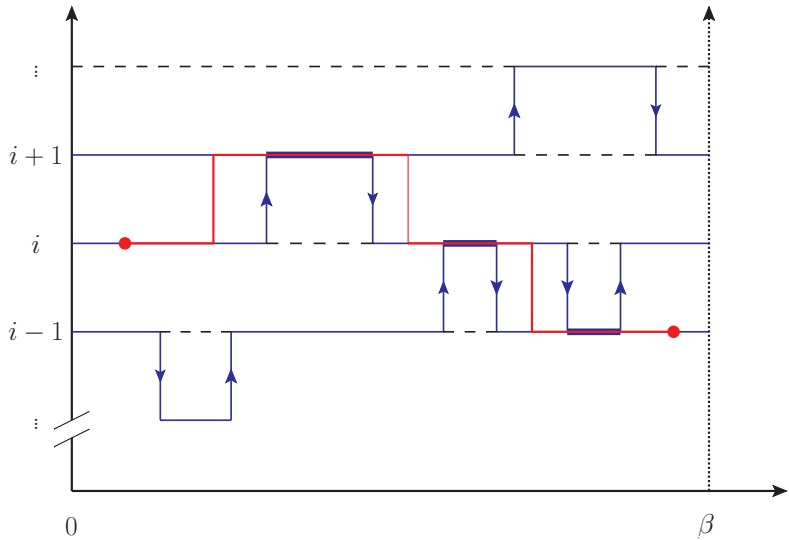}
\end{center}
\caption{Configuration of the $CP_{\rm g}$ space, the red disconnected line represents the worm.}
\label{FIG:Configuration_Worm}
\end{figure}
\end{center}

The Worm algorithm, originally developed by Prokof'ev, Svistunov and Tupitsyn \cite{BB:Worm,BB:Worm2}, works in an enlarged configuration space, in which
one allows one disconnected worldline, the worm, drawn as a red line in Fig. \ref{FIG:Configuration_Worm}. This is equivalent to work in the Grand-Canonical ensemble, as we shall discuss in Sec. \ref{SSEC:Updating}. The disconnected worldline allows to efficiently collect statistics for calculating the Matsubara Green function, defined as
\begin{equation}
\label{EQ:MatsubaraGreen}
G(j,\tau) = \langle \hat{\mathcal{T}}_\tau \hat{b}_{\rm i+j}(\tau_{\rm 0} + \tau)\hat{b}_{\rm i}^\dag (\tau_{\rm 0})\rangle,
\end{equation}
where $\hat{\mathcal{T}}_\tau$ is the time-ordering operator, $\tau_{\rm 0}$ and $\tau$ are two points in imaginary time, $i$ and $j$ are two sites of the lattice, and the symbol $\langle . \rangle$ stands for the statistical average of the expectation value of an operator. Due to space and imaginary time translational invariance of the system, the Green function Eq. (\ref{EQ:MatsubaraGreen}) does not depend on $i$ and $\tau_{\rm 0}$. The configuration space of the Matsubara Green function is called the $CP_{\rm g}$ space, and it is easy to see that the only difference between configurations contributing to the partition function $Z_{\scriptscriptstyle \rm eBH}$ and those contributing to the Green function $G$ is that, for the latter, one of the worldlines starts at $(i,\tau_0)$ and ends at $(i+j,\tau_0 + \tau)$, i.e. the worldline is disconnected.

\subsubsection{Updating procedures}{--- }
\label{SSEC:Updating}
Let us now discuss the updating procedure of the Worm algorithm, that is when the system is in a certain configuration $\nu$ and the algorithm
has to generate randomly a new configuration $\nu^\prime$ to collect statistics for evaluating the observables of interest.

Apart from the creation of a worm, which is done in the $CP$ space, all other
updates are done in the $CP_{\rm g}$ space through the two ends of the worm. Nearly all updates are done in pairs. One can
picture the updating scheme as sequence of 'drawing' and 'erasing' procedures, happening
at the end points of the worm. Below we list and describe the four types of updates the Worm algorithm uses to simulate single component Bose-Hubbard models.

\begin{center}
\begin{figure}[h]
\begin{center}
\includegraphics[width=0.9\linewidth]{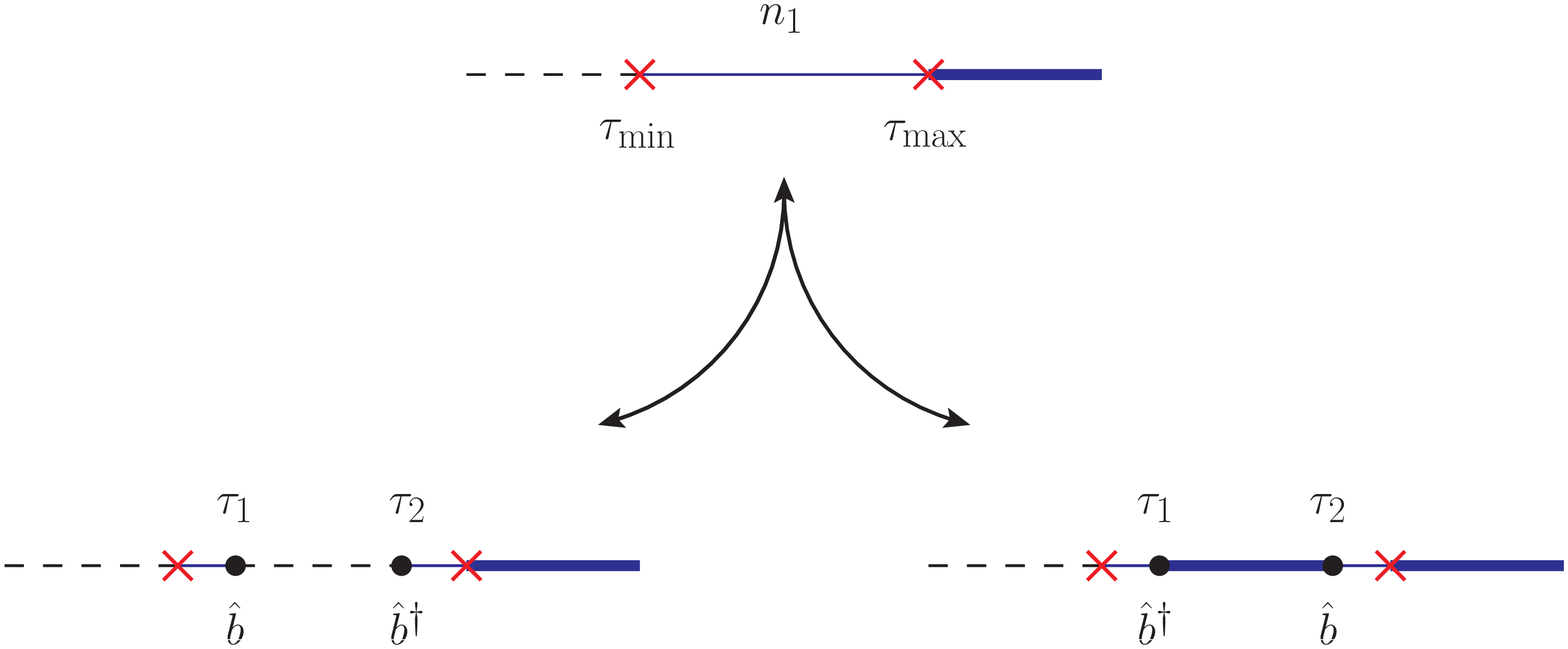}
\end{center}
\caption{Creation/Deletion of a worm. In the create worm update an interval is randomly selected (top), and two points $\tau_{\rm 1}$ and $\tau_{\rm 2}$, which will become the two extremities of the worm, are randomly chosen. Then one can either delete a piece of worldline (bottom left) or draw a piece of worldline (bottom right) with the constraint that the Hilbert space is observed.}
\label{FIG:WormCreation}
\end{figure}
\end{center}

\paragraph{Creation of a worm --- }
Creating a worm is the only update performed in the $CP$ space, therefore the starting point is a configuration $\nu$ belonging to $CP$. Each configuration can be thought as divided into intervals, each interval being delimited by kinks shown in Fig. \ref {FIG:WormCreation} as crosses (or worm extremities for configurations in the $CP_{\rm g}$ space).
In the create worm update one of the intervals of $\nu$, delimited by $\tau_{\rm min}$ and $\tau_{\rm max}$ (see interval $n_1$ in Fig. \ref {FIG:WormCreation}), is randomly selected. Then the algorithm suggests at random two points $\tau_{\rm 1}$ and $\tau_{\rm 2}$ within $n_1$, which will be the worm extremities (indicated by plain dots in Fig. \ref {FIG:WormCreation}), with the constraint $\tau_{\rm min} < \tau_{\rm 1} < \tau_{\rm 2} < \tau_{\rm max}$. With equal probability one suggests to draw or delete the piece of worldline delimited by $\tau_{\rm 1}$ and $\tau_{\rm 2}$, with the constraints that the resulting configuration belongs to the Hilbert space, i.e. it is not possible to erase from an empty interval or to draw on an interval which has reached the maximum occupation number allowed, if any. The worm is therefore created and all other updates will take place through its two extremities.
\paragraph{Deletion of a worm --- }
In analogy, the update opposite to creation of a worm is the deletion of a worm. It can only take place in the $CP_{\rm g}$ space and only if the two extremities of
the worm belong to the same interval.

\paragraph{Time shift --- }
This is the simplest of the updates and it consists of moving one of the extremities of the worm in the imaginary time direction, such as to lengthen or shorten the size of the worm.
The algorithm selects at random the imaginary time instant to which the extremity of the worm will be moved.

\begin{center}
\begin{figure}[h]
\begin{center}
\includegraphics[width=0.9\linewidth]{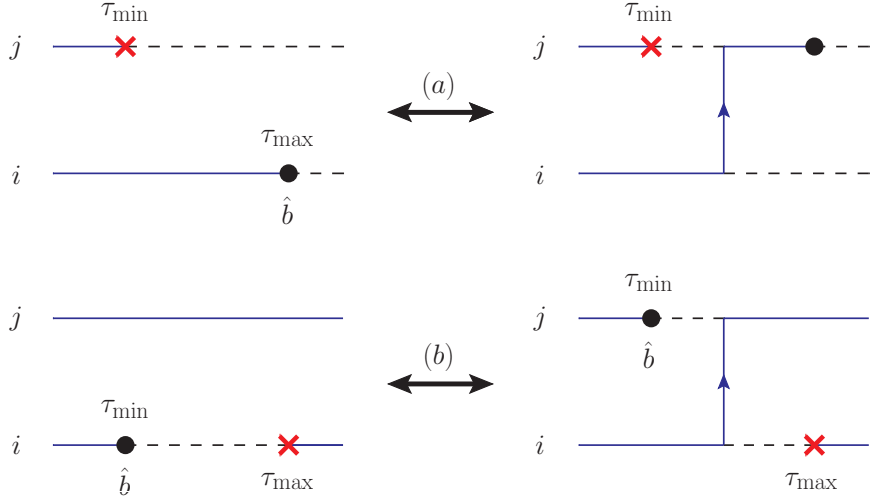}
\end{center}
\caption{Sketch of space shift updates that create or delete kinks. In the space shift left (a), one kink is created/deleted backward in the imaginary time, i.e. to the left of the worm extremity; in the space shift right (b) the kink is created/deleted to the right, i.e. forward in the imaginary time.}
\label{FIG:SpaceShift}
\end{figure}
\end{center}

\paragraph{Space shift --- }
This update changes the number of kinks and it consists of creating or deleting a kink to the left (space shift left) or to the right (space shift right) of the worm extremity. Fig. \ref{FIG:SpaceShift}(a) shows the creation/deletion of a kink backward in imaginary time, i.e. the space shift left. In the creation update a nearest neighbor of the site to which the worm extremity belongs is selected at random and the kink is inserted at an imaginary time instant within the interval delimited by $\tau_{\rm min}$ and $\tau_{\rm max}$ [see Fig. \ref{FIG:SpaceShift}(a)], i.e. with the requirement that the created (or deleted) kink does not interfere with any other interval.
The last update, the space shift right shown in Fig. \ref{FIG:SpaceShift}(b), is equivalent to the left one with the only difference that the kink is inserted or
deleted to the right of the worm extremity, i.e. forward in imaginary time.

These are all the updates performed by the WA. It is straightforward to see that the WA works in the Grand Canonical ensemble, i.e. allows to change the number of worldlines present in the configurations.  The chemical potential becomes an input parameter which fixes the average particle number. For example, suppose the algorithm starts with an initial configuration $\nu$ of zero particles
in the system. From this configuration the only possible update is to create a worm by drawing a piece of worldline. Then, trough the space shift and time shift updates, the worldline will eventually close on itself, corresponding to the insertion of one particle in the system.

\subsubsection{Advantages of the Worm algorithm}{--- }
\label{SEC:Advantages}
The updates described above are all \emph{local} and allow
to draw/erase any line, and create kinks between the sites.
Although only configurations belonging to the $CP$ space
contribute to the evaluation of the partition function,
by using the enlarged
configuration space $CP+CP_{\rm g}$ the intermediate configurations with one disconnected loop allow to
efficiently collect statistics for the Green function. For an
algorithm working in the $CP$ space only, instead, collecting
statistics for the Green function results computationally very
expensive.

Another advantage of the WA is that it does not suffer
from critical slowing down in the vicinity of a critical point. In
the critical region, a system develops long range
correlations, and in most cases an algorithm based on local updates
results very inefficient in simulating such a system for which the
relevant degrees of freedom are non-local. This results in the
divergence of the autocorrelation time with the system size.
Although the WA performs local updates, it overcomes this problem by using the drawing and erasing
updating procedures through the worm ends, which are directly linked to
the critical modes (long range order in $G(j,\tau)$).
As a result generating independent configurations in the critical region is very efficient.

\begin{figure}[h!]
\begin{center}
\includegraphics[width=1\linewidth]{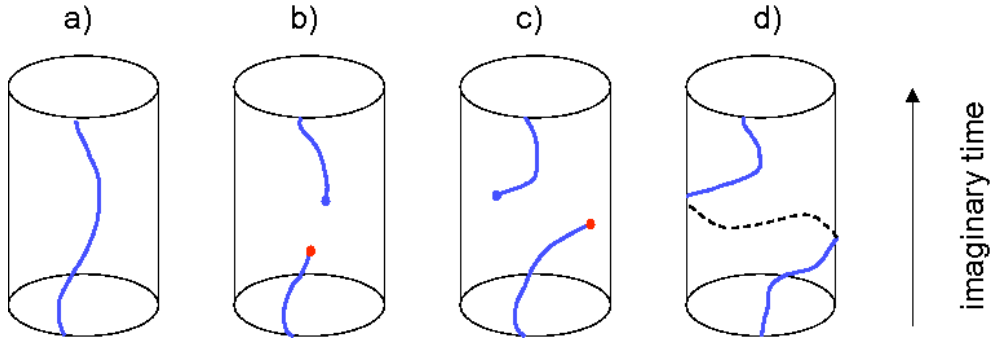}
\caption{One-dimensional system with (a) zero and
(d) one winding number(s). (b)-(c) sketch of how the WA is able to go from (a) to (d).}
\label{WN}
\end{center}
\end{figure}

The WA is also efficient in sampling
topologically different configurations and configurations which
are separated by an energy barrier. This property is a necessary condition in order to maintain ergodicity.
An example of two topologically different
configurations is shown in Fig. \ref{WN}, where a one-dimensional
system with one particle (worldline) is considered. Periodic
boundary conditions in time and space apply, i.e. the system is a
torus where the bottom and top facets of the cylinder are glued
together. Fig.~\ref{WN}(a) represents a configuration with zero
winding numbers, i.e. the worldline does not `wind' in space. 
Fig.~\ref{WN}(d), instead, represents a configuration with
one winding number, i.e. the worldline winds once in space. An algorithm based on local updates which only works in the
$CP$ space would not allow to sample configurations with different
winding numbers, unless a global
update which introduces a winding number at once, is introduced.
The WA, instead, can easily go from
configuration of Fig.~\ref{WN}(a) to configuration
Fig. \ref{WN}(d) (see a sketch in Fig. \ref{WN}(b)-(c)).

Being able to sample configurations with different winding
numbers is crucial in order to simulate SF systems. It was shown
in \cite{BB:Ceperley}, that the superfluid stiffness can be
extracted from the statistics of winding numbers
\begin{equation}
\rho_s=\frac{T\langle\textbf{W}^2\rangle}{dL^{d-2}}\; ,
\label{SF_formula}
\end{equation}
where $T$ is the temperature, $L$ the system size, $d$ the
dimensionality, and $\textbf{W}^2=\sum_{i=1}^dW_i^2$, with $W_i$
being the winding number in the coordinate $i$.

\section{Quantum Monte Carlo studies of dipolar gases}
\label{CH:PaperQuantumPhase}

Quantum Monte Carlo is one of the most powerful
methods we have to study equilibrium properties of strongly
interacting many-body quantum systems.
In the literature, there is a large amount of work devoted to the study of dipolar gases
with Quantum Monte Carlo techniques.
From self-assembled floating lattices, provided by trapped polar molecules \cite{BB:Pupillo01},
to the possibility of tuning, and shaping the long-range interaction potential
of polar molecules \cite{BB:Pupillo02}, to self-organized mesoscopic structures of matter waves
in zigzag chains \cite{BB:Greg01}, to the spectrum of the elementary excitation that can exhibit a roton minimum \cite{BB:Boronat02, BB:Boronat01},
to the emergence of an emulsion phase in triangular lattices \cite{BB:Pollet}.
The ones listed above are just a few of the outstanding properties of dipolar gases, which have been investigated
with various Monte Carlo techniques.

Based on the Path Integral Monte Carlo and the Worm algorithm, in \cite{BB:Barbara2}, we have studied the ground state properties of dipolar hard-core Bosons confined in a 2D square lattice of linear size $L$, satisfying periodic boundary conditions. The system is described by the extended Bose-Hubbard Hamiltonian (\ref{EQ:MCH}), where no cut-off in the dipolar interaction potential is used.
\begin{center}
\begin{figure}[t]
\begin{center}
\includegraphics[width=0.9\linewidth]{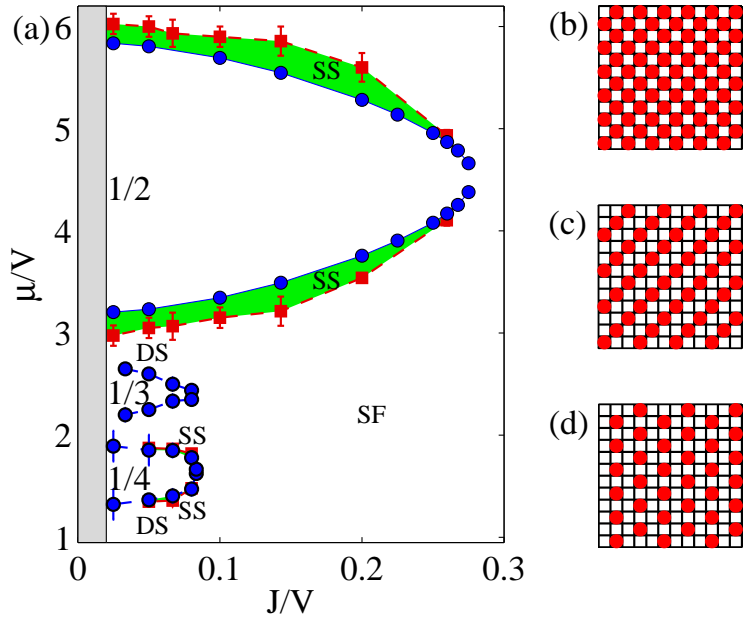}
\end{center}
\caption{Phase diagram corresponding to
the Hamiltonian Eq. (\ref{EQ:MCH}) as a function of $\mu$ and
$J$ at zero temperature. Lobes: Mott solids (densities indicated);
SS: supersolid phase; SF: superfluid phase. DS: parameter
region where devil' s staircase is observed. Panels
(b-d): sketches of the groundstate configuration for the Mott
solids in panel (a), with  $\rho = 1/2$, $1/3$ and $1/4$, respectively.}
\label{FIG:PhaseDiagramD}
\end{figure}
\end{center}

\subsection{Incompressible and supersolid phases}
\label{SEC:Incompressible}
The incompressible and supersolid phases are both characterized by
a finite value of the structure factor, defined as
\begin{equation}
S(\myvec{k}) = \sum_{\myvec{r},\myvec{r}^\prime} \frac{\langle n_{\myvec{r}} n_{\myvec{r}^\prime}\rangle}{N} e^{i \myvec{k}\dot (\myvec{r} - \myvec{r}^\prime)},
\end{equation}
with $\myvec{k}$ the reciprocal lattice vector, $n_{\myvec{r}}$ the density at position $\myvec{r}$, and $N$ the total number of particles.
While for the incompressible phases the superfluid fraction vanishes $\rho_s=0$, the supersolid phase is characterized by a finite value of $\rho_s$, indicating the presence of superfluid.

Our main results in the absence of harmonic confinement $\Omega = 0$, are summarized in Fig. \ref{FIG:PhaseDiagramD}, where we show the zero temperature phase diagram of the system, in the $J$ vs. $\mu$ plane,
in the range $J/V > 0.02$, and $1 < \mu/V < 6$ indicated by the unshaded area.
For finite $J$,
three main solid Mott lobes emerge with filling factor $\rho = 1/2$, $1/3$,
and $1/4$, named checkerboard (CB), stripe (ST), and star
(SR) solids, respectively. The corresponding groundstate
configurations are sketched in panels (b-d). We find that the CB solid is the most robust
against hopping and doping, and thus it extends furthest
in the $J$ vs. $\mu$ plane.

For large enough $J/V$, the low-energy phase is superfluid
(SF), for all $\mu$. At intermediate values of $J/V$,
however, we find that by doping the Mott solids either
with vacancies (removing particles) or interstitials
(adding extra particles) a supersolid phase (SS) can be
stabilized, with coexisting superfluid and crystalline orders.
 Instead, we find no evidence of SS in the absence of doping.
The green shaded area above and below the CB lobe boundaries
in Fig. \ref{FIG:PhaseDiagramD},
correspond to a SS obtained by doping the CB crystal with interstitials, and vacancies respectively.
Remarkably, the long-range interactions stabilizes the supersolid in a
wide range of parameters, in fact for one, or two nearest neighbors in the dipolar interaction range, no stable CB
SS was found for $\rho<1/2$ \cite{BB:Sengupta}.

\begin{center}
\begin{figure}[h!]
\begin{center}
\includegraphics[width=0.8\linewidth]{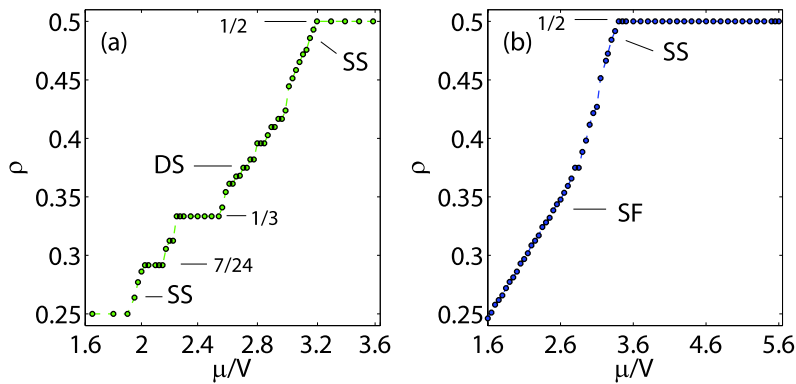}
\end{center}
\caption{$\rho$ vs. $\mu$. (a): Solids and SS for a
system with linear size $L = 12$ and $J/V = 0.05$. Some $\rho$ are
indicated. (b): SF and vacancy-SS for $L = 16$ and $J/V = 0.1$.}
\label{FIG:Devil}
\end{figure}
\end{center}

Interestingly, we find evidence for incompressible
phases in addition to those corresponding to the lobes
in Fig. \ref{FIG:PhaseDiagramD}. This is shown in Fig. \ref{FIG:Devil}, where the particle
density $\rho$ is plotted as a function of the chemical potential
$\mu$.
In Fig. \ref{FIG:Devil}, a continuous increase of $\rho$
as a function of $\mu$ signals a compressible phase, while a
solid phase is characterized by a constant $\rho$ for increasing
$\mu$. Panel (a), corresponding to $J/V = 0.05$, shows
a series of large constant-density plateaux connected by
a progression of smaller steps and regions of continuous
increase of  $\rho$. Here, the main plateaux correspond to the
Mott lobes of Fig. \ref{FIG:PhaseDiagramD}, while the other steps correspond to
incompressible phases, with a fixed, integer, number of
particles. This progression of steps is an indication of a
Devil's-like staircase in the density, which was
discussed in \cite{BB:Cooper} for a one dimensional system.
Instead, for $J/V = 0.1$ in panel (b), no evidence of such a phase is found.

\section*{Acknowledgments}
This tutorial was supported by Spanish MEC (FIS2008-00784, QOIT),
EU projects AQUTE and NAMEQUAM, and ERC grant QUAGATUA. M.L.
acknowledges also Alexander von Humboldt Stiftung and Hamburg
Prize for Theoretical  Physics.  It is a great pleasure for us to
thank all the people in the quantum optics theory group of ICFO
for interesting discussions. We are especially grateful to K.
Rz{\c a}{\.z}ewski, who invited us to write this paper. C. T.
thanks C. Menotti, and M. Lewenstein for their constant support,
and patience, they showed him during these years, the result of
which is not only in this work. C. T. thanks B. Capogrosso-Sansone
and G. Pupillo, for their guide in the Quantum Monte Carlo work,
and Peter Zoller for the kind hospitality in Innsbruck. This work
was partially written at the Indian Association for the
Cultivation of Science, in Calcutta, while C. T. was visiting K.
Sengupta. The warm hospitality he showed him is unique, and C. T.
thanks him together with all the people of the theoretical physics
department.

\appendix

\section{Parametrization}
\label{SEC:Parametrization}
Through the variation of $(q,P)$, we aim at describing the transition between the states $\ket{\Phi}_{\rm initial}$ and $\ket{\Phi}_{\rm final}$, as
for example the one schematically represented in Fig. \ref{FIG:Connection}(c). During this process there are sites initially occupied that become empty, like the blue-framed site of
Fig. \ref{FIG:Connection}(c), which we call the (B)-sites, and vice versa, like the site on the left of B, which is initially empty and occupied at the end, and we name the (A)-sites.
When the initial and final states are non-degenerate, as for example the case sketched in Fig. \ref{FIG:Actions}(d), there are also sites that do not change and remain either full (F) or empty (E).

During this process, the Gutwiller amplitudes (\ref{EQ:GWAmplitudesStability}) have to be normalized at each site, and the total number of particles has to be conserved, namely
\begin{eqnarray}
\label{EQ:Normalization}
|f_{\rm 0}^{(i)}|^2 + |f_{\rm 1}^{(i)}|^2 &=& 1, \; \forall i \\
\sum_{i=1}^{N_S} |f_{\rm 1}^{(i)}|^2  &=& N, \label{EQ:Conservation}
\end{eqnarray}
where $N_S$ is the total number of sites.
We choose $(q,P) \equiv (q_{\rm 0}^{\scriptscriptstyle \mathrm{B}},P_{\rm 0}^{\scriptscriptstyle \mathrm{B}})$ to be the variational parameters of the blue-framed site,
and the normalization condition (\ref{EQ:Normalization}) together with the conservation of the number of particles (\ref{EQ:Conservation}) between A and B give us three coupled equations
\begin{equation}
\label{EQ:ConditionsCC}
\eqalign{
 q^2 - P^2 + (q_{1}^{\scriptscriptstyle \mathrm{B}})^2 - (P_{1}^{\scriptscriptstyle \mathrm{B}})^2 &= 2 \\
 (q_{0}^{\scriptscriptstyle \mathrm{A}})^2 - (P_{0}^{\scriptscriptstyle \mathrm{A}})^2 +
 (q_{1}^{\scriptscriptstyle \mathrm{A}})^2 - (P_{1}^{\scriptscriptstyle \mathrm{A}})^2 &= 2 \\
 (q_{1}^{\scriptscriptstyle \mathrm{B}})^2 - (P_{1}^{\scriptscriptstyle \mathrm{B}})^2 +
 (q_{1}^{\scriptscriptstyle \mathrm{A}})^2 - (P_{1}^{\scriptscriptstyle \mathrm{A}})^2 &= 2.
 }
\end{equation}
As explained in \cite{BB:Trefzger01}, we make use
of the following Ansatz
\begin{equation}
\label{EQ:Ansatz}
\eqalign{
q_{1}^{\scriptscriptstyle \mathrm{A}} &= q \\
P_{1}^{\scriptscriptstyle \mathrm{A}} &= P \\
P_{0}^{\scriptscriptstyle \mathrm{A}} &= P_{1}^{\scriptscriptstyle \mathrm{B}} = -P\\
q_{0}^{\scriptscriptstyle \mathrm{A}} &= q_{1}^{\scriptscriptstyle \mathrm{B}}.
}
\end{equation}
It is clear that a situation where site A is empty and site B is full corresponds to the values $(q,P) = (0,0)$, while at $(q,P) = (\sqrt{2},0)$ the contrary is true.
For degenerate initial and final states as in the case of Fig. \ref{FIG:Connection}(c), the remaining sites they either behave like A or B, which implies another set of conditions summarized as follows
\begin{equation}
\label{EQ:ConditionsCC02}
\eqalign{
q_0^{(i)}  &= q_{0}^{\scriptscriptstyle \mathrm{A(B)}} \\
P_0^{(i)} &= P_{0}^{\scriptscriptstyle \mathrm{A(B)}} \\
q_1^{(i)}  &= q_{1}^{\scriptscriptstyle \mathrm{A(B)}} \\
P_1^{(i)} &= P_{1}^{\scriptscriptstyle \mathrm{A(B)}} \\
}
\end{equation}
depending on whether the site $i$ is initially empty (A) or occupied (B). Instead, when the initial and final states are non-degenerate, as is the case considered in Fig. \ref{FIG:Actions}(d), there are also sites that we assume not to change and remain either full (F) or empty (E). For these sites the constraints are respectively given by
\begin{equation}
\label{EQ:ConditionsCC03}
\eqalign{
q_{0}^{\scriptscriptstyle \mathrm{F}} &= 0 \\
q_{1}^{\scriptscriptstyle \mathrm{F}} &= 2 \\
P_{0}^{\scriptscriptstyle \mathrm{F}} &= P_{1}^{\scriptscriptstyle \mathrm{F}} = 0,
}
\end{equation}
and
\begin{equation}
\label{EQ:ConditionsCC04}
\eqalign{
q_{0}^{\scriptscriptstyle \mathrm{E}} &= 2 \\
q_{1}^{\scriptscriptstyle \mathrm{E}} &= 0 \\
P_{0}^{\scriptscriptstyle \mathrm{E}} &= P_{1}^{\scriptscriptstyle \mathrm{E}} = 0.
}
\end{equation}
All these conditions (\ref{EQ:ConditionsCC})-(\ref{EQ:ConditionsCC04}), which we name $\mathcal{C}_c$, enter explicitly into the calculation of Hamiltonian (\ref{EQ:HamiltonianStationary}).

\section*{References}
\bibliographystyle{iopart-num}
\bibliography{Tutorial_v4}

\end{document}